# Project X Broader Impacts

**July 2013**

**Edited by:**

David Asner/Pacific Northwest National Laboratory
Pushpa Bhat/Fermi National Accelerator Laboratory
Stuart Henderson/Fermi National Accelerator Laboratory
Robert Plunkett/ Fermi National Accelerator Laboratory

# I INTRODUCTION

The proposed Project X proton accelerator at Fermilab, with multi-MW beam power and highly versatile beam formatting, will be a unique world-class facility to explore particle physics at the intensity frontier. Concurrently, however, it can also facilitate important scientific research beyond traditional particle physics and provide unprecedented opportunities in applications to problems of great national importance in the nuclear energy and security sector.

The high intensity proton beam from Project X with an optimally designed target station can serve as an experimental facility for studies of electric dipole moments of neutrons and nuclei, fusion and fission energy R&D, irradiation testing of materials, and production of certain radioactive isotopes. High power proton accelerators are also envisioned as drivers for subcritical nuclear reactor systems for the transmutation of nuclear waste and generation of electrical power; these are significant and vital applications. Project X can be used as a platform to demonstrate some of the key concepts and technologies of Accelerator Driven Systems (ADS). The high power proton beams can also be used to produce intense neutron beams or low energy muons for materials science research.

The rich and diverse research program with Project X, in particle physics and beyond, has been described in detail in Part II of this book. In this Part, we describe the broader impacts – applications in the nuclear energy sector including R&D for accelerator driven systems, nuclear physics, and materials science. We briefly outline them here.

### *Applications in Fusion Energy Science & Nuclear Energy*

While the primary driver for the development of Project X is particle physics research, the high beam power and intensity in conjunction with a flexible target station offers the potential to address important questions in nuclear energy (irradiation testing of fast reactor structural materials, integral testing of fast reactor fuel rodlets, separate-effects testing of fission reactor fuel materials), fusion energy science (irradiation testing of fusion structural materials), and nuclear physics (cold neutrons, isotope production, nuclear electric dipole moment research).

Advanced nuclear energy systems have the potential to deliver significant improvements in sustainability, safety, reliability and proliferation-resistance relative to the conventional, Generation II nuclear power systems. National needs in Advanced Nuclear Energy Systems have been articulated in a number of recent reports. The Department of Energy's Nuclear Energy Roadmap [1] outlines the four main objectives of the Nuclear Energy R&D Program, as follows:

1. Develop technologies and other solutions that can improve the reliability, sustain the safety, and extend the life of current reactors;
2. Develop improvements in the affordability of new reactors to enable nuclear energy to help meet the Administration's energy security and climate change goals;
3. Develop sustainable nuclear fuel cycles;
4. Understand and minimize the risks of nuclear proliferation and terrorism.

An important cross-cutting technical capability which is needed to advance the goals of the program is that of materials irradiation at relevant neutron fluxes, energy spectra, and volumes.

The Fusion Energy community has emphasized as a central thrust the goal of developing the materials science and technology needed to harness fusion power. A report on research needs [2] highlights the need to "establish a fusion-relevant neutron source to enable accelerated evaluations of the effects of radiation-induced damage to materials."

The materials science community articulated in *Basic Research Needs for Advanced Nuclear Energy Systems* [3] the fundamental challenge of understanding and controlling chemical and physical phenomena "…from femtoseconds to millennia, at temperatures to 1000°C and for radiation doses to hundreds of displacements per atom. This is a scientific challenge of enormous proportions, with broad implications in the materials science and chemistry of complex systems."

These research needs highlight the tremendous challenges in developing materials that can withstand high temperatures and extreme radiation environments. Figure I-1 shows the operating regions in material temperature and displacement damage (measured in lattice displacements per atom) for current fission reactors and future fission and fusion reactors. Meeting the challenges of developing novel materials for these environments will require new, very intense, neutron sources as essential tools.

There is substantial interest in the U.S. and around the world in the development of accelerator-based irradiation sources. Modern high-power superconducting continuous wave linear accelerators are capable of producing prototypical conditions that simulate the steady-state operation of fission and fusion reactors at relevant neutron fluxes. The high beam power available from Project X provides neutron fluxes that rival or exceed those obtained at reactor-based irradiation sources. An optimized Target Station at Project X, driven by a MW-class beam can support Nuclear Energy initiatives in Fuel Cycle Technologies, Nuclear Reactor Technologies, and Advanced Modeling and Simulation. The materials irradiation testing capabilities of such a Target Station could enable, for example, efforts to ensure the sustainability and safety of the current fleet of reactors for lifetime extensions, development of new higher performance and safer reactor fuels and materials, development of innovative economical small reactors, development of new advanced reactor concepts such as those using liquid metal or molten salt coolants, development of transmutation fuels for reducing legacy wastes requiring deep geologic storage.

The continuous wave (CW) proton beam from Project X will be a unique facility in the world that can provide an unprecedented experimental and demonstration facility for fission and fusion energy and nuclear physics R&D, filling an important gap in the irradiation testing needs of the U.S. and the world. Such a facility would provide a unique opportunity for cooperation between the DOE Office of Nuclear Energy and the DOE Office of Science through sharing of capital infrastructure and resources.

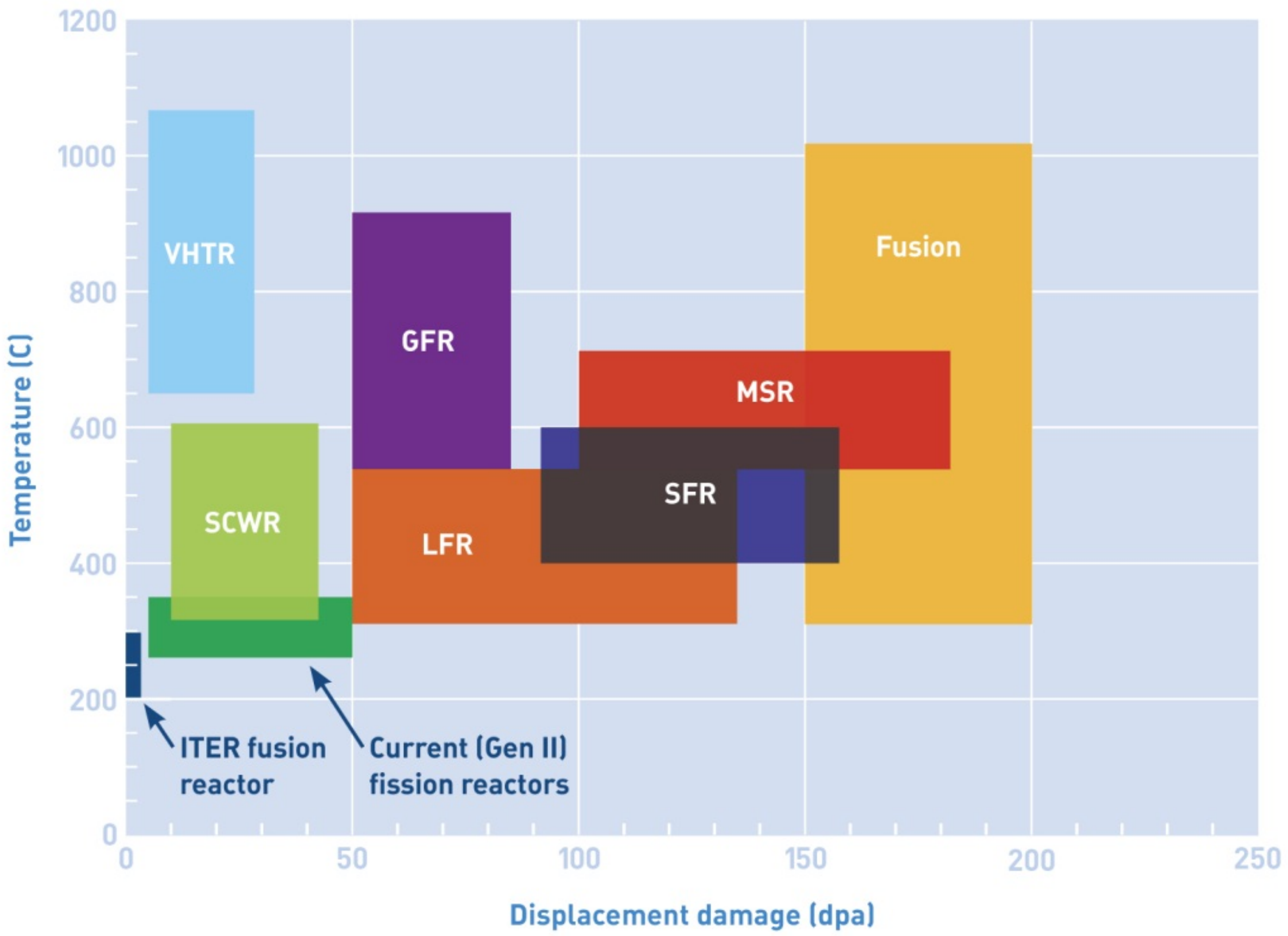

**Figure I-1**: Operating regions in material temperature and displacement damage, measured in lattice displacements per atom (dpa), for current fission reactors and future fission and fusion reactors. Fission reactors include very-high-temperature reactors (VHTR), supercritical water-cooled reactors (SCWR), gas-cooled fast reactors (GFR), lead-cooled fast reactors (LFR), sodium-cooled fast reactors (SFR), and molten-salt reactors (MSR). From: S.J. Zinkle, OECD/NEA Workshop on Structural Materials for Innovative Nuclear Energy Systems, Karlsruhe, Germany, June 2007.

A related technology thrust is in the development of advanced externally driven reactor systems as a means for transmutation of waste from power reactors. Such accelerator-driven system (ADS) concepts have been considered in the U.S. in the past and are under study worldwide, principally in China, Europe and India. An important topic in the Nuclear Energy Roadmap [1] is R&D in sustainable fuel cycle options, for which externally driven transmutation systems are one element. Should the priorities in the U.S. change in the future in favor of developing externally driven reactor technologies, Project X would be an ideal R&D platform for ADS.

An accelerator-driven subcritical reactor system requires a high power, highly-reliable, proton accelerator coupled to a subcritical core through a spallation target system. A technical assessment of accelerator and spallation target technology [4] was recently carried out in order to assess the readiness of the technology for ADS applications. The study concluded that the technical developments carried out over the last ~15 years place this technology at a point where it is ready to support an ADS demonstration mission. In fact, the European ADS program has reached similar conclusions and is embarking on the construction of a MW-class accelerator-driven subcritical system to principally serve an irradiation mission [5].

Project X provides an ideal test-bed for the development of ADS concepts and technologies. The MW-class continuous wave beam in the GeV-range is precisely the beam which is needed for ADS development. An optimized target station at Project X will be capable of supporting the development and demonstration of ADS and its associated technologies and concepts. ADS is discussed in more detail in Appendix -1.

.

### *Applications in Nuclear Physics*

The continuous wave proton beam from Project X along with a versatile target station can provide unique opportunities for highly sensitive experiments with neutrons to test fundamental symmetries and models of physics beyond the standard model (SM) – especially valuable if new physics is discovered at the LHC. Copious quantities of cold and ultra cold neutrons and rare radioactive isotopes such as $^{225}$Ra, $^{223}$Rn, $^{211}$Fr can be produced and used in searches for electric dipole moments (EDM) of neutrons and nuclei. EDMs, particularly the neutron and heavy atom EDMs are highly sensitive probes for sources of CP violation beyond those present in the SM.

Project X spallation facility also presents an opportunity to probe neutron-antineutron (NNbar) oscillations with free neutrons, with unprecedented sensitivities. Improvements would be achieved by creating a unique facility, combining a dedicated high intensity cold neutron source with advanced neutron optics technology and special detectors with demonstrated capability to detect antineutron annihilation events with minimal or no background. Existing neutron sources at research reactors and spallation sources cannot provide the required space and the necessary access to the cold source that would enable set-up of a highly sensitive NNbarX oscillation search experiment. A dedicated source devoted exclusively to fundamental neutron physics at Project X represents an exciting tool to explore fundamental nuclear and particle physics questions accessible through cold neutrons.

### *Applications in Materials Science*

Materials Science needs have been articulated in the extensive Basic Research Needs reports which are based on a series of DOE Basic Energy Sciences (BES) workshops [3].

Project X will offer the possibility of addressing problems in materials science through the creation of a facility to exploit polarized, low-energy muons created via the decay of pions, which can be copiously produced by Project X beams. Such a facility would be unique in the U.S., and significantly increase extremely limited global capacity for the technique, known as Muon Spin Rotation (μSR).

μSR is a powerful probe of materials which has made important contributions to a wide range of topics of interest to a broad cross-section of the scientific community. Strongly represented in the user community are researchers studying superconductivity, quantum magnetism and chemistry, and $\mu$SR has been successfully used to make considerable advances in these fields. Important advances have also been made in the study of semiconductors, biological and soft-matter systems, and quantum diffusion. In recent years, the advent of "ultralow energy" $\mu$SR beams has allowed for stunning contributions in the study of thin films, multi-layers and surface science.

μSR research facilities exist in Canada, Switzerland, the UK (ISIS/RAL) and Japan (first at KEK and now at J-PARC). There is a plan to construct a facility in South Korea (RISP). All existing facilities are over-subscribed and must reject many good proposals. There has been no capability for experiments utilizing μSR in the United States since the closure of the LAMPF muon facility in the 1990's. Such a program utilizing the high power beams provided by Project X would be a cost-effective approach to establishing world-leading μSR capability in the U.S., in support of the research needs of the materials science community.

## References

[1] Nuclear Energy Research and Development Roadmap Report To Congress, April 2010, US DOE: http://energy.gov/sites/prod/files/NuclearEnergy_Roadmap_Final.pdf

[2] Research Needs for Magnetic Fusion Energy Sciences: http://burningplasma.org/web/ReNeW/ReNeW.report.web2.pdf

[3] Basic Research Needs Workshop Reports: http://science.energy.gov/bes/news-and-resources/reports/basic-research-needs/

[4] Accelerator and Target Technology for Accelerator Driven Transmutation and Energy Production: http://science.energy.gov/~/media/hep/pdf/files/pdfs/ADS_White_Paper_final.pdf

[5] H. Ait Abderrahim et. al., Nuclear Physics News, vol. 20, no. 1, 2010, p. 24.

# II PROJECT X SPALLATION AND IRRADIATION FACILITY

**D.W. Wootan, D.M. Asner, M.A. Peterson, D. Senor**
**Pacific Northwest National Laboratory**

# Acronyms and Abbreviations

| | |
|---|---|
| ADS | accelerator-driven subcritical systems |
| ANL | Argonne National Laboratory |
| ATR | Advanced Test Reactor |
| CW | continuous wave |
| DOE | U.S. Department of Energy |
| DOE-NE | U.S. Department of Energy Office of Nuclear Energy |
| EIS | Environmental Impact Statement |
| ESS | European Spallation Source |
| FES | Office of Fusion Energy Sciences |
| FFMF | Fission Fusion Materials Facility |
| FNAL | Fermi National Accelerator Laboratory |
| INL | Idaho National Laboratory |
| ISIS TS1 | ISIS Target Station 1 |
| JAEA | Japan Atomic Energy Agency |
| J-PARC | Japan Proton Accelerator Research Complex |
| LANL | Los Alamos National Laboratory |
| LANSCE | Los Alamos Neutron Science Center |
| LBE | liquid lead bismuth eutectic |
| LBNE | Long Baseline Neutrino Experiment |
| linac | Project X Linear Accelerator |
| MEGAPIE | Megawatt Pilot Experiment |
| MTS | Materials Test Station |
| ORNL | Oak Ridge National Laboratory |
| PIE | post-irradiation examination |
| PNNL | Pacific Northwest National Laboratory |
| PSI | Paul Scherrer Institute |
| PXES | Project X Energy Station |
| R&D | research and development |
| SINQ | Swiss Spallation Neutron Source |
| SNS | Spallation Neutron Source |
| TEF-T | Transmutation Experimental Facility-ADS Target Test Facility |
| UK | United Kingdom |

## II.1 Summary

Project X is a high intensity continuous wave (CW) proton beam accelerator [1] proposed to be built at Fermi National Accelerator Laboratory (FNAL) in the next decade as described in the RDR (Part I) of this book. The recent papers “Accelerator and Target Technology for Accelerator Driven Transmutation and Energy Production” [2], “Accelerators for America’s Future” [3], “Proceedings of the Workshop on Applications of High Intensity Proton Accelerators” [4], and “Fermilab Project-X Nuclear Energy Application: Accelerator, Spallation Target and Transmutation Technology Demonstration“ [5] have endorsed the idea that the next generation particle accelerators could enable technological breakthroughs for nuclear physics and fission and fusion energy applications. The MW scale CW proton beam from Project X as described in the RDR can serve a variety of functions beyond those of traditional particle physics research.

While the primary driver for development of Project X is particle physics research, the high beam power and intensity offers the potential for a flexible target station to address important questions in other fields such as nuclear energy (irradiation testing of fast reactor structural materials, integral testing of fast reactor fuel rodlets, separate-effects testing of fission reactor fuel materials), fusion energy science (irradiation testing of fusion structural materials), and nuclear physics (cold neutrons, and isotopes for nuclear electric dipole moment research). Accordingly, the Project X target station concept has been developed to evaluate the ability of this unique accelerator facility to pursue world-leading materials science and nuclear physics research.

The continuous wave (CW) proton beam from this accelerator will be a unique facility in the world and due to the CW nature of the beam it would provide a reactor-like irradiation environment that can provide an unprecedented experimental and demonstration facility for nuclear physics, fusion and fission energy R&D that can fill an important gap in the irradiation testing needs of the U.S. and the world. Such a facility would provide a unique opportunity for cooperation by sharing resources and capital infrastructure between the DOE Office of Nuclear Energy and the DOE Office of Science.

### II.1.1 Mission Support

A versatile Project X target station could support the DOE Office of Nuclear Energy missions to develop used nuclear fuel management strategies and technologies, develop sustainable fuel cycle technologies and options, and develop new and advanced reactor designs and technologies that advance the state of reactor technology to improve competitiveness and advance nuclear power as an energy resource. The materials irradiation testing capabilities of such a Target Station could enable, for example, efforts to ensure the sustainability and safety of the current fleet of reactors for lifetime extensions, development of new higher performance and safer reactor fuels and materials, development of innovative economical small reactors, development of new advanced reactor concepts such as those using liquid metal or molten salt coolants, development of transmutation fuels for reducing legacy wastes requiring deep geologic storage, and potential investigation of neutron source driven systems as a means for transmutation of waste from power reactors. [6]

A versatile Project X target station could support DOE Office of Science Fusion Energy Science mission goal [22] of developing the scientific understanding required to design and deploy the materials needed to support a burning plasma environment, a key step in the research and development of practical fusion energy applications. " The pursuit of fusion energy embraces the challenge of bringing the energy-producing power of a star to earth for the benefit of mankind. This pursuit is one of the most challenging programs of scientific research and development that has ever been undertaken." The promise is an energy system whose fuel is nearly inexhaustible and results in modest radioactivity and zero carbon emissions to the atmosphere. There is currently no facility available anywhere in the world that can provide fusion-relevant neutron flux and material radiation damage rates in a reasonable volume [7].

A flexible, integrated Project X target station could also support the Office of Nuclear Physics in the DOE Office of Science mission to develop a roadmap of matter that will help unlock the secrets of how the universe is put together. The Project X Target Station will enable a new generation of symmetry-testing experiments with the goal of advancing the understanding of basic nuclear physics phenomena that support fundamental searches for physics beyond the Standard Model. The MW scale CW proton beam from Project X can produce copious quantities of cold neutrons (CN), very cold neutrons (VCN) and ultra-cold neutrons (UCN), as well as special short-lived Ra, Fr, and Rn isotopes that can be utilized in sensitive searches for physics beyond the Standard Model [8]. Examples of potential experimental facilities that could be integrated into the Target Station include searches for neutron-antineutron oscillations (NNbarX) and nuclear electric dipole moments (EDMs). A permanent EDM violates both time reversal symmetry and parity. The existence of an EDM can provide the "missing link" for explaining why the universe contains more matter than antimatter [9].

A possible additional application of the Project X target station could support one of the missions of the Office of Nuclear Physics in the DOE Office of Science – to develop and produce radioactive isotopes. The Project X program could produce isotopes only where there is no U.S. private sector capability or production capacity is insufficient to meet U.S. needs. It is not envisioned that the Target Station would be used in a production mode, with the associated schedule, separations, and yield issues, but rather in a mode to facilitate production of research quantities of unique isotopes that cannot be obtained without the very high neutron energy spectrum or high power proton beam, on a schedule consistent with normal Project X accelerator operations [9].

The Project X Target Station can provide a test bed for accelerator physics and accelerator technology by addressing accelerator target issues such as:

- Beam-Target Interface
- Radiation damage/transmutation reaction impurities
- Remote Handling
- Liquid metal target components
- Solid/rotating metal target components

Materials selected for accelerator targets, diagnostics, structures, and mechanical systems are typically based on limited data from fission reactors and the fusion community. There is potential for a much different response in a high power accelerator due to high flux of high energy particles. The Target Station could allow irradiation testing in environments more prototypical of accelerator targets.

Table II-1 summarizes the capabilities of the Project X Target Station to support the various DOE programs in the Office of Nuclear Energy, and the Office of Science.

| | Fission Reactor Materials (SFR, LFR, HTGR, LWR, MSR) | Fusion Materials | Nuclear Physics Isotopes | Cold Neutrons VCN, UCN |
|---|---|---|---|---|
| **Office of Nuclear Energy**<br>• Fuel Cycle Technologies<br>o Used Fuel Disposition R&D<br>o Fuel Cycle R&D<br>• Advanced Modeling & Simulation<br>• Nuclear Reactor Technologies<br>o LWR Sustainability Program<br>o Advanced Reactor Technologies<br>o Small Modular Reactors<br>o Space Power Systems | x | | | |
| **Office of Science**<br>• Nuclear Physics<br>o Low Energy Nuclear Physics Research<br>o Theoretical Nuclear Physics Research<br>o Isotope Development and Production for Research and Applications | x | | x | x |
| **Office of Science**<br>• Fusion Energy Science<br>o Fusion Materials and Technology | x | X | | |

**Table II-1**: Summary of Programs Benefiting from Project X Target Station

### II.1.2 **Target Station Options**

The basic concept that is proposed for the Project X target station is one beam line of about 1 MW power directed either horizontally or vertically to a liquid lead, liquid lead-bismuth, or solid tungsten spallation target. The proposed Project X Target Station accelerator beam parameters are 0.91 MW beam power, 1 mA beam current, and 1 GeV beam energy. The Project X Stage 1 beam timing for the 1 GeV beam incorporates a 60 msec beam-off period every 1.2 seconds, resulting in a 95% duty factor for the otherwise continuous wave (CW) beam, as described in Appendix-I of the RDR. Two options for the Target Station are being considered, one is an Energy Station focused on DOE Nuclear Energy missions and Fusion Energy Science materials irradiation testing, and the second is an Integrated Target Station that combines the materials testing mission with Nuclear Physics experiments such as ultra-cold neutrons for NNbar and neutron-EDM experiments and production of unique isotopes for atomic-EDM experiments. Various configurations of liquid and solid spallation targets and surrounding irradiation testing configurations are being evaluated. The spallation target produces copious neutrons with an energy spectrum similar to a liquid metal cooled fast reactor. Neutrons produced in the spallation region escape into the surrounding solid lead matrix region cooled by helium gas, liquid water, liquid sodium, liquid lead, or lead-bismuth. This matrix region contains several modules for various types of experiments such as fusion materials irradiation testing, fission reactor materials irradiation, production of nuclear physics isotopes for atomic-EDM searches, or production of cold neutrons for NNbar and n-EDM nuclear physics experiments. Each module contains independent cooling loops and materials that produce appropriate nuclear environments.

The neutron spectrum in these different modules could be tailored to the requirements by using moderating or filtering materials. Preliminary investigations indicate that fairly large volumes (~600 liters) of high neutron flux ($>10^{14}$ n/cm$^2$/sec) can be created that rival or surpass the limited test volumes available in existing high power test reactors. Multiple test modules are envisioned in the target region, surrounding the spallation target, each with an independent test region and coolant loop. Each test module could be designed to be removed and reinstalled independently of the others. Vacuum insulation layers can be used to isolate modules with extreme temperature variations from the matrix region. These reconstitutable modules provide tremendous flexibility in designing tests to meet evolving needs. Extensive instrumentation and temperature control are also key attributes that can be used to provide a testing environment tailored to particular program needs.

Since the moderator and cooling regions of UCN sources utilize only a fraction of the spallation neutrons produced in the target, this capability could be included as one of the Project X Target Station irradiation modules and still allow several other fast spectrum test modules. Relative priorities of research needs could determine the makeup and configuration of test modules.

One advantage of an accelerator based system, for an experiment, over a reactor is the proximity of the instrumentation to the experiment. This makes gas handling and other aspects of instrumented tests a lot easier to design and operate. This also allows real-time measurement capabilities (e.g. direct pressure measurement) that can't be done when the experiment is 33 meters or more away from the instrumentation.

The CW beam produces a high neutron flux and the high duty factor provides a neutron irradiation capability to accumulate fluence comparable to large research reactors, but with the volume and flexibility to tailor the neutron spectrum, temperature, coolant, and structural materials to match a wide variety of both thermal and fast spectrum reactor types. Except for the Swiss Spallation Neutron Source (SINQ) cyclotron accelerator at PSI, the other existing neutron spallation facilities are pulsed systems. They are all designed for producing neutron beams, such as for scattering studies, and not necessarily for materials irradiation. The Los Alamos Neutron Science Center (LANSCE at LANL) and Spallation Neutron Source (SNS at ORNL) facilities are comparable in power to the proposed Target Station, but have pulse frequencies of 20 or 60 Hz, lower duty factors, and are not designed with the flexibility for tailored irradiation testing that is envisioned for the Project X-driven Target Station [11].

All of the other planned neutron spallation facilities are pulsed systems, except for perhaps the Indian ADS system, compared to the continuous wave Target Station beam. The Japan Proton Accelerator Research Complex (J-PARC) and Indian systems are planned to be oriented to ADS R&D and would have subcritical fuel regions surrounding the spallation target. The SNS second target station and the European Spallation Source (ESS) are planned to be mainly neutron beam facilities, but could be used for limited materials irradiation [11]. The spallation target station in Project X offers an opportunity to leverage and benefit from the design efforts over the years on the proposed Materials Test Station (MTS) at LANL [8].

Because existing cold neutron beamlines, such as SNS and LANSCE were designed for generic neutron scattering, there was no opportunity to optimize the target moderator for nuclear/particle physics research, or to establish a dedicated ultra-cold neutron source. The Spallation target in Project X provides a unique possibility to optimize the performance of the target for the sensitivity needed by each experiment.

## II.2 Introduction

Fermilab is developing a design of a High Intensity Proton Accelerator complex, known as Project X, to support future Particle Physics Programs at the Intensity Frontier. Fermilab's accelerator research and development (R&D) program is focused on the superconducting radio-frequency technologies for the proposed Project X. Pacific Northwest National Laboratory (PNNL) and Argonne National Laboratory (ANL), two U.S. Department of Energy (DOE) National Laboratories with experience in Nuclear Energy, are supporting Fermilab by focusing on developing and evaluating the concept of a high-intensity continuous wave proton beam target station for nuclear physics, fusion energy, and fission energy applications. PNNL developed a report on the Project X Energy Station [15] that explored the potential opportunities related to supporting DOE Nuclear Energy R&D.

That report was the impetus for the Project X Energy Station Workshop in January 2013 [12]. The objective of the workshop was to identify and explore the nuclear relevant R&D that would be possible in a Nuclear Energy Station associated with the Project X Accelerator and identify the design requirements for conducting the research. Previous workshops have focused on the nuclear and particle physics research associated with Project X [8]. Participants at this workshop were U.S. researchers working on accelerator-based applications, nuclear and materials science, applications of high- intensity proton beams and targets, advanced nuclear reactor concepts, advanced nuclear fuel cycles, light-water reactor sustainability, enhanced and accident tolerant fuels, and isotope production. The workshop identified the synergy and benefits that the Project X could bring to the nuclear research community.

The U.S. Nuclear Energy mission will always require the use of test reactors, but one should investigate whether an Integrated Target Station associated with Project X could accelerate and enhance the ability to test and evaluate early research concepts for nuclear physics, fusion energy science, and nuclear energy. In the following section (section II.3), we describe the nuclear environment and testing needs for the nuclear energy, fusion energy science, and nuclear physics R&D missions. Section II.4 describes the options for configuring the Project X Target Station, and how it might meet the mission needs. Section II.5 provides the conclusions.

## II.3 Mission Needs

### II.3.1 Conclusions from Project X Energy Station Workshop

The Project X Energy Station Workshop provided a good forum for bringing together ideas, issues, and expertise from the accelerator, particle physics, and nuclear energy communities. The participants developed a better understanding of the nuclear materials testing needs and how those needs can be satisfied in a Project X target facility [12].

In particular, the Workshop identified unique mission priorities that a Project X Target Station could provide:

- Fusion structural materials irradiation
- Fast reactor structural materials irradiation
- Fast reactor fuels integral effects testing

- Fuels in-situ separate effects testing

High-value mission needs were identified that could take advantage of the unique characteristics of the Project X beam to conduct research of interest to DOE-NE. The guiding principles behind identification of the mission needs included the following:

- Does the Project X beam provide unique conditions of interest to the materials and fuels community?
- What niche materials and fuels applications are enabled by the Project X beam conditions?
- What materials and fuels applications are complementary to (not duplicative of) existing reactor- and accelerator-based irradiation facilities (with an emphasis on domestic capabilities)?

The Energy Station Workshop identified the highest priority mission needs relevant to the target station, in rough order of priority [12]:

- Fusion reactor structural materials – there is no facility available anywhere in the world that can provide fusion-relevant neutron flux and achieve a minimum of 20 displacements per atom (dpa) per calendar year in a reasonable irradiation volume.
- Fast reactor structural materials – there are limited numbers of fast reactors internationally, and none in the United States. Therefore, the Project X Target Station would be very valuable for study of these materials as is for fusion reactor materials. The fast spectrum and high dpa rates provided by the Target Station would be a significant improvement over thermal reactor irradiations, with tailored flux which can achieve close to the right spectrum, but at relatively low dpa rates. In addition to materials relevant to conventional fast reactors, there are newer fast reactor concepts (e.g., the TerraPower traveling wave reactor) to be tested that require ultra-high doses to simulate very long service lifetimes (e.g., 400+ dpa in cladding alloys).
- In-situ, real-time measurements of various separate-effects phenomena in fuels or materials (e.g., microstructural evolution, pellet-clad chemical interactions, fission gas release) – Such in-situ measurements are, in principle, more feasible in an accelerator-based system than in a reactor, and they are very valuable for modelers, but sensor technology will require concurrent development. In-situ measurements are relevant for fusion materials and fast reactor fuels and materials, and could be useful for thermal reactor fuels and materials as well, because of the difficulty of obtaining such information in test reactors. Separate-effects investigations of this type would likely require encapsulated fuel pellet samples.
- Integral effects testing of fast reactor fuels, including driver fuel, minor actinide burning fuel, and transmutation of spent fuel: These tests would provide valuable information regarding fuels for many of the same reasons described above for fast reactor materials. Integral effects would likely require rodlet-scale testing.

A possible additional application of the Project X Target Station is production of unique research isotopes that cannot be obtained without the very high neutron energy spectrum. Examples include $^{32}Si$ and $^{225}Ac$. It is not envisioned that the Target Station would be used in a production mode, with the associated schedule, separations, and yield issues, but rather in a mode to facilitate production of research quantities of isotopes on a schedule consistent with normal Project X accelerator operations.

A variety of other potential materials- and fuels-related areas of study were discussed that did not seem to offer such compelling cases for use of the Project X Target Station when considered in the context of existing reactor- and accelerator-based facilities. Some of the areas discussed in this category included irradiation of thermal reactor materials and fuels (with the one exception mentioned above), neutron- or synchrotron-based materials science, high-energy neutron cross-section measurement, and transient testing. In addition, the question of direct irradiation in the proton beam was considered. In general, the difficulties of relating proton to neutron irradiation, particularly at high neutron energies that cannot be benchmarked by comparison with reactor data, seem to outweigh the potential advantages of reaching high dpa rates by directly using the proton beam for irradiation. Additional difficulties include very high and non-prototypic He generation rates and H implantation.

There is a range of sample sizes for structural materials of interest, from very small (millimeter-scale) to relatively large (maybe 10 cm) that should be accommodated in the irradiation facility. The smaller end of the size range is appropriate for fundamental studies of irradiation damage mechanisms, while the larger end of the range is appropriate for bulk samples needed for engineering property measurements. Thus, the irradiation volume must be designed to accommodate the full range, i.e., there must be areas with relatively uniform (and high) flux over centimeter-scale dimensions. Also, the irradiation facility should include some replaceable large modules, as well as fixed, perhaps smaller, irradiation positions to accommodate specimens for long-term irradiations to achieve high dose (200+ dpa).

For both materials and fuels irradiation testing, active temperature control of test specimens during irradiation is an absolute requirement. While relatively straightforward during steady-state operation, the issue of beam trips and downtime (both planned and unplanned) must be addressed. These transients exist on both short time scales (beam trips and downtime during normal operation) and longer time scales (planned and unplanned extended outages). Some of these events could have consequences for irradiation damage mechanisms (e.g., cascade annealing, atomic diffusion, phase transformations), particularly for samples located in the highest-flux regions in the vicinity of the proton beam and spallation target. Farther away from the target, the short time scale events are likely to be smeared out and less consequential. Extended temperature transients can introduce significant uncertainty in irradiation data interpretation. Therefore, there needs to be specifications related to temperature control during off-normal events. It was suggested that perhaps SNS or the International Fusion Materials Irradiation Facility offer a potential model for some of the specifications associated with beam trips and down time.

The issue of beam availability is a significant one for materials or fuels irradiation testing. For materials, maximizing the dose rate per calendar year is desirable, while for fuels maximizing the irradiation time per calendar year is desirable. For both, higher availability is desirable. Project X Target Station availability of 70% is probably needed to provide the desired dose and fission rates.

Implications of the high-energy tail resulting from spallation will require further consideration, but there are potentially positive as well as negative implications. For fusion materials, the high energy tail offers the potential to achieve a variety of dose and He generation rates, which could significantly enhance the understanding of irradiation damage mechanisms and effects in a regime that has received very little attention (due to lack of fusion-relevant neutron sources with high dose rates). For fission reactor materials, on the other hand, the high-energy tail could be problematic due to non-prototypical high He generation and, possibly, transmutation rates. Ultimately, the impact of the high energy tail for both fusion and fast reactor materials will need to be assessed on an alloy-by-alloy basis.

In general, it appears that the Project X Target Station will need to accommodate at least rodlet- sized fuel pins to be useful to the fuels community for evaluating fast reactor fuels. There also was consensus that it does not make sense to consider equipping Project X Target Station to perform post- irradiation examination (PIE) on fuels or materials because existing infrastructure and capabilities are maintained already in the DOE complex at great cost. However, at a minimum, the Project X Target Station facility will have to have the capability to handle irradiated materials (and potentially fuels) and properly package those samples for shipment to other DOE sites for PIE. This is a non-trivial capability that needs to be considered carefully. In addition, it is highly desirable for the facility to have the capability to receive, as well as ship, irradiated materials and fuels. For example, it would be beneficial to irradiate previously irradiated materials to reduce the time necessary to reach high dose. Similarly, for research on spent fuel transmutation, the facility must be able to receive previously irradiated and properly packaged spent fuel.

Another capability that must exist, either in-house at FNAL or cooperatively arranged with other DOE labs (e.g., INL, PNNL, ORNL), is experiment and module design expertise. Even if existing capabilities are utilized at the DOE laboratories, FNAL will require safety analysis and design review expertise to evaluate experiment and module designs submitted by users. In addition, FNAL will need to evaluate bounding safety cases for experiment and module design. As an example, FFTF and Experimental Breeder Reactor II both had user's guides that outlined all the requirements that experiments had to meet to be accepted by the reactor facility. FNAL may want to consider development of such a user's guide for the target station. Finally, it is strongly recommended that FNAL involve safety, security, environment, and quality assurance organizations early in development of the target station design to identify and resolve issues while they are still manageable.

The neutron flux and spectrum will need to be benchmarked (e.g., with flux wires or equivalent) soon after the facility becomes operational to facilitate accurate neutronics modeling for subsequent experiments.

### II.3.2 Nuclear Energy Requirements

A versatile Project X target station could support the DOE Office of Nuclear Energy missions. The DOE Office of Nuclear Energy Research and Development Roadmap, April 2010 [6] listed four main research and development objectives:

- Develop technologies and other solutions that can improve the reliability, sustain the safety, and extend the life of current reactors;
- Develop improvements in the affordability of new reactors to enable nuclear energy to help meet the Administration's energy security and climate change goals;
- Develop sustainable nuclear fuel cycles;
- Understanding and minimization of risks of nuclear proliferation and terrorism.

In order to meet each of these four objectives, the DOE NE roadmap focuses on:

- Aging phenomenon and degradation of system structures and components such as reactor core internals and reactor pressure vessels, as well as fuel reliability and safety performance issues, develop and test advanced monitoring and NDE technologies, improve materials data such as composite cladding;
- Fundamental nuclear phenomena and development of advanced fuels and materials to improve the economics and safety of advanced reactors such as corrosion resistant materials, radiation resistant alloys for fast spectrum concepts;
- Development of a suite of sustainable fuel cycle options that improve uranium resource utilization, maximize energy generation, minimize waste generation, improve safety, and limit proliferation risk, down-selecting fuels for once-through fuel cycles, modified open fuel cycles, and closed fuel cycles;
- Development of the tools and approaches for understanding, limiting, and managing proliferation risks, such as options that enable decreasing the attractiveness and accessibility of used fuel and intermediate materials, and transmuting materials of potential concern.

The main Nuclear Energy initiatives that have R&D needs that could benefit from the Project X Target Station are Fuel Cycle Technologies, Nuclear Reactor Technologies, and Advanced Modeling and Simulation. Table II-2 lists the major focus areas under these initiatives that could benefit from the Project X Target Station [6]. Table II-3 lists the various neutronic environments needed for irradiation testing to support the different fission reactor concepts.

| Initiative | Focus Areas | Testing Needs |
|---|---|---|
| Fuel Cycle Technologies | Used Fuel Disposition R&D<br>Fuel Cycle R&D | Structural material properties as a function of dpa and temperature<br>Small scale tests to provide proof or validation of system elements |
| Advanced Modeling & Simulation | Nuclear Fuels<br>Advanced Nuclear Reactors | Experimental data to validate state-of-the-art computer modeling and simulation of reactor systems and components |
| Nuclear Reactor Technologies | LWR Sustainability Program<br>Advanced Reactor Technologies<br>Small Modular Reactors<br>Space Power Systems | Basic physics<br>Material research and testing<br>Integral tests of fuel, structural materials<br>Feature tests of components<br>Fuel performance with minor actinides |

**Table II-2**: Summary of DOE Nuclear Energy Initiatives that Could Benefit from the Project X Target Station

| Parameter | LWR | SFR | LFR | MSR | HTGR |
|---|---|---|---|---|---|
| Temperature range (°C) | ~300 | ~550 | 500-800 | 700-800 | 600-850 |
| Max damage rate (dpa) | 50-100 | 100-200 | 100-200 | 100-200 | 5-30 |
| Max helium conc (appm) | ~0.1 | ~40 | ~40 | ~3 | ~3 |
| Max Neutron Energy (MeV) | <1-2 | <1-3 | <1-3 | <1-2 | <1-2 |
| Coolant | water | sodium | lead or lead-bismuth eutectic | Molten salt | Helium |

**Table II-3**: Summary of Nuclear Energy Fission Reactor Testing Environments

The Fuel Cycle Research and Development (FCRD) program conducts long term science based R&D for fuel cycle technologies. This includes 1) developing technologies to improve the sustainability of current reactors, 2) developing improvements in affordability of new small modular reactors and high temperature reactors through improved structural materials and fuels, 3) Developing sustainable nuclear fuel cycles, and 4) minimizing proliferation risks.

Fuel Cycle Research and Development Areas include structural materials, nuclear fuels, reactor systems, instrumentation and controls, power conversion systems, process heat transport systems, dry heat rejection, separations processes, waste forms, risk assessment methods, computational modeling and simulation, and small scale tests to provide proof or validation of system elements. Three variations of fuel cycles are being investigated, Once-through, Modified open, and Full recycling (transmutation).

The FCRD program is developing transmutation fuel technologies to reduce the quantity of high level nuclear waste for deep geologic disposal. Plutonium and minor actinides such as neptunium and americium are included in the fuel matrix where they are burned along with the other fuel isotopes in fast spectrum reactors.

These transmutation fuels cannot be qualified for use until candidate fuels have been irradiated and tested in a prototypic environment. Gaining access to fast spectrum irradiation testing facilities is very difficult, since there are only a few facilities in Asia that can do this type of testing.

A key challenge facing the nuclear fuel cycle is reducing the radiotoxicity and lifetime of spent nuclear fuel. Partitioning or sorting of nuclear waste isotopes and accelerator-based transmutation combined with geological disposal can lead to an acceptable societal solution to the problem of managing spent nuclear fuel. Accelerators can also drive next-generation reactors that burn non-fissile fuel, such as thorium, that can be burned with the use of particle beams. Both or either of these approaches could lead to an increase in power generation through greenhouse gas emission-free nuclear energy and could provide a long-term strategy for the growth of nuclear power in the U.S. (See Appendix—1 for more details on this topic.)

The following considerations apply to irradiation testing capability to support DOE NE reactor and fuel cycle R&D programs:

- Stable, well characterized test spaces
- Capable of testing fuels and materials from coupon size up to assembly sizes (~100 liters)
- Neutron environment characteristic of both thermal and fast spectrum nuclear reactors
- Fuel pin coolant environments of water, sodium, lead, gas, molten salt
- Flexibility of fuels to be tested – homogeneous/heterogeneous, LEU, Pu/Th bearing

- Instrumentation capable of characterizing fuel, clad coolant temperatures from ambient up to 300 to 1000°C
- Neutron flux range of up to $5x10^{15}$ $n/cm^2/s$
- Damage rates that range from a few dpa up to 50 dpa per year
- Associated post irradiation examination or shipping capabilities

Next-generation reactors, whether based on any of these technologies, require materials that are much more radiation resistant than those used in today's reactors. Next generation reactor materials will also have to survive in the high temperature, potentially reactive environments. Accelerators can spur the development of these next-generation materials by producing radiation environments similar to those found in future reactors, providing a platform for materials development that does not currently exist.

### II.3.3 **Fusion Energy Science Requirements**

A versatile Project X target station could support DOE Office of Science Fusion Energy Science mission goal [22] of developing the scientific understanding required to design and deploy the materials needed to support a burning plasma environment, a key step in the research and development of practical fusion energy applications. " The pursuit of fusion energy embraces the challenge of bringing the energy-producing power of a star to earth for the benefit of mankind. This pursuit is one of the most challenging programs of scientific research and development that has ever been undertaken." The promise is an energy system whose fuel is nearly inexhaustible and results in modest radioactivity and zero carbon emissions to the atmosphere. There is currently no facility available anywhere in the world that can provide fusion- relevant neutron flux and material radiation damage rates in a reasonable volume [7].

The Office of Fusion Energy Science has four strategic goals [22]:

- Advance the fundamental science of magnetically confined plasmas to develop the predictive capability needed for a sustainable fusion energy source;
- Pursue scientific opportunities and grand challenges in high energy density plasma science to explore the feasibility of the inertial confinement approach as a fusion energy source, to better understand our universe, and to enhance national security and economic competitiveness;
- Support the development of the scientific understanding required to design and deploy the materials needed to support a burning plasma environment; and
- Increase the fundamental understanding of basic plasma science, including both burning plasma and low temperature plasma science and engineering, to enhance economic competiveness and to create opportunities for a broader range of science based applications.

To support goal number 3 above, a fusion materials irradiation capability is needed to address critical gaps in irradiation capability needed to qualify materials for future science missions. Materials development and performance is a long-standing feasibility issue and is a critical factor in realizing the environmental and safety potential of fusion. Many materials related details in burning fusion plasma facilities are needed to provide high confidence in their design. The conditions of radiation damage, thermal heat flux, and high energy particle bombardment are the most extreme that exist on earth. Nuclear fission reactors and ion beam irradiation facilities can be used to make incremental, at best, scientific progress, since those facilities lack the volume, flux, and spectral characteristics to perform experiments on materials and subcomponents in a simulated fusion environment. A test platform capable of isothermal irradiation effects testing with neutron flux equivalent to the first wall of a DT fusion power reactor. "Mature technologies for a fusion-relevant neutron source include proton beam/high-Z target spallation sources like SNS, MTS, Project X, and D+-Li stripping sources like IFMIF. Viable facility options that meet the mission needs described above include US participation in IFMIF, MTS, and Project X." [7]

Specific requirements include [7]:

- >0.4 liter high-flux volume for irradiation testing with equivalent 14 MeV neutron flux $>10^{14}$n/cm$^2$/s
- 20 dpa/year so that degradation from volumetric swelling, irradiation creep, phase instabilities, helium embrittlement, and solid transmutation can be observed in a reasonable time
- Medium and low flux irradiation volumes to test subcomponent assemblies and partially integrated experiments exposed to temperature, mechanical loads, and corrosive media
- >70% availability to provide exposures >100 MW-year/m$^2$ in a few years
- Relevant temperature ranges controlled to within 5%
- Flux gradients <20% per cm to provide consistent exposures over the volume

Success in these efforts will spread beyond fusion needs into a broad range of capabilities in material nanostructure, predictive behavior, and custom engineered material properties. Tailoring materials at the microstructural level might allow mitigation of neutron degradation properties while maintaining high performance macroscopic properties and margins of safety. This materials irradiation capability combined with an associated modeling program could improve the confidence in predicting how a particular material will perform under the conditions it is exposed to as well as designing new materials to optimize performance for specific locations. The materials knowledge base developed could be applicable for many harsh-environment applications.

The DOE fusion program is part of an international effort to develop magnetically confined nuclear fusion reactors such as the International Thermonuclear Experimental Reactor (ITER) under development in France by an international consortium, and subsequent demonstration and commercial power plants. Materials must be developed that can survive the fusion environment. Low activation materials are required to allow maintenance. There is a parallel program developing inertial fusion concepts. Materials surrounding the fusion ignition region must also survive in a demanding environment.

Technology gaps for fusion reactor research and development requiring materials qualification include:

- Plasma facing components
- Low activation materials
- Solid breeder materials
- Safety

Both low activation structural materials and tritium-producing blanket materials are being developed for fusion applications. Structural material properties are needed as a function of dpa and temperature with a cumulative ~150-200 dpa and a temperature range of 550-1000 °C. Prototypic neutron energies are predominantly at 14 MeV. Maximum helium and hydrogen concentrations are ~1500 appm and ~6750 appm.

First wall and structural materials in a future fusion power plant will be exposed to a 14 MeV neutron flux which cannot be created in a test reactor. The design, licensing, and safe operation of a fusion reactor will require materials to be qualified in a neutron source that simulates fusion-relevant neutron spectra and temperatures. An accelerator is the only way to generate a neutron flux environment that approaches fusion reactor first-wall conditions.

A neutron source for the qualification of fusion reactor materials should meet the following criteria:

- Neutron spectrum with neutrons up to the energies corresponding to the first wall/blanket conditions in a future fusion reactor
- Continuous mode operation with high availability
- 20-50 dpa/fpy in high flux region allowing accelerated testing
- Irradiation volume on the order of 0.5-1 liter in the high flux region

The fusion program at the ITER facility plans to test at least seven blanket types in test blanket modules

- Helium-cooled Lithium-Lead blanket
- Dual-Coolant (He and LiPb) type Lithium-Lead (DFLL and DCLL) blankets

- Dual-Coolant (He and LiPb) Lithium-Lead Ceramic Breeder (LLCB) blanket
- Helium-cooled Ceramic/Beryllium blanket
- Water-cooled Ceramic/Beryllium blanket

A fusion materials irradiation testing capability for the Project X Test Station could be used for testing each of these materials under prototypic conditions.

### II.3.4 Nuclear Physics Requirements

A *Forum on Spallation Sources for Particle Physics* was held at FNAL in March, 2012 [8] Much of the discussion of Nuclear Physics requirements and projections of how the Project X Target Station could support Nuclear Physics was extracted from the materials presented at that forum. A versatile Project X Target Station could support the Office of Nuclear Physics in the DOE Office of Science mission to develop a roadmap of matter that will help unlock the secrets of how the universe is put together. The Project X Target Station could enable a new generation of symmetry-testing experiments with the goal of advancing the understanding of basic nuclear physics phenomena that support fundamental searches for physics beyond the Standard Model. Examples of potential experimental facilities that could be integrated into the Target Station include searches for neutron- antineutron oscillations (NNbar) and nuclear electric dipole moments (EDMs). The MW scale CW proton beam from Project X can produce copious quantities of cold neutrons (CN), very cold neutrons (VCN) and ultra-cold neutrons (UCN), as well as special short-lived Ra, Fr, and Rn isotopes ($^{219}$Rn, $^{223}$Rn, $^{211}$Fr, $^{221}$Fr, $^{223}$Fr, $^{223}$Ra, $^{225}$Ra, $^{225\text{-}229}$Ac) to support fundamental searches for physics beyond the Standard Model. These particular radon, francium, and radium isotopes have favorable nuclear and atomic properties for enhanced EDM searches, and the Project X target station could potentially supply these isotopes in abundance. For example, the projected EDM of $^{225}$Ra is 1000 times larger than the EDM of mercury. The potential electron EDM of Francium is greatly enhanced due to relativistic effects. A permanent EDM violates both time reversal symmetry and parity. “The existence of an EDM can provide the “missing link” for explaining why the universe contains more matter than antimatter.”

#### *Ultra-Cold Neutrons*

Ultra-cold neutrons have the following properties:

- Can be stored in material bottles for hundreds of seconds and piped around corners
- Typical velocities 0-8 m/s (0-350 neV) (kT<4 mK)
- Wavelengths >50 nm
- 100% polarizable with magnetic fields
- Lifetime < 1000 seconds

Primary spallation neutrons are too fast to be useful for most nuclear physics applications, such as a NNbar search or nEDM search. Creation of ultra-cold neutrons (UCN) would involve moderation to thermal energies with a moderator such as heavy water, cooling to very cold neutrons (VCN) using cryogenic material such as solid methane, and then to UCN temperatures using liquid helium. Radiative heating of moderator and heat removal are design challenges. Optimization of moderator configuration is needed to provide maximum yield of cold neutrons, which can then be enhanced for VCN and UCN production. Channeling of the VCN-UCN to the NNbarX detector system might utilize high-m super reflectors and graphite. Table II-4 lists UCN projects operating or under construction around the world.

| Source | Type | Ec(neV) | UCN/cm$^3$ | Status | Purpose |
|---|---|---|---|---|---|
| LANL | Spallation/D2 | 180 | 35 | operating | UCNA/Users |
| ILL | Reactor/turbine | 250 | 40 | operating | n-EDM/users |
| Pulstar | Reactor/D2 | 335 | 120 | construction | Users |
| PSI | Spallation/D2 | 250 | 1,000 | construction | n-EDM |
| TRIUMF | Spallation/HE-II | 210 | 10,000 | planning | n-EDM/users |
| Munich | Reactor/D2 | 250 | 10,000 | R&D | gravity |
| SNS | N beam/HE-II | 130 | 400 | R&D | n-EDM |

**Table II-4**: World's UCN Projects [8]

The Project X Target Station accelerator will have relatively low heating compared to reactors, and should have a favorable duty cycle (>70%). UCN production will likely be limited by energy deposition in the moderator. The ratio of UCN flux to volumetric heating improves at larger distances from the spallation target, but at the expense of cold neutron flux to source power.

UCN can be produced using a $D_2O$ moderator tank, thermal radiation shields to maintain temperatures of ~4K, and a cold source such as liquid $H_2$, liquid $^4He$, solid $D_2$, or solid $CH_4$ to get temperatures ~0.8K. $CH_4$ is the brightest known cold neutron moderator but is not usable at high power sources due to radiation damage. Other reflectors, such as high albedo materials like diamond nanoparticles might be used as radiation-hard reflectors near the moderator to improve cold/VCN brightness [13]. Multilayer mirrors might also improve UCN populations provided to experiments.

An example of an existing ultra-cold neutron source is at the SINQ facility, shown in Figure II-1. In the SINQ UCN source, spallation neutrons are produced inside a cannelloni-style target (lead inside Zircaloy tubes, 21 cm diameter and 55 cm long) and then thermalized in an ambient-temperature heavy-water moderator. The thermalized neutrons are further down-scattered in a 10-liter volume of solid $D_2$ (50 cm diameter, 15 cm thick) cooled to 5K. UCN escape from the $D_2$ volume into a storage tank, from which they pass through guides to experimental areas. [8]

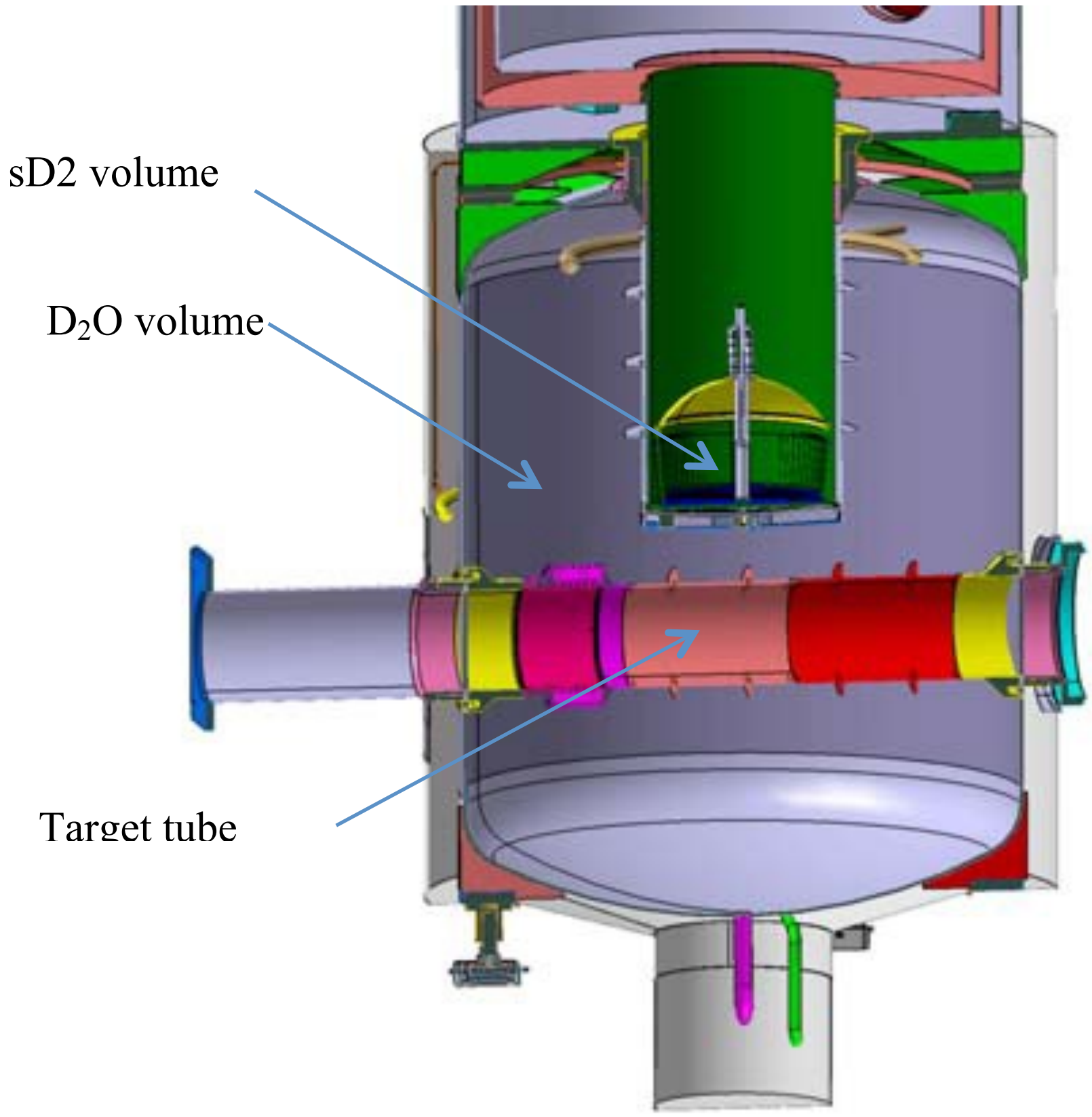

**Figure II-1**: SINQ Spallation UCN Source

LANL has developed concepts for producing UCN in LANSCE involving two tungsten spallation targets with 40L He /$H_2$ in between or a cylindrical proton target with the beam rastered around the circumference with UCN moderator material (potential pre-moderators of heavy water or beryllium to reduce heating) in the center, all imbedded in a bismuth (300K) matrix [8].

### *Neutron-Antineutron Oscillation (NNbar) Search*

The idea for a NNbar oscillation search (described in Part 2) is to 1) observe a sample of neutrons in a vacuum in the absence of a magnetic field for as long as possible, 2) observe as many neutrons as possible, 3) detect the neutron-antineutron transition by the annihilation reaction, and 4) measure the probability of appearance or set a limit. A capability for searching for neutron-antineutron transitions could be new physics for Project X with little competition. There is a strong experimental motivation for developing this capability since there is a possibility of increasing the detection probability by a factor of >100 with minimal background. There is a strong theoretical motivation in that new theories of neutrino mass, low-scale quantum gravity, low scale baryogenesis models, etc., accommodate neutron-antineutron transition probabilities that could be testable by a new experiment. The spallation target in Project X is a unique possibility to optimize the performance of the target for the sensitivity needed by a single experiment. Optimization was not possible before due to constraints (construction and regulational) at reactors or to having multi-user facilities at SNS, PSI, etc. A previous NNbar search was conducted at ILL in 1989-1991 [14]. Improvements that might be possible with the Project X spallation source include [8]:

- Larger number of neutrons
- Lower neutron velocities, larger time (VCN, UCN)
- Larger source to target flight distance
- Use more of the $4\pi$ geometry
- Use diffusive reflection for fast neutrons
- Use super-mirror reflectors
- Use neutron focusing ellipsoidal mirrors
- Use gravity for neutron manipulation

The primary issue for a NNbar search using the Project X target station is optimizing the source/transport geometry to provide a robust source of UCN and to transport those UCN to a detector. The SINQ design for a UCN source is an example of a spallation system oriented towards UCN production.

### *Neutron Electric Dipole Moment (n-EDM)*

Recent NSAC Review Priorities stated that n-EDM research is the highest priority for neutron science [8]:

"The successful completion of a n-EDM experiment, the initiative with the highest scientific priority in US neutron science, would represent an impressive scientific and technical achievement for all of nuclear physics, with ramifications well beyond the field".

Table II-5 describes some of the n-EDM experiments around the world. A next generation n-EDM experiment (described in Part 2) that focuses on an integrated design of source and detector could make large gains in UCN density (e.g.$10^4$-$10^5$/cm$^3$). The lower end of this range overlaps with the TRIUMF goal of $10^4$/cm$^3$. While the goal of n-EDM at SNS is to be statistics limited, follow-on studies at Project X could exploit the statistically limited techniques developed at SNS [8].

| Experiment | UCN Density | Description |
|---|---|---|
| Sussex-RAL-ILL (past experiment) | 0.7 UCN/cm$^3$ | Room temperature, in vacuo, $d_n$<3x10$^{-26}$ e-cm |
| Sussex-RAL-ILL CryoEDM | 1000 UCN/cm$^3$ | Superfluid 4He |
| SNS | 430 UCN/ cm$^3$ | Superfluid 4He |
| PSI | 1000 UCN/cm$^3$ | In vacuo |
| TRIUMF | 10,000-50,000 UCN/cm$^3$ | |

**Table II-5**: Past and Future n-EDM Efforts [8]

### *Atomic Electric Dipole Moment (EDM)*

The search for static electric dipole moments in atoms and electrons (EDM) (described in Part 2) could be flagship experiments to search for physics beyond the Standard Model. The search for EDMs requires high yields of radioactive isotopes of Rn, Fr, and Ra. Project X could extend yields of such isotopes by factors of 100-10000 over existing facilities. Project X will enable a new generation of symmetry-testing experiments and bring exciting opportunities for discovering physics beyond the Standard Model. An EDM search probes energy scales beyond the LHC. The existence of an EDM can possibly help explain the mystery of matter dominance in the universe [8].

Parameters of existing ISOL spallation targets (CERN/ISOLDE, TRIUMF/ISAC, Oak Ridge HRIBF, Legnaro INFN/SPES) can be extrapolated to much higher beam power at the 1 GeV Project X Target Station. Issues to address include [8]:

- Effusion delays from large target chamber.
- Thermal conductivities and temperature limits of refractory thorium compounds, $ThC_2$, $ThO_2$, ThN.
- Thermal simulations coupling beam power deposition with thermal conduction, radiation, and stress effects.
- Operating temperatures ~2000°C to release isotopes

## II.3.5 **Nuclear Physics Isotopes Requirements**

A possible additional application of the Project X Target Station could be support to the Office of Nuclear Physics in the DOE Office of Science mission to develop and produce radioactive isotope products. The program produces isotopes only where there is no U.S. private sector capability or other production capacity is insufficient to meet U.S. needs. It is not proposed that the Target Station would be used in a production mode, with the associated schedule, separations, and yield issues, but rather in a mode to facilitate production of research quantities of unique isotopes that cannot be obtained without the very high neutron energy spectrum or high power proton beam on a schedule consistent with normal Project X accelerator operations.

There are limited isotope production capabilities in the US. Two examples of isotopes are $^{238}Pu$ and medical/industrial isotopes. For $^{238}Pu$ production, there is no current domestic source for NASA to use as a power supply for deep space missions. DOE has identified some potential for production in HIFR and ATR, and completed some preliminary tests, but has not initiated that option. In the past, this isotope was purchased from Russia, but this source is no longer available. Medical/industrial isotopes are produced in limited amounts from ATR, HFIR, University reactors, and cyclotrons to meet current needs. Examples of potential research radioisotopes identified in the Project X Energy Station Workshop include $^{32}Si$ and $^{225}Ac$ [12]. Whether or not production of research quantities of particular isotopes using the Project X Target Station direct 1 GeV proton beam or spectrally tailored neutron flux regions should be pursued will depend on case-by-case evaluations of product quantity, purity, detrimental reactions, expense, etc.

### II.3.6 **Common Issues in Materials Irradiation Testing**

Materials are an immediate priority for both the fission and fusion communities. Extending the lifetime of the current fleet of light water reactors depends on understanding how the materials fail as they age. New generations of power reactors may operate at higher temperatures. New fuel types may be able to burn more efficiently, thereby extending the time between outages and extracting more energy from the fuel, thereby extending our energy resources. Fuel burnup in reactors is limited to about 20% primarily because the cladding mechanical integrity is reduced by radiation damage and elevated temperature. For fast reactor and fusion applications, helium accumulation from $(n, \alpha)$ reactions causes embrittlement. Table II-6 lists some common materials issues for fission, fusion, and accelerator spallation facilities.

| Materials Compatibility Issues | Materials properties issues | Integrated performance issues |
| --- | --- | --- |
| • Coolant<br>• Cladding<br>• Target<br>• Moderator<br>• Transmutation products<br>• Decomposition products<br>• Thermal and Irradiation Stability Issues<br>• Target material swelling<br>• Cladding swelling<br>• Moderator swelling<br>• Target spalling<br>• Irradiation damage effects<br>• Chemical stability<br>• Phase stability | • Thermal conductivity<br>• Heat capacity<br>• Melting point<br>• Emissivity<br>• High temperature strength<br>• Creep behavior<br>• Thermal expansion<br>• Vapor pressure | • Effects of fission or transmutation product buildup<br>• Stoichiometry changes during irradiation<br>• Gas release<br>• Thermal performance<br>• Target material restructuring<br>• Power-to-Melt behavior<br>• Changes in properties with burnup<br>• Fabricability<br>• Target burn efficiency |

**Table II-6**: Common Materials Issues for Fission, Fusion, and Spallation R&D

## II.4 Target Station

### II.4.1 Target Station Concept

The concept for the Project X Target Station is a beam line of 1 MW power directed to a 10-cm diameter liquid lead or lead-bismuth spallation target. The spallation target produces copious neutrons at fusion- and fission-relevant energies. Neutrons produced in the spallation region escape into the surrounding target region, which contains several test modules with independent coolant loops. These test modules could be interchangeable, allowing the facility to accommodate multiple users. The neutron spectra in the test modules could be tailored by using moderating or filtering assemblies, as necessary. Preliminary calculations indicate that large volumes are available (~600 liters with neutron flux >$1x10^{14}$ n/$cm^2$/sec) that rival or surpass the limited test volumes in existing high power test reactors. Further, unlike fission reactors, the Project X Target Station provides significant high-energy neutron flux at positions within and near the spallation target to achieve high dose rates (20-40 dpa per 365 operating days at 1 MW beam power) with fusion-relevant He generation rates. The highest dose rates would be associated with sample volumes on the order of a few liters.

Figure II-2 shows a cross-sectional schematic depiction of the initial notional concept of how the Project X Target Station could be configured. Figure II-3 shows potential configurations and representative neutron spectra of the independent test modules. More details can be found in the workshop presentation materials [12] (select the presenter for each day to access the presentation) and in the PNNL whitepaper [15]. The proton beam from the Project X accelerator is extracted at a proton beam energy of 1 GeV and a beam current of 1 mA, for a total beam power of 1 MW. This beam is directed on a spallation target to produce neutrons. For these initial studies, the proton beam is assumed to be spread uniformly over the target diameter, because the exact mechanism of spreading the beam (such as rastering or defocusing) has not been determined.

The proton beam directed on the heavy metal liquid spallation target creates fairly large volumes of neutron flux that rival or surpass the limited test volumes available in existing test reactors. The initial concept for the spallation target is a 10-cm diameter flowing liquid lead bismuth eutectic (LBE) target that produces approximately 30 neutrons per proton. The 1 GeV protons penetrate approximately 50 cm into the LBE target. The melting point of LBE is ~126°C, so a 200°C inlet temperature, 300°C outlet temperature, and maximum of 2 m/s flow velocity (based on erosion and corrosion concerns) appear reasonable. The optimum target diameter is one that provides adequate heat removal while maximizing the neutron flux. Smaller diameters produce higher neutron flux levels close to the target, but the beam power is deposited over a smaller volume. For example, reducing the target diameter from 10 cm to 5 cm increases the peak neutron flux from $0.6 \times 10^{15}$ to $1 \times 10^{15}$ n/cm$^2$/sec. A similar LBE spallation target technology was demonstrated in the Megawatt Pilot Experiment (MEGAPIE) in 2006 [16]. The neutrons produced in the spallation reaction have an energy spectrum similar to a fission spectrum, but with a high energy tail extending to the beam energy. Use of a solid spallation target, such as tungsten, has also been considered, and this results in a higher peak flux but shorter axial extent. This accelerator beam and spallation target arrangement could be developed in either a vertical or a horizontal layout. A horizontal layout is shown in Figure II-2, which offers benefits for the accelerator design, because it would eliminate the need for a 90-degree bend in the beam.

Two options for the Target Station are being considered, one is an Energy Station focused on DOE Nuclear Energy missions and Fusion Energy Science materials irradiation testing, and the second is an Integrated Target Station that combines the materials testing mission with Nuclear Physics experiments such as ultra-cold neutrons for NNbarX and EDM isotopes. Various configurations of liquid and solid spallation targets are being evaluated. The spallation target produces copious neutrons with an energy spectrum similar to a liquid metal cooled fast reactor. Neutrons produced in the spallation region escape into the surrounding solid lead matrix region, which is can be cooled by helium gas, liquid water, liquid sodium, liquid lead, or lead-bismuth. This matrix region contains several modules for various types of experiments such as fusion materials irradiation testing, fission reactor materials irradiation, production of nuclear physics isotopes for EDM searches, or production of cold neutrons for NNbar nuclear physics experiments. Each module contains independent cooling loops and materials that produce appropriate nuclear environments.

The neutron spectrum in these different modules could be tailored to produce different spectra by using moderating or filtering materials. Preliminary investigations indicate that fairly large volumes (~600 liters of high neutron flux (>$10^{14}$ n/cm$^2$/sec) can be created that rival or surpass the limited test volumes available in existing high power test reactors. Multiple test modules are envisioned in the target region, surrounding the spallation target, each with an independent test region and coolant loop. Each test module could be designed to be removed and reinstalled independently of the others. Vacuum insulation layers can be used to isolate modules with extreme temperature variations from the matrix region. These reconstitutable modules provide tremendous flexibility in designing tests to meet evolving needs. Extensive instrumentation and temperature control are also key attributes that can be used to provide a testing environment tailored to particular program needs.

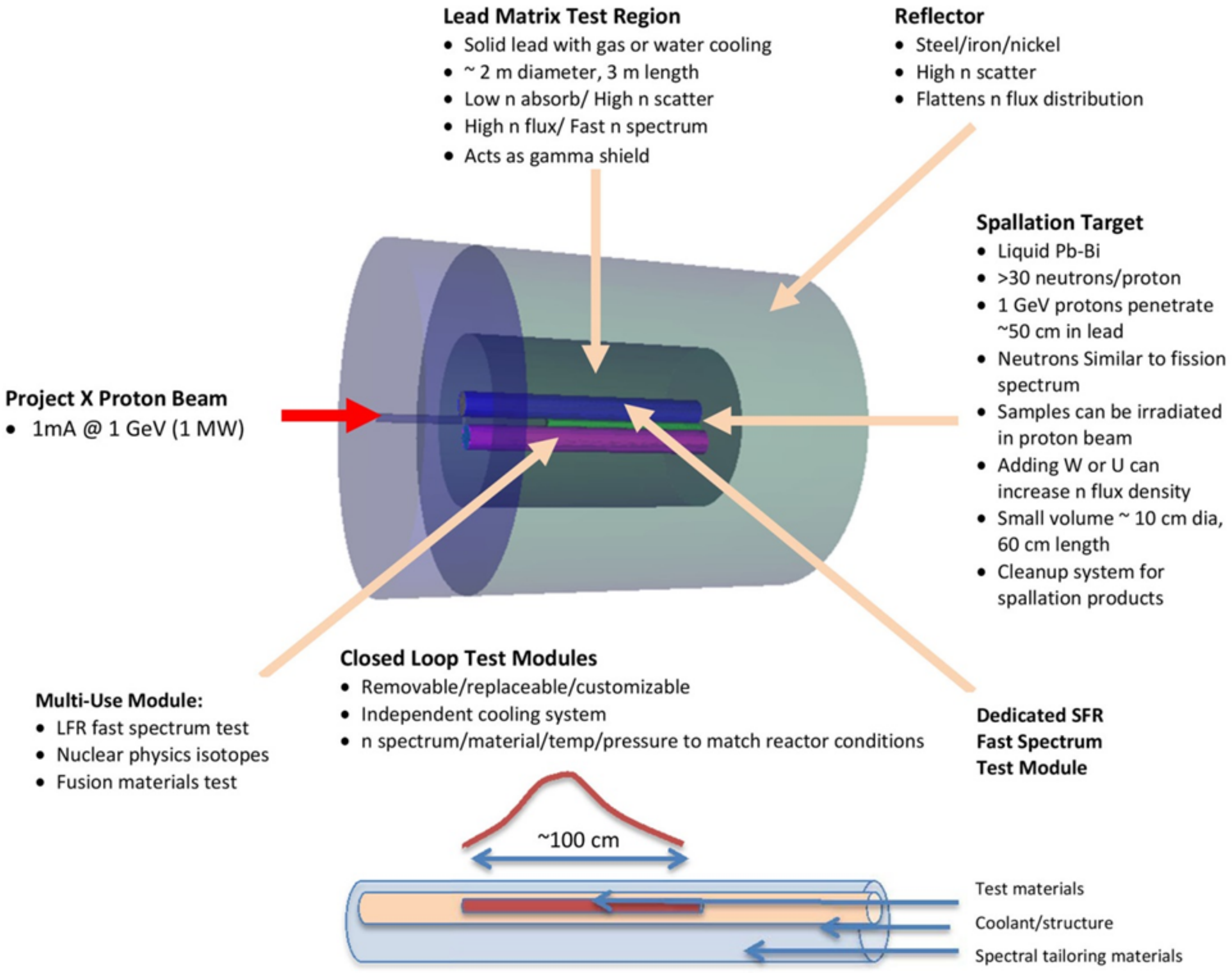

**Figure II-2**: Target Station Concept

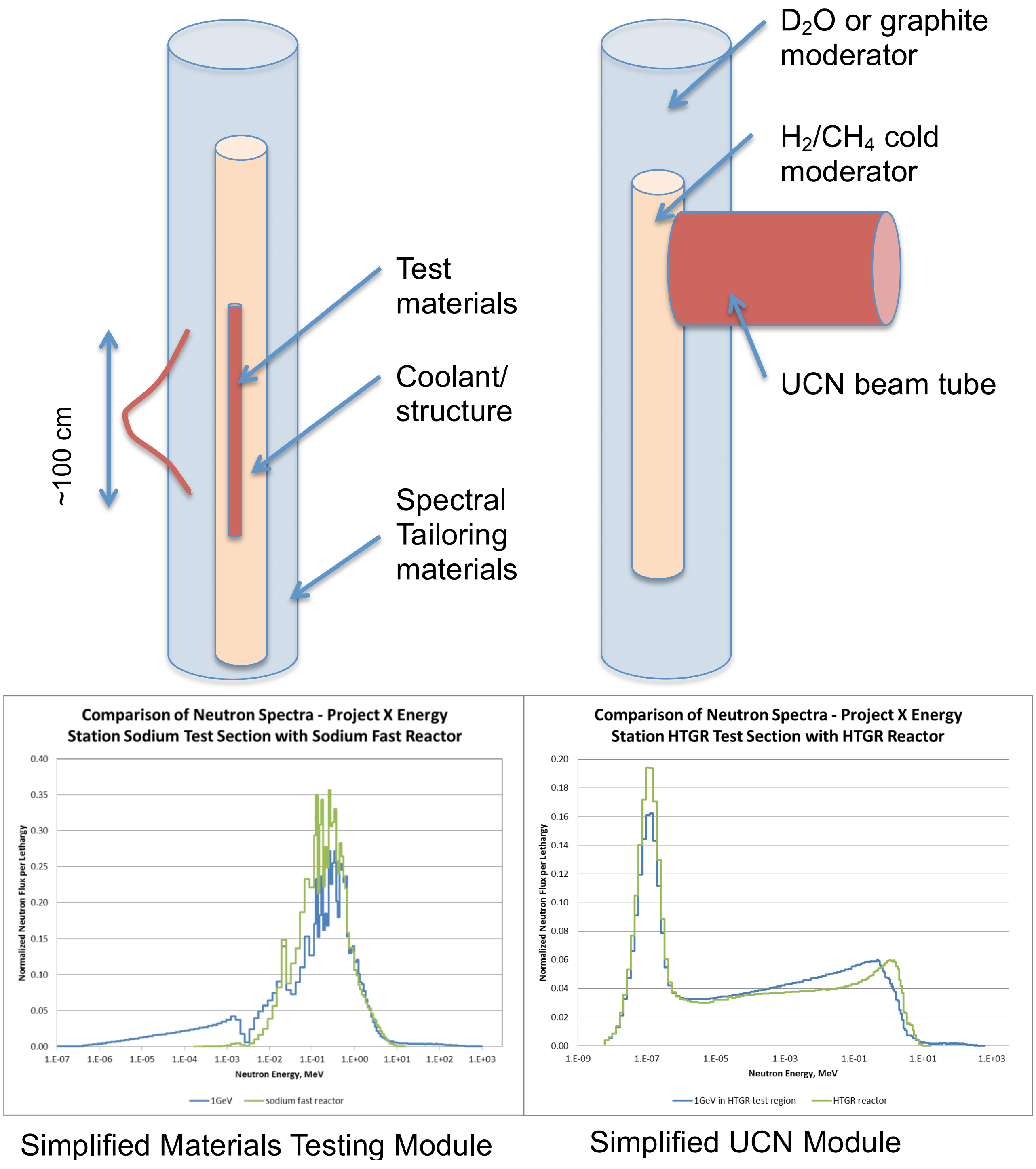

**Figure II-3**: Simplified Target Station Test Modules and representative neutron spectra for proposed Project X test modules (blue traces) and SFR and HTGR reactors (green traces).

The spallation target is surrounded by a high-scatter, low-absorption test matrix that reduces the radial leakage of neutrons from the system. A solid lead matrix 200 cm in diameter and 300 cm long, with the surface of the spallation target recessed 100 cm from the front matrix surface was used in the reference case. The heat deposited in the matrix can be removed by air or gas coolant channels, or water around the periphery. This matrix has holes to provide space for test loops and other fixed irradiation spaces. The distribution of neutron flux in the test matrix is shown in Figure II-4. Simulations of the neutron and proton tracks in the spallation target and lead matrix are shown in Figure II-5. The lead can be seen to be an effective neutron scatter material and the protons are nearly all confined to the flowing lead spallation target region. Volumes at various neutron flux levels are shown in Table II-7. The region with a neutron flux greater than $1x10^{14}$ $n/cm^2/s$ extends axially over 100 cm, allowing long samples to be irradiated. Peak dpa rates in iron range up to 20 dpa/year. Other materials considered for the test matrix included Zircalloy, which has better strength at high temperatures compared to lead, but is not as effective at scattering, resulting in lower neutron flux levels.

The various closed-loop test modules are arranged in the test matrix around the spallation target. The number of test modules can be varied depending on demand. The target station could start with one module, and then additional modules could be added as needed. Figure II-6 shows some potential configurations for the Project X Target Station. Options considered include liquid or solid spallation targets and inclusion of $D_2O$ or graphite moderator regions to generate UCN. The option of including target capsules directly in the proton beam have also been considered, which might be useful for thorium carbide targets for generating EDM isotopes. The native neutron spectrum in the target matrix is similar to that in a lead fast reactor, so little modification of the spectrum would be needed to test that environment. Modules for testing other fast reactor environments, such as sodium fast reactors or gas fast reactors, would require minimal tailoring of the neutron spectrum. Thermal reactor environments, such as pressurized water reactors, boiling water reactors, graphite reactors, or molten salt reactors could also be reproduced, if needed, in a module of less than 30-cm diameter. Modules can be tailored for a variety of environments, such as fusion reactor materials testing, isotope production, or cold neutrons for physics tests. The size of these modules will depend on the amount of room required to reproduce specific reactor operating conditions of temperatures, pressures, materials, and neutron spectrum. The optimum distance of the module from the spallation source depends on the combination of neutron spectrum, dpa rates, and He and H generation rates desired. These modules could be arranged in a vertical or horizontal arrangement around a horizontal beam spallation target. Multiple test modules are envisioned, each with an independent test region and coolant loop. Each test module can be removed and reinstalled independently of the others. These reconstitutable assemblies can provide tremendous flexibility in designing tests that meet client needs, which will evolve over time. Extensive instrumentation and temperature control are also key attributes that can be used to provide a testing environment tailored to particular program needs. Effects of any differences in neutron spectra between those simulated by flux tailoring in the Target Station modules and the individual reactor concepts can be evaluated through comparable materials irradiations and interpretation of the results. Closed-loop modules have been utilized in test reactors such as the Fast Flux Test Facility (sodium), BOR-60 (sodium, lead), and ATR (pressurized water).

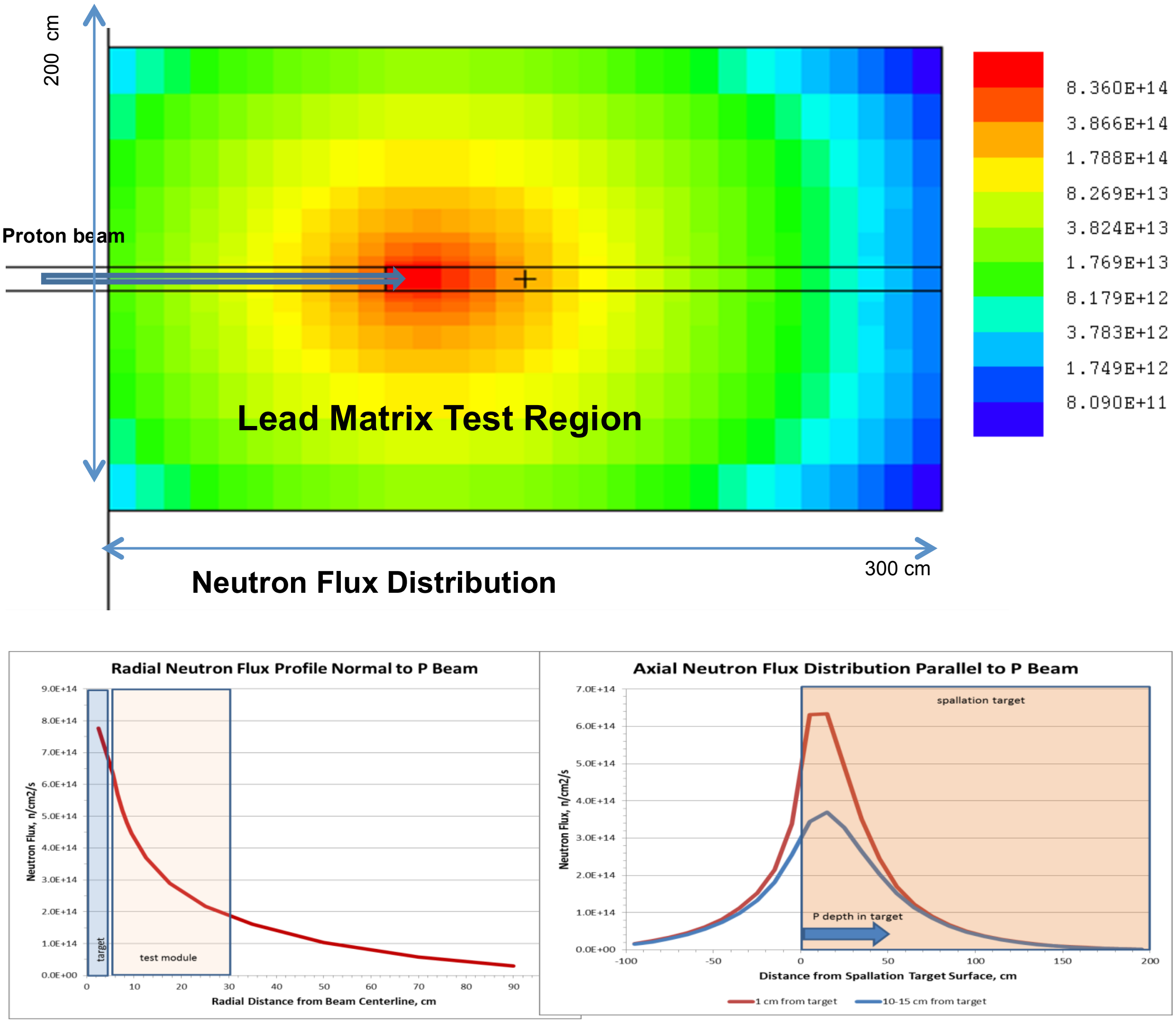

**Figure II-4**: Neutron Flux Distribution in Lead Matrix Test Regions

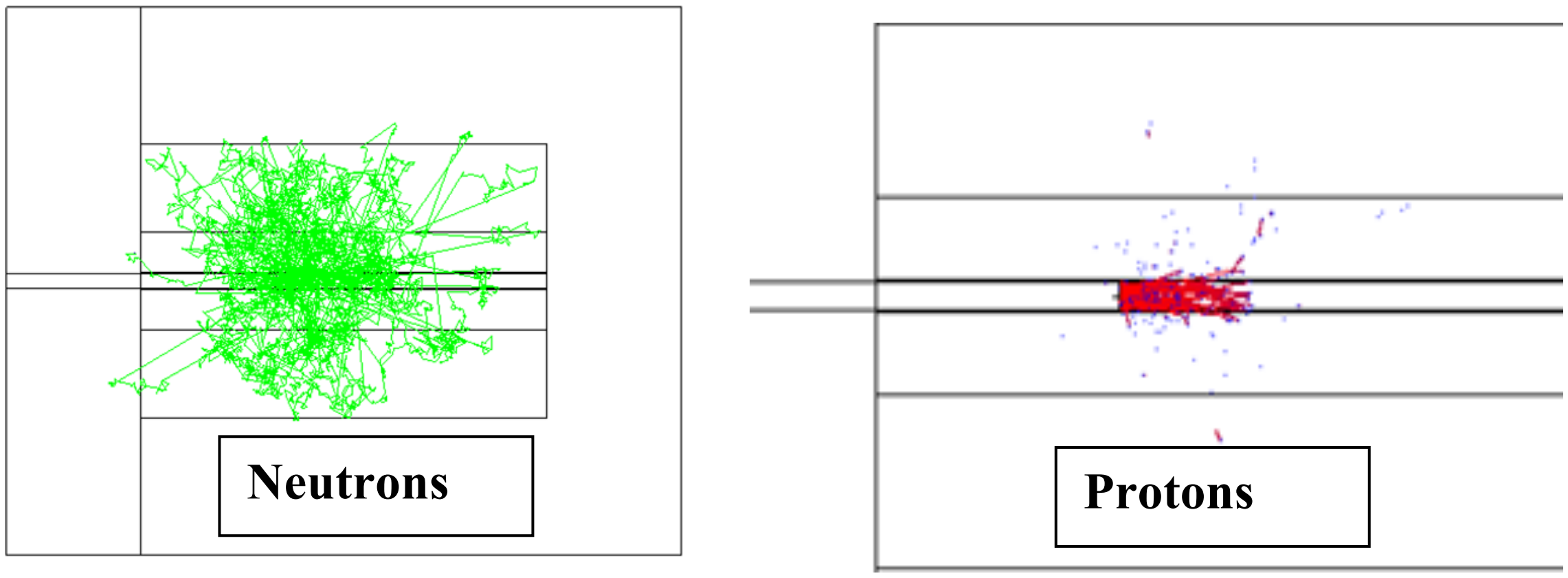

**Figure II-5**: Neutron and Proton Tracks from 1 GeV protons in Spallation Target

| Neutron Flux Range (n/cm2/s) | Axial Extent (cm) | Outer Extent (cm) | Volume (liters) |
|---|---|---|---|
| >5e14 | 30 | 8 | ~2.8 |
| >3e14 | 50 | 15 | ~23 |
| >1e14 | 110 | 60 | ~600 |
| >5e13 | 160 | 80 | ~2000 |
| >1e13 | 250 | 100 | ~9000 |

**Table II-7**: Neutron Flux Volumes in Lead Matrix Test Region

Figure II-6 shows some of the potential options for configurations that provide different environments, including combining the irradiation testing and nuclear physics missions.

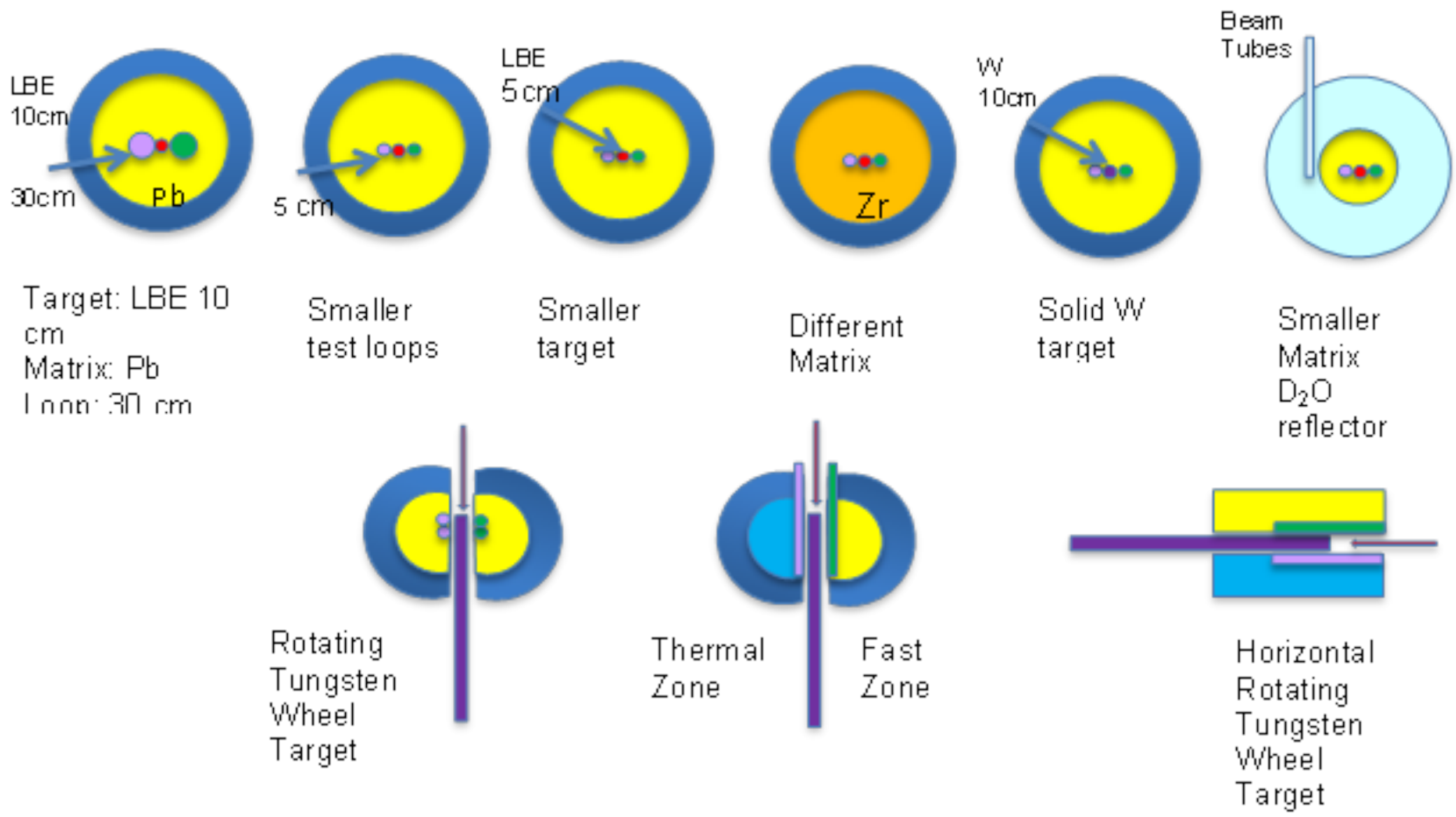

**Figure II-6**: Options for Configuring Project X Target Station

## II.4.2 Existing and Planned Spallation Neutron Source Facilities

Table II-8 compares the Project X Target Station concept with existing proton-induced spallation neutron source facilities. The proposed Target Station accelerator beam parameters used in the current studies are a continuous wave proton beam of 1 MW beam power, 1 mA beam current, and 1 GeV beam energy. The RDR parameters for the 1 GeV Project X beam are 0.91 MW beam power, 1 mA beam current, and 1 GeV beam energy. The Project X Stage 1 beam timing for the 1 GeV beam incorporates a 60 msec beam-off period every 1.2 seconds, resulting in a 95% duty factor for the otherwise continuous wave (CW) beam. This CW beam produces a high neutron flux and the high duty factor provides a neutron irradiation capability to accumulate fluence comparable to large research reactors, but with the volume and flexibility to tailor the neutron spectrum, temperature, coolant, and structural materials to match a wide variety of both thermal and fast spectrum reactor types. Except for the Swiss Spallation Neutron Source (SINQ) cyclotron accelerator, the existing neutron spallation facilities are pulsed systems. They are all designed for producing neutron beams, primarily for scattering studies, and not optimized for materials irradiation. The Los Alamos Neutron Science Center (LANSCE) and Spallation Neutron Source (SNS) facilities are comparable in power to the proposed Target Station, but have pulse frequencies of 20 or 60 Hz, lower duty factors, and are not designed with the flexibility for tailored irradiation testing that is envisioned for the Project X-driven Target Station.

| | Project X Target Station | LANSCE Lujan – LANL[a] | SNS – ORNL[b] | SINQ MEGAPIE – PSI[c] | SINQ Solid Target – PSI[d] | ISIS TS1 – UK[e] |
|---|---|---|---|---|---|---|
| Initial Operation | ~2021 | 1972 | 2006 | 2006 | 1996 | 1984 |
| Target | LBE | W | Hg | LBE | Pb/Zr | W |
| Beam Current, mA | 1 (0.91) | 1.25 | 1.4 | 1.25 | 2.3 | 0.2 |
| Beam Energy, GeV | 1 | 0.8 | 1 | 0.59 | 0.59 | 0.8 |
| Beam Power, MW | 1 (0.91) | 1 | 1.4 | 0.8 | 1.2 | 0.16 |
| Beam Frequency, Hz | CW, 40 MHz | 20 | 60 | CW | CW | 50 |
| Pulse Length, μs | | 0.25 | 0.7 | | | 0.1 |
| Duty Factor, % | >50 | | 6 | | | 2.5 |
| Neutron Flux, $n/cm^2/s$ (vol, liters) | 6e14 peak >3e14 (23L) >1e14 (600L) | beam | beam | beam | beam | beam |

(a) Los Alamos Neutron Science Center–Lujan – Los Alamos National Laboratory
(b) Spallation Neutron Source – Oak Ridge National Laboratory
(c) Swiss Spallation Neutron Source, Megawatt Pilot Experiment – Paul Scherrer Institute
(d) Swiss Spallation Neutron Source, Solid Target – Paul Scherrer Institute
(e) ISIS Target Station 1 – United Kingdom

**Table II-8:** Comparison of the Target Station with Existing Spallation Sources (Source: http://pasi.org.uk/Target_WP1 )

Table II-9 compares the Project X Target Station concept with proposed or planned proton spallation accelerator neutron source facilities. All of the other planned neutron spallation facilities are pulsed systems, except for perhaps the Indian ADS system, compared to the continuous wave Target Station beam. The Japan Proton Accelerator Research Complex (J-PARC) and Indian systems are planned to be oriented to ADS R&D and would have subcritical fuel regions surrounding the spallation target. The SNS second target station and the European Spallation Source (ESS) are planned to be mainly neutron beam facilities, but could be used for limited materials irradiation. The spallation target station in Project X offers an opportunity to leverage and benefit from the design efforts over the years on the proposed Materials Test Station (MTS) at LANL [8].

| | Project X Target Station | MTS/FFMF[(a)] – LANL | SNS Long Pulse – ORNL | J-PARC TEF-T[(b)] – JAEA[(c)] | ESS - Sweden | ADS – India |
|---|---|---|---|---|---|---|
| Initial Operation | ~2021 | Not Scheduled | ? | ? | ~2018 | ? |
| Target | LBE | W (dual) LBE cooled | W/Ta or Hg | LBE | W He cooled | Pb, LBE, W |
| Beam Current, mA | 1 (0.91) | 1.25 | 1.15 | 0.4 | 50 | 10-30 |
| Beam Energy, GeV | 1 | 0.8 | 1.3 | 0.6 | 2.5 | 1 |
| Beam Power, MW | 1 (0.91) | 1 | 1.5 | 0.2 | 5 | 10-30 |
| Beam Frequency, Hz | CW, 40 MHz | 120 | 20 | 25-50 | 14 | CW |
| Pulse Length, μs | | 1000 | 1000 | 500 | 2860 | |
| Duty Factor, % | >50 | 7.5 | | 1.25 | 5 | |
| Neutron Flux, n/cm$^2$/s (vol, liters) | 6e14 peak >3e14 (23L) >1e14 (600L) | 1.6e15 (0.2 L) 40 fuel rodlets; 0.45 L materials | beam | ADS | 2.2e15(0.4 L) target 1.2e15 (5L) reflector | ADS |

(a) Fission Fusion Materials Facility
(b) Transmutation Experimental Facility-ADS Target Test Facility
(c) Japan Atomic Energy Agency

**Table II-9**: Comparison of Target Station with Proposed/Planned Spallation Sources (Source: http://pasi.org.uk/Target_WP1 )

Because existing cold neutron beamlines, such as SNS and LANSCE were designed for generic neutron scattering, there was no opportunity to optimize the target moderator for nuclear/particle physics research, or to establish a dedicated ultra-cold neutron source. The Spallation target in Project X is a unique possibility to optimize the performance of the target for the sensitivity needed by each experiment.

### II.4.3 **Assumptions of Project X Operations**

The Target Station could be structured as a National User Facility (NUSF) similar to what has been done at ATR. This would maximize collaboration between DOE, Universities, industry, and even allow foreign participation. Potential users would propose tests that would be evaluated by a committee. Utilization would also be open to international testing.

The use of reconstitutable assemblies lessens testing costs, providing for a potentially broader utilization, especially by universities. The facility could grow to become a unique and valuable University educational resource for teaching as well as research.

MW scale CW proton beams can serve a variety of functions beyond those of traditional particle physics research. It is assumed that the Project X Target Station can be operated as a joint facility to simultaneously support experimental research and development in the following areas:

- Office of Science - nuclear physics ultra-cold neutrons (NNbar and n-EDM) for standard model tests
- Office of Science – ISOL nuclei (isotopes for atomic EDM studies standard model tests),
- Office of Science - research isotopes
- Office of Science – fusion energy science materials irradiation testing
- Office of Nuclear Energy – fission energy applications

Copious production of special short-lived isotopes could be generated to support fundamental searches for physics beyond the Standard Model.

### II.4.4 Target Station Capabilities & Challenges

There was consensus at the Project X Energy Station Workshop at Fermilab that combining the particle/nuclear physics mission with the nuclear energy mission into a single target station would be preferable to having two separate target stations competing for proton beam current (assuming each mission receives approximately half of the available current).

Both liquid and solid targets in various configurations were discussed at the workshop and are continuing to be explored. The concept of the Target Station, some potential configurations and the surrounding lead matrix region that would contain multiple modules for experiments have been discussed earlier in Section II.4.1. The modules will host various types of experiments such as materials irradiation testing, neutron or nuclear EDM searches; and each module will have independent cooling loops and materials to produce appropriate environment needed by specific experiments.

The envisioned Target Station capabilities include:

- Flexible design allowing support to multiple missions for NE, NP, FES
- Benefits of test reactor volumes and neutron fluxes without reactor issues – licensing, fuel supply, safety, waste
- Robust technology and design that allows evolution to tomorrow’s technology
- Continuous wave, high availability, high beam current provides potential for

irradiation tests to high fluence

- Energy distribution of spallation neutrons similar to fast reactor fission spectrum but with high energy tail up to proton energy
- Ability to tailor neutron spectrum from fast to thermal as well as the gamma to neutron flux ratio
- H and He generation in materials higher than in reactor allowing accelerated aging testing
- Potential for research isotope production and/or neutron beams simultaneous with irradiation test

### *Engineering Challenges*

One of the primary engineering challenges with the Target Station design is providing for continuous operations with remote change out of test modules. For multiple test modules or closed loops, it would be preferable to be able to change out one module with minimal effects on other modules. Continuation of cooling/heating for separate modules while one module is replaced is a design goal. The approach is to provide separate cooling of the lead matrix surrounding the test modules and vacuum insulation to enable independent temperature environments in each module. The shielding design will need to be robust enough to account for radiation streaming when one module is removed, necessitating a remote change-out capability. A lead shield plug may be required to be inserted at the same time as a module is removed.

Another important engineering challenge is to accommodate concurrent multi-mission capabilities with much different environments, ranging from cryogenic moderators for ultra-cold neutrons, fission and fusion neutron energies, heavy water, liquid metals, and temperatures up to 1000°C conditions. Interaction of these various environments will need to be carefully considered in the configuration of the integrated Test Station.

### *Use of Proton Beam for Material Damage Studies*

Researchers have long studied the use of charged particle irradiation as a supplement or surrogate for neutron irradiation. As far back as the 1960s, heavy ion irradiation experiments were conducted to achieve high damage levels (dpa) in a short amount of time in support of fast reactor structural materials development [17]. More recent work has included investigating irradiation damage mechanisms for fusion reactor first wall and structural materials [18], irradiation assisted stress corrosion cracking in light water reactor cladding materials, and embrittlement of reactor pressure vessel steels [20]. Other researchers have used neutron irradiation to simulate charged particle damage as a way of assessing the performance of materials intended for use in space, including semiconductor materials used for solar cells. [19]

The principal challenge of translating charged particle irradiation to neutron irradiation is addressed by identifying the appropriate temperature offset such that the microstructural feature of interest produced by irradiation is comparable. In general, this means ensuring a comparable irradiation-induced interstitial-to-vacancy annihilation rate between neutron and charged particle irradiation. With a proton beam of 1-3 MeV, such microstructural equivalency has been demonstrated in stainless steels, pressure vessel steels, and Zr-base alloys with a 50 to 100°C shift in temperature (higher temperature for proton irradiation versus neutron irradiation) [20].

There is potential to use the 1 GeV proton beam from the Project X accelerator directly for fission or fusion reactor materials irradiation testing in the Target Station spallation target or the beam dump. Such an approach would more efficiently use the proton beam than conversion to neutrons via spallation. It would also provide more volume for test specimens and experiment hardware, and reduce shielding requirements.

While structural materials are typically irradiated in a test reactor, the use of charged particle irradiation does offer some advantages beyond those listed above. Very high damage rates (dpa/s) are possible with a proton beam, typically a factor of 100 to 1000 higher than fission neutrons. This allows relatively short irradiation times to achieve high damage levels. Short irradiation times also translate to relatively inexpensive irradiation experiments compared to test reactors. Protons do not penetrate materials as deeply as neutrons, due to electronic energy loss during scattering, so irradiation of bulk samples for mechanical or thermal property testing is probably not feasible. However, protons penetrate deeply enough such that sufficient interaction volume is available to assess fundamental scientific questions associated with irradiation damage mechanisms and microstructural evolution. For example, a 30 MeV proton beam will penetrate on the order of a millimeter into stainless steel with a relatively uniform damage profile over that depth [21]. Activation tends to be slightly less with proton irradiation than with neutron irradiation at comparable energies. However, at 1 GeV, there will be some activation of target materials.

It is unlikely there will be sufficient proton flux or irradiation volume in the Project X beam dump to accommodate irradiation of large test specimens that are typically used to measure engineering properties. However, it is possible that an irradiation facility could be designed into the beam path to accommodate direct proton irradiation of millimeter-scale samples suitable for microstructural evaluation and perhaps measurements such as microhardness. The proton beam energy is sufficient to irradiate a volume of material entirely suitable for fundamental scientific investigations of irradiation damage mechanisms and microstructural evolution during irradiation. Such a capability would provide a valuable complement to neutron irradiation testing facilities currently available by allowing relatively short and inexpensive irradiation experiments to elucidate fundamental mechanisms while allowing large test matrices and providing the ability to down-select the most promising materials for subsequent test reactor irradiations for evaluating irradiation effects on bulk material properties. There are some limitations to the use of proton irradiation, including lack of constituent element transmutation and implantation of potentially high concentrations of hydrogen. Therefore, it is important to carefully consider, on a case-by-case basis, the test matrix and experiment objectives to ensure they are consistent with the use of proton irradiation. For some applications, these factors might not be detrimental, and in some cases they might be advantageous (e.g. hydrogen implantation in fusion reactor first wall materials).

A number of computational studies are needed to fully explore the capability of materials irradiation testing using the Project X proton beam. These include calculations of the expected penetration depth and damage rate from the 1 GeV proton beam at 1 mA current in typical fission or fusion structural materials. This would lead to an estimate of the temperature offset required for the proton beam to achieve the same microstructural effects as fission neutrons. An activation calculation would be needed to evaluate the adequacy the beam dump shielding. Together, these calculations would provide an assessment of the feasibility of performing meaningful radiation materials science research in the Project X beam.

### II.4.5 **Technical Feasibility**

The technical feasibility of the Project X Target Station can be addressed in two parts, 1) the accelerator, and 2) the spallation target station.

#### ***Project X Accelerator***

The Project X R&D Program is being undertaken by a collaboration of twelve national laboratories and universities, and four Indian laboratories (Project X Collaboration). A comprehensive R&D program is underway, aimed at mitigating the primary technical and cost risk associated with the Project X accelerator. The existing Reference Design is supported by detailed electromagnetic and beam dynamics modeling and simulations, and provides the context for the R&D program. The primary supporting technologies required to construct the Project X accelerator exist today. Fermilab, with national and international collaborators, has an extensive development program in superconducting radio frequency acceleration. This program has produced both spoke resonator and elliptical accelerating structures that meet the requirements of Project X. The Project X Integrated Experiment (PXIE) will be demonstrating the accelerator front end components and is the focus of an intensive development and systems testing program. Proof-of-concept components exist. While Project X is currently pre-CD-0, a preliminary, bottoms- up, cost estimate exist and the state of development is sufficient to support an expeditious move to construction (CD-3), in parallel with ongoing development, over the next three-four years. The Project X accelerator is ready to construct.

### *Project X Integrated Target Station*

The Project X Target Station is currently in the pre-conceptual design phase, but no significant technological challenges have been identified and specifying of mission and technical requirements is ongoing. The design of the target station can leverage significant experience in liquid metal reactor technologies, such as the fast Flux Test Facility (FFTF); design and conduct of reactor irradiation testing at DOE NE reactors such as FFTF at Hanford (with reconstitutable irradiation rigs and fabricated closed loops), ATR in Idaho, and HFIR at ORNL; design and operation of the 1MW liquid metal target at the SNS at ORNL, design concepts for the MTS at LANL, the demonstrated 1 MW liquid lead bismuth spallation target MEGAPIE, the fusion materials irradiation testing design of the IFMIF and predecessors, and the demonstrated UCN source at the SINQ accelerator facility. Closed-loop modules have been utilized in test reactors such as the Fast Flux Test Facility (sodium), BOR-60 (sodium, lead), and ATR (pressurized water).

The Target Station would operate as a national user facility, presumably with the cost shared between organizations that utilize it.

## II.4.6 Meeting the Mission Needs

### *Fusion Energy Science*

At the Energy Station workshop held at Fermilab, attended by accelerator physicists and engineers as well as nuclear materials experts from the fission and fusion communities, the most compelling application identified for the Project X target station is irradiation of fusion reactor structural materials. It was noted that there is currently no facility available anywhere in the world that can provide fusion-relevant neutron flux and achieve a minimum of 20 dpa per calendar year in a reasonable irradiation volume.

Unlike fission reactors, the Project X Target Station provides significant high-energy neutron flux at positions within and near the spallation target to achieve high dose rates (20-40 dpa per 365 operating days at 1 MW beam power) with fusion-relevant He generation rates. The highest dose rates would be associated with sample volumes on the order of a few liters. Figure II-7 below shows a representative neutron spectrum with a significant population of fusion-relevant 14 MeV neutrons that cannot be achieved in a fission reactor.

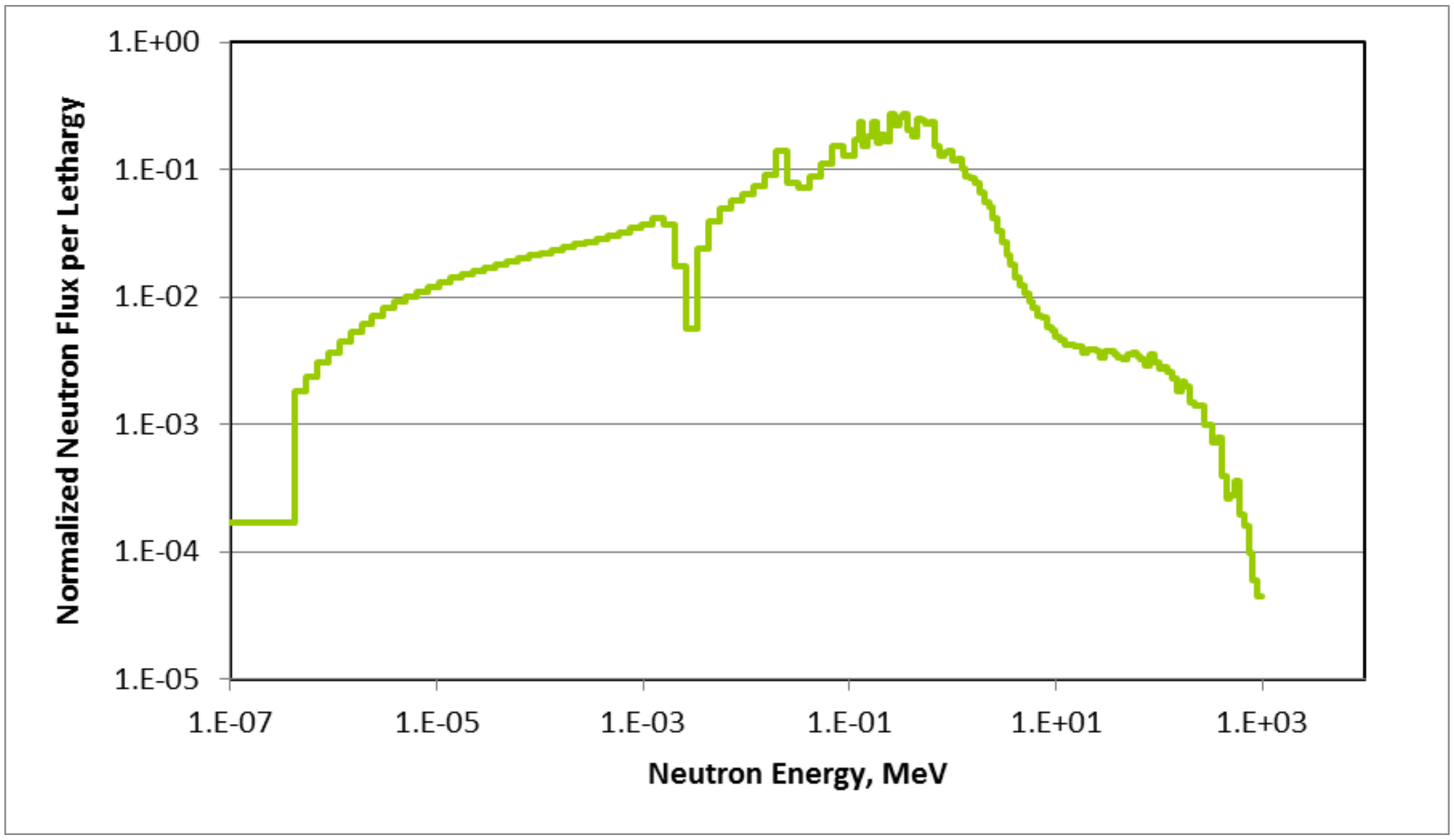

**Figure II-7**: Project X Target Station Spallation Neutron Spectrum with 1 GeV Incident Proton Beam [15].

To ensure maximum relevance to the fusion materials community, the proposed Project X Target Station can accommodate a range of sample sizes for structural materials of interest, from very small (mm-scale) to relatively large (maybe 10 cm). The smaller end of the size range is appropriate for fundamental studies of irradiation damage mechanisms, while the larger end of the range is appropriate for bulk samples needed for engineering property measurements. Because of its ample irradiation volume (due to the intensity of the incident proton beam) the Project X Target Station can accommodate the full range of sample sizes. Further, the target and incident beam can be designed to ensure there are areas of relatively uniform (and high) flux over cm-scale dimensions. At the same time, the irradiation facility can include not only replaceable large modules as described above, but also fixed, irradiation positions to accommodate specimens for long-term irradiations to achieve high dose (150+ dpa).

One question that must be addressed to further evaluate the feasibility of performing fusion materials irradiation tests are the effects of beam transients. For any irradiation experiment, active temperature control of test specimens during irradiation is an absolute requirement. While relatively straightforward during steady-state operation, the issue of incident proton beam trips and downtime (both planned and unplanned) must be addressed. These transients exist on both short time scales (beam trips and downtime during normal operation) and longer time scales (planned and unplanned extended outages). It is possible that some of the events could have consequences for irradiation damage mechanisms (e.g. cascade annealing, atomic diffusion, phase transformations), particularly for samples located in the highest-flux regions adjacent to the proton beam and spallation target. Recent experiences at SNS provide examples of real-world accelerator operation that offer an indication of the degree of reliability to be expected in such a system.

As shown in the figure above, there is a high-energy tail resulting from spallation that is not prototypic of a fusion neutron spectrum. This is an issue that will require further consideration, but there are potentially good as well as bad implications. For fusion materials, the high-energy tail offers the potential to achieve a variety of dose and He generation rates in irradiation experiments, which could significantly enhance the understanding of irradiation damage mechanisms and effects in a regime that has received very little attention (due to lack of fusion-relevant neutron sources with high dose rates).

Specific capabilities of the Project X Target Station relative to Fusion Energy Science include:

- Dedicated fusion loop for materials testing with high energy neutron spectrum test environment at relevant temperatures
- Room for separate lead, helium, water loops that can be used to simultaneously test materials interactions
- The dpa accumulation and high energy neutron spectrum component can simulate fusion environments better than reactors
- Potential testing in the proton beam can provide high dpa and high equivalent neutron energy effects
- H and He generation rates for corresponding damage accumulation could allow testing of fusion materials
- Candidate fusion blanket materials can be irradiated in prototypical conditions of coolant, temperature, and high neutron flux
- Temperature is a critical parameter in materials irradiation and precise temperature control will be a key aspect of the Target Station Test Module design

### *Nuclear Energy*

The materials irradiation testing capabilities of the Target Station could enable, for example, efforts to ensure the sustainability and safety of the current fleet of reactors for lifetime extensions, development of new higher performance and safer reactor fuels and materials, development of innovative economical small reactors, development of new advanced reactor concepts such as those using liquid metal or molten salt coolants, development of transmutation fuels for reducing legacy wastes requiring deep geologic storage, and investigation of accelerator driven systems as a means for transmutation of waste from power reactors.

The Project X Target Station could provide both thermal spectrum and fast spectrum test environments at the relevant temperatures. While thermal spectrum irradiation test volumes can be accessed at ATR and HFIR, there are currently no available U.S. fast reactor spectrum irradiation volumes. A key advantage of the versatile Target Station concept is that there is room for separate sodium, lead, helium, molten salt, and water loops with independent cooling systems that could provide the environments needed to simultaneously test materials for each concept. As future irradiation testing needs develop, the Target Station could be adapted to a variety of configurations, unlike dedicated reactor facilities. Metal hydrides such as zirconium hydride or calcium hydride have been demonstrated effective in providing spectral tailoring in a sodium fast reactor environment, so they could also be used for spectral tailoring as needed for each concept. The higher proportion of high energy neutrons would provide higher H and He generation rates for the same corresponding dpa accumulation, which could allow accelerated aging testing of materials.

Candidate fuel and cladding materials can be irradiated in a prototypic environment of coolant, neutron spectrum, and temperature. The temperature is a critical parameter in materials irradiation, and precise temperature control will be a key aspect of the Target Station design. The peak neutron flux in the test regions approach those achieved in existing fast test reactors. The independent closed loop modules provide the flexibility to support multiple simultaneous irradiation test regions and maximize irradiation volumes.

For general materials irradiation testing, the Project X Target Station can be used to investigate:

- Correlating charged particle damage with existing reactor neutron damage data
- Proton, neutron and gamma induced reactions
- Gas generation
  - Hydrogen implantation, migration, blistering
  - Helium generation from alpha production reactions
- Reactor materials testing
  - Cladding and structural materials
  - Control rod absorber materials
  - Reactor vessel and internal component materials
  - Pre-screening tests for reactor irradiation testing
- Accelerator materials testing
  - Chopper, scraper, absorber, dump materials
  - Beam window materials
  - Low activation materials
  - Radiation resistant materials (effects of impurities, heat treatment, temperatures)
  - Can reach 20-40 dpa/yr

## *Nuclear Physics*

The Project X Target Station UCN source could be configured similar to the SINQ UCN source, with spallation neutrons produced in the target (liquid lead or solid tungsten) and then thermalized in an ambient-temperature heavy-water moderator, and then further down-scattered in a volume of solid $D_2$ cooled to 5K. UCN can be directed from the $D_2$ volume into a storage tank, from which they can pass through guides to experimental areas.

Project X (1mA @1GeV protons on thorium target) predicted yields of important EDM isotopes: $^{219}$Rn>$10^{14}$, $^{223}$Rn~$10^{11}$, $^{211}$Fr~$10^{13}$, $^{221}$Fr>$10^{14}$, $^{223}$Fr>$10^{12}$, $^{223}$Ra>$10^{14}$, $^{225}$Ra>$10^{13}$, $^{225-229}$Ac>$10^{14}$. This compares to a 1 mCi $^{229}$Th source yield of 4x$10^7$ $^{225}$Ra/s. Project X might yield 1x$10^{13}$ $^{225}$Ra/s, which is a 1-2 order of magnitude increase in projected EDM sensitivity. Project X might yield $10^{11}$ $^{223}$Rn/s, which is 3-4 orders of magnitude greater than TRIUMF ISAC. High efficiency extraction of important isotopes could be accomplished by chemical separation of irradiated targets for isotopes with half-lives of days. Extraction of isotopes with half-lives of minutes might require online extraction from hot carbide spallation targets. [8] The goal of nEDM@SNS is to be statistics limited. Project X is a following opportunity that would exploit the statistically limited techniques developed at SNS .[8]

## *Isotope Production*

The Project X Target Station could support isotope production by providing irradiation environments spectrally tailored for isotope production at the relevant temperatures and coolant for the targets. Spectrum tailoring can be used to enhance production of specific isotopes using a variety of moderators such as $D_2O$, graphite, beryllium, and metal hydrides. A rabbit system can be used for rapid insertion and removal of short half- life radioisotopes. Rather than just allow neutrons to leak out of the various test regions to be captured in shield materials, the option of use of these leakage neutrons for isotope production, such as $^{238}$Pu or $^{60}$Co could be considered.

Specific characteristics of the Project X Target Station relevant to isotope production include:

- Low activation target and structural materials specific to isotope production can be developed.
- Target/capsule material compatibilities can be tested for isotope production in other facilities.
- Dedicated isotope production loop with capability to vary neutron spectrum test environment and temperatures to optimize for isotopes of interest
- Room for separate loops that can be used to simultaneously produce and test a variety of isotopes

- Testing in the proton beam can provide accelerator produced isotopes
- Higher neutron energies than reactors can enhance production of isotopes only produced by fast neutrons
- Temperature is a critical parameter in some isotope target irradiations and precise temperature control will be a key aspect
- A rabbit system can be integrated into the test module for rapid insertion and retrieval of short lived radioisotopes
- Reflector region can utilize “waste” neutrons for isotope production such as $^{60}Co$ or $^{238}Pu$
- Spallation reactions produce broad range of reaction products and the target cleanup system could be designed to separate particular isotopes of interest

### *How Target Station Matches the Mission Needs*

The following tables (Table II-10) provide a summary of the Project X Target Station capability to support the various mission needs.

| | Mission Needs | Testing Environments | Target Station Considerations |
|---|---|---|---|
| **Office of Nuclear Energy**<br>• Fuel Cycle Technologies<br>o Used Fuel Disposition R&D<br>o Fuel Cycle R&D<br>• Advanced Modeling & Simulation<br>• Nuclear Reactor Technologies<br>o LWR Sustainability Program<br>o Advanced Reactor Technologies<br>o Small Modular Reactors<br>o Space Power Systems | • Fuels and materials up to ~100 liters<br>• Instrumentation capable of characterizing fuel clad, coolant temperatures up to 1000 °C<br>• Associated PIE or shipping capabilities | • Stable, well characterized test environments<br>• Fast reactors SFR, LFR, GFR<br>• Fuel pin coolant environments of sodium, lead, gas, molten salt, water<br>• Neutron flux up to $5 \times 10^{15}$ n/cm$^2$/s<br>• Damage rates up to 50 dpa/year | • Multiple independent closed loops for irradiation testing with sodium, lead, etc.,<br>• Fast reactor neutron spectrum<br>• Tailoring of neutron spectrum, materials, temperatures to match other reactor environments<br>• Ability to irradiate fuel pin rodlets under prototypic conditions<br>• Remote handling and shipping |

| | Mission Needs | Testing Environments | Target Station Considerations |
|---|---|---|---|
| **Office of Science**<br>• Fusion Energy Science<br>o Fusion Materials and Technology | • Fusion materials testing environment<br>• 14 MeV neutrons<br>• Structural materials properties as a function of dpa and temperature<br>• Plasma facing components<br>• Low activation structural materials<br>• Solid breeder materials<br>• Safety | • Stable, well characterized test environments<br>• Materials surrounding fusion ignition region<br>• >0.4 liter volume<br>• $>10^{14}$ equivalent<br>• 14 MeV neutron flux<br>• >20 dpa/year<br>• Medium and low flux volumes exposed to temperature, mechanical loads, corrosive media<br>• Exposures $>100MW\text{-}y/m^2$<br>• flux gradients <20%per cm | • Dedicated fusion material irradiation testing loop<br>• 20 dpa/yr<br>• Temperatures up to 1000 °C |

| | Mission Needs | Testing Environments | Target Station Considerations |
|---|---|---|---|
| **Office of Science**<br>• Nuclear Physics<br>  o Low Energy Nuclear Physics Research<br>  o Theoretical Nuclear Physics Research<br>  o Isotope Development and Production for Research and Applications | • Source of Ultra cold neutrons for n-EDM, NNbarX<br>• Source of isotopes for ISOL atomic EDM<br>• Capability for R&D for research isotopes | • Stable, well characterized test environments<br>• UCN n velocities <4mK<br>• UCN density $>3x10^4$ $UCN/cm^3$ | • Separate closed loop with heavy water, Be, metal hydride, moderator region,<br>• Cryogenic cooled He, $H_2$,HE-2, $CH_4$ volume for producing Ultra cold neutrons, reflected CN beam transport to n-EDM, NNbarX experiments<br>• Capability for irradiating Thorium spallation target capsules in proton beam region to produce ISOL isotopes<br>• Capability of inserting and removing short- lived isotope production targets (rabbit system) |

**Table II-10:** Comparison of Project X Target Station Capabilities with Mission Needs

## II.5 Conclusions and Recommendations

The broader impacts of the Project X Target Station include providing enabling technologies for the advancement of R&D for:

- Office of Science Fusion Energy Science - fusion materials irradiation testing
- Office of Science Nuclear Physics - providing an ultra-cold neutron source for NNbarX and n- EDM searches
- Office of Science Nuclear Physics – specialty isotopes for atomic EDM and similar studies
- Office of Science Nuclear Physics – research isotopes
- Office of Nuclear Energy – fission reactor fuels and materials irradiation testing

Broader impacts of the Project X Target Station beyond the Particle Physics missions include:

- The Target Station, based on using Project X high energy protons impinging on a heavy metal target to produce spallation neutrons, could provide a new continuous neutron source to complement materials testing at aging US research reactors
- No major technical challenges to designing and building such a facility (continuous spallation source operating in Switzerland for years)
- The Target Station is well suited for fission/fusion materials studies, isotope production, transmutation studies, and a cold neutron source
- The CW beam from the Project-X linac will be a unique facility in the United States to address key physics and technology demonstration

The key advantage of the Project X Target Station is that it would be a single powerful facility with the flexibility to meet a variety of needs as currently envisioned. It will also have the flexibility to adapt to changing needs in the future.

The configuration and design of the Project X Target Station is still at a pre-conceptual level and additional work is recommended to evaluate the following fundamental parameters that can significantly affect the layout of the target station:

- Vertical versus horizontal proton beam alignment on spallation target
- Beam window or windowless spallation target design
- Solid rotating spallation target versus liquid heavy metal target
- Multiple spallation targets with beam split between them
- Limiting beam power density by expanding beam diameter or beam rastering over larger surface
- Radiation heating/temperature limitations for coupling cold neutron capability in spallation target

It is recommended that the integrated approach to the Target Station be pursued vigorously through further studies of configurations, combining the materials irradiation testing for fusion and fission environments/needs with the nuclear physics experimental needs.

## II.6 References

1. Fermilab Project X website (http://projectx.fnal.gov/)
2. Accelerator and Target Technology for Accelerator Driven Transmutation and Energy Production, http://www.science.doe.gov/hep/files/pdfs/ADSWhitePaperFinal.pdf
3. Accelerator for America's Future, W. Henning and C. Shank, editors, http://www.acceleratorsamerica.org/files/Report.pdf
4. R. Raja, S. Mishra, Applications of High Intensity Proton Accelerators – Proceedings of the Workshop, World Scientific, July 30, 2010.
5. Y. Gohar, D. Johnson, T. Johnson, S. Mishra, Fermilab Project X Nuclear Energy Application: Accelerator, Spallation Target and Transmutation Technology Demonstration, https://indico.fnal.gov/getFile.py/access?resId=1&materialId=8&confId=3579
6. Nuclear Energy Research and Development Roadmap Report To Congress, April 2010, US DOE Office of Nuclear Energy (http://energy.gov/sites/prod/files/NuclearEnergy_Roadmap_Final.pdf)
7. Report of the FESAC Subcommittee on the Prioritization of Proposed Scientific User Facilities for the Office of Science, March 21, 2013 http://science.energy.gov/~/media/fes/fesac/pdf/2013/FESAC_Facilities_Report_Final.pdf
8. Project X Forum on Spallation Sources for Particle Physics, March 2012, Fermilab https://indico.fnal.gov/getFile.py/access?contribId=2&sessionId=1&resId=0&materialId=slides&confId=5372
9. "U.S. Particle Physics: Scientific Opportunities, A Strategic Plan for the Next Ten Years", http://science.energy.gov/~/media/hep/pdf/files/pdfs/p5_report_06022008.pdf
10. Isotopes for the Nation's Future, A Long Range Plan, NSAC Isotopes Subcommittee, August 27, 2009 (http://science.energy.gov/~/media/np/nsac/pdf/docs/nsaci_ii_report.pdf)
11. http://pasi.org.uk/Target_WP1
12. DM Asner, et al., 2013, *Project X Energy Station Workshop Report, PNNL-22390*, Pacific Northwest National Laboratory, Richland, Washington.; Also see https://indico.fnal.gov/conferenceTimeTable.py?confId=5836#20130129.detailed
13. R Cubitt, et al., NIMA 622, 182-185 (2010)
14. M. Baldo-Ceolin, et al., Z. Phys. C63 (1994) 409
15. Wootan DW and DM Asner. 2012. *Project X Nuclear Energy Station*. *PNNL-21134*, Pacific Northwest National Laboratory, Richland, Washington. Available at https://indico.fnal.gov/materialDisplay.py?materialId=6&confId=5836

16. Wagner W, F Gröschel, K Thomsen and H Heyck. 2008. "MEGAPIE at SINQ – The First Liquid Metal Target Driven by a Megawatt Class Proton Beam." *Journal of Nuclear Materials* 377(1):12-16.

17. Garner, FA. 1983. "Impact of the Injected Interstitial on the Correlation of Charged Particle and Neutron-Induced Radiation Damage," Journal of Nuclear Materials, 117:177-197.

18. Mazey, DJ. 1990. "Fundamental Aspects of High-Energy Ion-Beam Simulation Techniques and their Relevance to Fusion Materials Studies," Journal of Nuclear Materials, 174:196-209.

19. Ruzin, A, G Casse, M Glasner, A Zanet, F Lemeilleur and S Watts. 1999. "Comparison of Radiation Damage in Silicon Induced by Proton and Neutron Irradiation," IEEE Transactions on Nuclear Science, 46(5):1310-1313.

20. Was, GS. 2007. Fundamentals of Radiation Materials Science. Berlin: Springer.

21. Was, GS and TR Allen. 2007. "Radiation Damage from Different Particle Types," in Radiation Effects in Solids, KE Sickafus, EA Kotomin, and BP Uberuaga, Eds. Berlin: Springer.

22. http://science.energy.gov/fes/

# III MUON SPIN ROTATION AT PROJECT X

## A powerful addition to the Science Portfolio of the Department of Energy and Fermilab

**R. Plunkett, R. Tschirhart, A. Grassellino, A. Romanenko**
**Fermi National Accelerator Laboratory**

**G. MacDougall**
**University of Illinois at Champaign-Urbana**

**R. H. Heffner**
**Los Alamos National Laboratory**

Acknowledgments

The authors would like to acknowledge the following contributors to the Project X Muon Spin Rotation Forum: P. Oddone, S. Henderson, S. Holmes, A. Suter, E. Morenzoni, G. Luke, R. Kiefl, P. Percival, S. Kilcoyne, R. Cywinski, Y. Miyake, C. Polly, S. Striganov.

Thanks to V. Lebedev , N. Mokhov, E. Won, and B. Kiburg for useful conversations, and to P. Bhat for many useful suggestions.

## III.1 Summary

This section discusses the wider application of Project X to problems in material science, specifically through the creation of a facility to exploit polarized, low-energy muons created via the decay of pions, which can be copiously produced by Project X beams. Such a facility would be unique in the United States and can significantly increase extremely limited global capacity for the technique, known as Muon Spin Rotation (μSR in the following).

Most commonly, the μSR technique uses a spin-polarized beam of muons with kinetic energies of 4 MeV or less. Such muons are produced by decay of pions at rest, and utilize conservation of angular momentum in the two-body decay with a left-handed neutrino partner to produce a polarized beam (Figure III-1). Muons are implanted into samples of materials, and decay after precessing in the local magnetic field. Detection of the angular distribution of decay positrons (for μ+) as a function of time gives information on the local static field distribution and the field fluctuation rate. Experiments can be carried out in either zero of finite applied fields. Simple in concept, μSR has found wide application in characterization of materials, particularly in studies of magnetism and superconductivity.

Methods exist for lowering the energy of the polarized muons further, creating a beam of quasi-thermal muons with kinetic energies of typically 1-30 keV. These "LEM" beams (for Low Energy Muons) present an exceptional scientific opportunity for probing phenomena within the first 200 nm of the surface of materials, and are a subject of intense research interest.

We propose utilizing the unique time structure of Project X to create a facility to exploit the μSR technique via a number of different avenues. Several experimental halls will receive dedicated pulses from Project X. Other beamlines will utilize muons created by a thin target inserted into the proton stream sent to the spallation target area, giving adequate intensities to produce one or more LEM beams. In this way, the great flexibility of Project X can be used to create a facility with unique capabilities for programs utilizing the whole range of muon spin rotation techniques.

There is no μSR facility currently in operation in the United States. The materials science research community relies heavily on off-shore facilities which are in very high demand. Therefore, the proposed facility at Fermilab will be highly beneficial to the US research community and a powerful addition to the science portfolio of DOE.

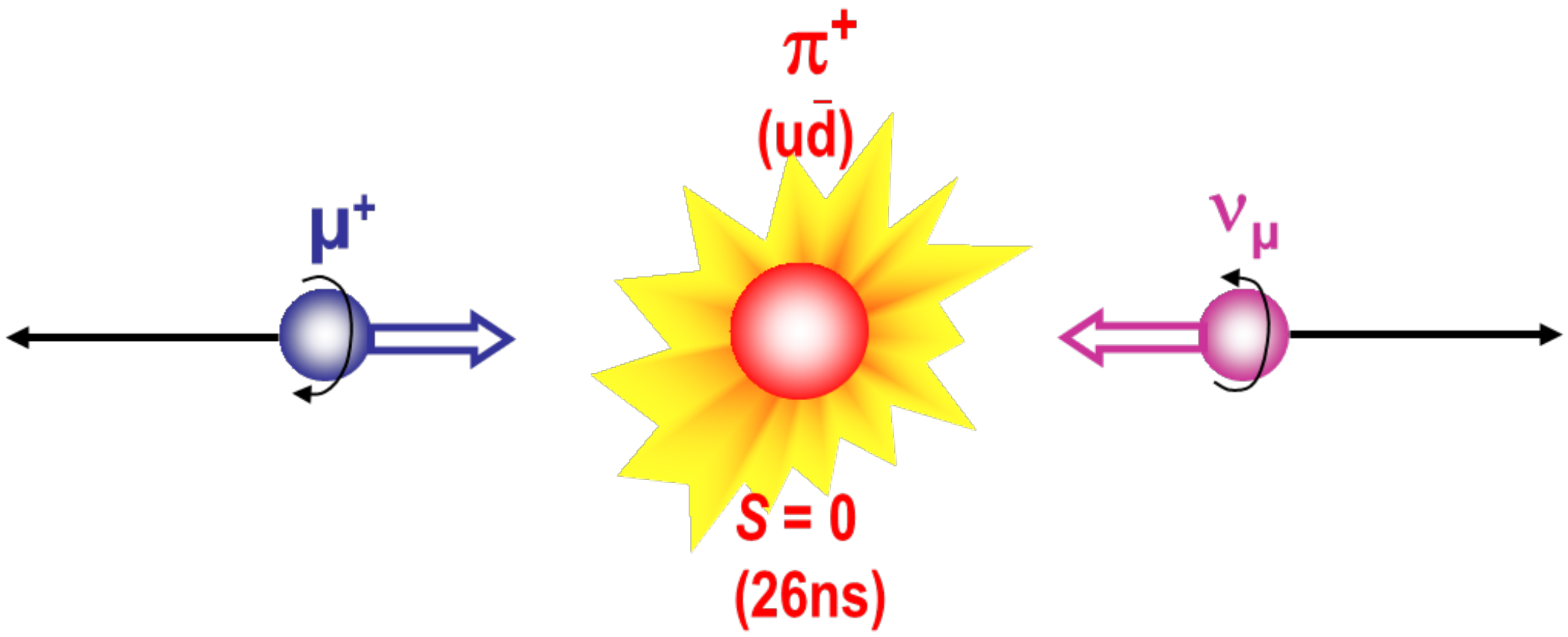

**Figure III-1**: Decay of a charged pion, in its rest frame, showing the polarization of the decay products. The neutrino is always left-handed, constraining the polarization of the muon.

## III.2 The science case for μSR

μSR has made important contributions to a wide range of topics of interest to the condensed-matter science community, including such topics as superconductivity, quantum magnetism and chemistry. Important advances have also been made, by employing μSR, in the study of semiconductors, biological and soft-matter systems, and quantum diffusion. In recent years, the advent of "ultra-low energy" μSR beams has produced unique contributions in the study of thin films, multi-layers and surface science.

In this section, we discuss some of the past and continuing successes of μSR as a probe of matter, with an eye to demonstrating some of its unique characteristics. Because μSR is highly complementary to other widely-used probes - including NMR, x-ray scattering and neutron scattering- there is a very large potential user base in North America and abroad.

### III.2.1 Materials science

In materials science, μSR is primarily a probe of static and dynamic magnetic correlations inside materials of interest. Like other resonance probes, such as NMR and ESR, μSR gains this information in real-space by monitoring the time-evolution of the muon magnetic moment. Unlike other probes which polarize host atoms with strong magnetic fields, μSR injects naturally spin-polarized muons to probe local magnetism.

Also, instead of using rf-fields to gather information, a $\mu$SR experimenter collects daughter electrons or positrons resulting from the symmetry-violating decay of the muon itself.

Like all probes of matter, $\mu$SR has advantages and disadvantages. In particular, its unique characteristics make a powerful *complement* to several other probes of matter, in some cases providing information unobtainable in any other way. Some of these characteristics are outlined below:

- $\mu$SR is a real-space probe, which gives information similar to NMR but distinct from reciprocal-space probes such as neutron scattering and resonant x-ray scattering. It sacrifices detailed information about long-wavelength correlations, but has the ability to directly detect phase separation and measure volume fractions of competing phases.

- As a foreign probe of magnetism, $\mu$SR can measure the properties of any system in which a muon can be stopped. There are no particular requirements or restrictions regarding constituent atoms or material properties. This also affords the flexibility to use a wide range of sample environments.

- $\mu$SR does not require the presence of external magnetic fields, allowing for the measurement of glassiness or magnetism in materials in which the applied fields used in NMR have a non-negligible effect.

- $\mu$SR is sensitive to the presence of dynamic fluctuations in the10's of picosecond to microsecond timescale. This is slower than fluctuations probed by traditional neutron spectrometers, but much quicker than those probed by NMR or AC susceptibility.

- The supreme sensitivity of the muon spin is regularly used to detect local fields as small as 0.1G in crystals as small as $2mm^3$. This has been leveraged to detect the existence of local moments as small a $0.001\mu_B$.

- Through the detection of vortex lattices, $\mu$SR is one of the few techniques (along with small angle neutron scattering) capable of measuring superconducting penetration depths and coherence lengths in the vortex phase of type-II superconductors. Low energy $\mu$SR is now providing some of the first measures of penetration depths in the Meissner state of superconducting materials.

- The relatively high signal-to-noise in modern $\mu$SR instruments allows for spectra to be collected in less than one hour. This allows experimenters to detail material response to fields or to determine detailed phase diagrams of a new material in a matter of days, consistent with the time allotted for a single $\mu$SR experiment.

- The advent of novel low-energy (keV range) muon beams now allows for the depth-range implanting of probe magnetic moments, opening up new avenues for research of surface and interface effects in new magnetic materials.

***Magnetism***

The study of magnetism is the most common area of application of μSR, due to the sensitivity of the muon and its capability to probe both static and dynamic local fields. Magnetic volume fractions are determined by measuring the amplitude of the precessing signal in a zero-field μSR experiment, or alternatively, by measuring the non-decaying amplitude of the signal in a weak transverse field experiment. In the first method, the frequency of the oscillation will be proportional to the ordered moment size, and the amplitude proportional to ordered volume fraction. In the second, that portion of the signal due to the ordered volume will decay in a fraction of a microsecond, allowing one to unambiguously associate the remaining precessing signal with the paramagnetic volume fraction.

Historically, the ability of μSR to study materials in low or zero applied magnetic field, and the unique fluctuation range to which it is sensitive, have allowed the technique to make a large impact in the study of spin glasses [Uemura1980; Uemura1985; Pinkvos1990; MacLaughlin1983; Heffner1983] and geometrically frustrated magnets [Carretta2008]. The high signal-to-noise ratio has been utilized to map out phase diagrams and study quantum phase transitions [Dalmas de Reotier 1997; Niedermeyer1998; Uemura2007; Carlo2009; Luetkens2009; Pratt2011], and to make careful measurements of order parameters to extract critical exponents [Dalmas de Reotier 1997; Pratt2011].The high sensitivity has been useful in the study of spin singlets or highly renormalized moments in reduced dimensional [Kojima1995; Kadono1996, Kojima1997; Matsuda1997], frustrated [Carretta2008] or heavy fermion magnetic materials [Amato1997]. Parallel to ongoing developments in each of these fields, recent work with low-energy muons has also produced insights into the effect of surfaces or interfaces in thin-film magnets and magnetic heterostructures [Shay2009; Suter2011; Boris2012; Hofman2012].

*Frustrated Magnetism, Phase Diagrams and Critical Exponents*

Several of these benefits were displayed in a recent high-profile study of the quasi-2D organic antiferromagnet κ-(BEDT-TTF)$_2$Cu$_2$(CN)$_3$ by Pratt *et al*. [Pratt2011] – see Figure III-2. This material is a largely isotropic Mott insulator which contains spin-1/2 $Cu^{2+}$ moments in weakly interacting two-dimensional planes. The moments are in a plane with a frustrated triangular configuration, and κ-$(BEDT\text{-}TTF)_2Cu_2(CN)_3$ is one of the leading candidates to be the long-sought *quantum spin liquid*, wherein magnetic order is suppressed to zero temperature and the ground state is a singlet comprising a macroscopically large number of entangled spins. Thermodynamic probes have established that this material contains no well-defined magnetic order down to 20mK, despite a sizeable nearest-neighbor exchange of J~250K, but at the time of this study, at least one NMR study had demonstrated the emergence of inhomogeneous moments below 4K and with applied fields above 2T [Shimizu2006].

This zero-field μSR study has confirmed and expanded upon these initial NMR results. Pratt *et al* showed that the magnetic field distribution is consistent with what is expected from randomly oriented nuclear dipole moments- thus ruling out the existence of electronic moments larger than ~$0.001\mu_B$. Through careful weak transverse field (WTF) μSR measurements, they were able to observe a sudden increase in the root-mean-squared field as applied fields were increased above $H_0 = 14$mT. They associated this as a transition to a weak (small moment) antiferromagnetic state with field $B_{rms} < 0.5$mT. A series of WTF measurements further allowed them to track the phase boundary in parameter space and determine that it scaled according to expectations for a Bose-Einstein condensate in two-dimensions. The measured critical field was consistent with a spin gap $\Delta_s \sim 3.5$mK - five orders of magnitude smaller than J and indicative of an emergent low-energy scale for the relevant physics.

The authors measured the temperature dependence of the magnetic fluctuation rate, identifying four distinct temperature regimes at zero field. By extending this analysis to finite fields, a second quantum critical point was identified at $H_1$=4T and a a gap involving vortex-like excitations called visons, with $\Delta_v >> \Delta_s$ was inferred, consistent with previous NMR results and thermal conductivity data [Yamshita2008]. Combining their μSR results with NMR data, the authors constructed a detailed H-T phase diagram extending to fields as high as 10 T and temperatures as high as 10 K, complete with critical exponents for comparison with theory. The data allowed Pratt *et al.* to conclude that this system is best described by a spin-liquid model where the $H_0$ is associated with the Bose condensation of magnons, and $H_1$ transition is associated with a deconfinement transition, where S=1 spin wave excitations fragment into freely propagating S=1/2 spinons.

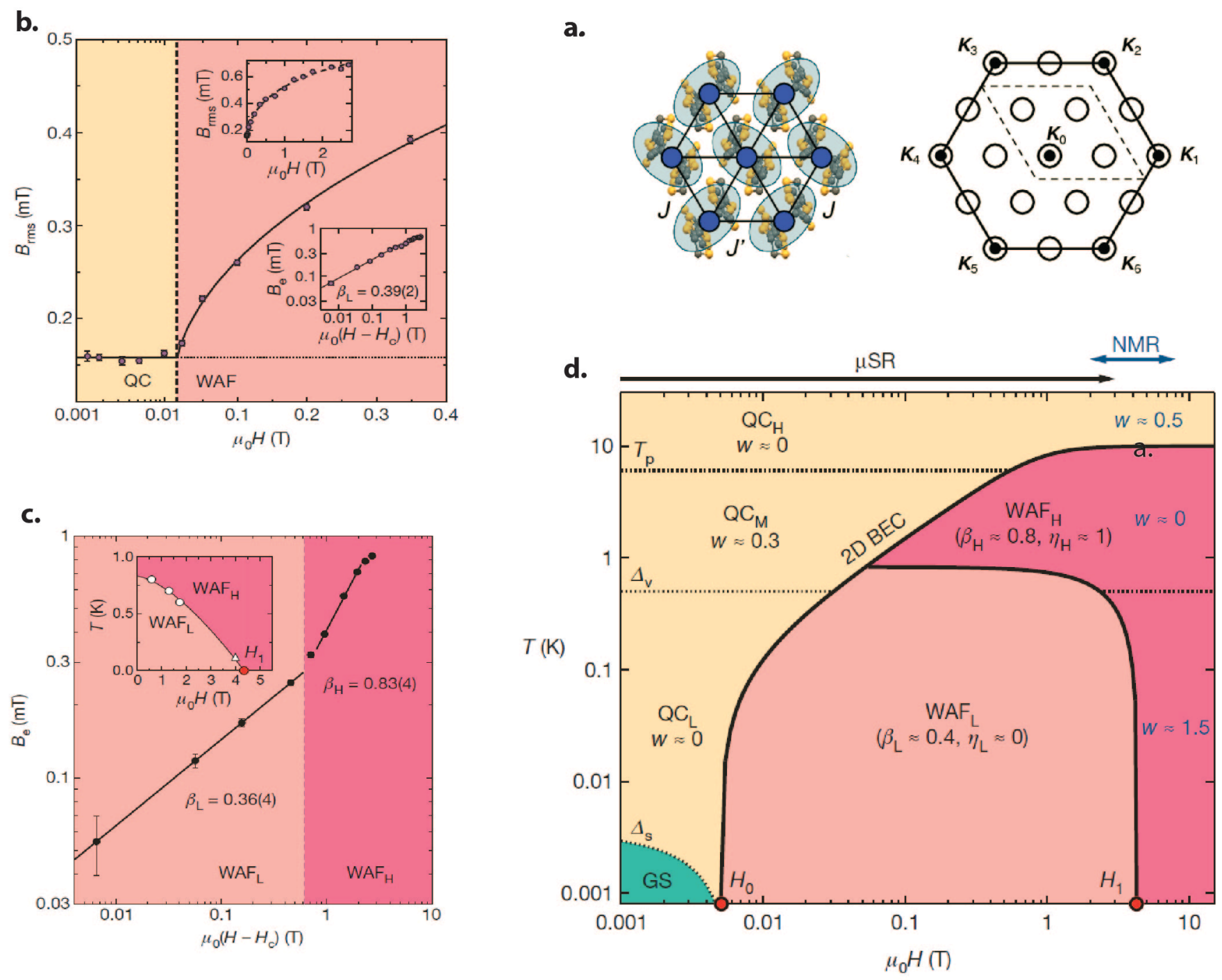

**Figure III-2**: Phase diagram and critical behavior of κ-$(BEDT\text{-}TTF)_2Cu_2(CN)_3$ - (a.) Sketches of real and reciprocal space lattice structures, demonstrating the symmetry of the S=1/2 BEDT-TTF dimers in the two-dimensional plane. (b.) Field dependence of the $\mu$SR linewidth, measured to 120mK. The increase above $H_0$(120mK)= 14mT represents a transition from a quantum critical phase to a small moment antiferromagnetic phase at this temperature. Upper inset shows higher field behaviour, and lower inset shows critical scaling analysis which led to the determination of $H_0$. (c.) Similar analysis investigating a second quantum critical transition at $H_1$=4T, between two qualitatively different antiferromagnetic phases. (d.) A phase diagram extracted from analysis similar to (b.) and (c.) at several different fields and temperatures. Included are critical scaling exponents, extracted from analysis of fluctuation rates in zero field, and information from a separate NMR study. Adapted from [Pratt2011].

Through similar measurements μSR has been used to explore the properties of other frustrated triangular [Olariu2006] and Kagome [Ofer2009] lattice systems. In three dimensions μSR has long been used to explore the statics and dynamics of so-called pyrochlore antiferromagnets, wherein magnetic moments lie on a network of corner sharing tetrahedra. The observed behavior in these systems varies but is frequently unconventional. Examples include spin-glass-like behavior in the absence of disorder in $YMo_2O_7$ [Dunsiger1996] and ground states with partially or entirely dynamic spin fluctuations persisting to the lowest measurable temperatures [Gardner1999; Dunsiger2006; Dalmas de Reotier2006; Dalmas de Reotier2012]. In recent years, μSR has played a major complementary role to neutron scattering in exploring order in frustrated systems with neutron absorbing 5*d* transition metal ions. Examples include the identification of magnetically ordered and disordered states below the metal insulator transitions in $Eu_2Ir_2O_7$ [Zhao2011] and $Nd_2Ir_2O_7$ [Disseler2012], respectively, or the confirmation of the order state in a multi-probe review of magnetism in $Na_2IrO_3$ [Choi2012].

*Quantum Phase Transitions and Phase Separation*

The above study of Pratt shows the usefulness of measuring critical exponents to discriminate between various theoretical models of magnetic systems. However, μSR has also played a key role in the study of *first-order* phase transitions, where the critical behavior is avoided. One recent example is a study by Uemura *et al.* of the first-order quantum phase transitions in MnSi and $(Sr_{1-x}Ca_x)RuO_3$ [Uemura2007]. MnSi has long been of interest as a prototype for weak moment itinerant antiferromagnetism and, more recently, for the partially ordered state seen at high pressures and fields. $SrRuO_3$ is a correlated metal, which is an end member of the famous Ruddleson-Popper series $Sr_{n+1}Ru_nO_{3n+1}$ ($n\rightarrow\infty$). It has a ferromagnetic transition at $T_c$=150K, which can be destroyed by replacing $Sr^{2+}$ cations with isovalent $Ca^{2+}$. Uemura *et al.* studied the order-disorder transition in MnSi by selecting muons of the correct velocity to penetrate the front wall of a hydrostatic pressure cell and implant in a single-crystal sample. Through an examination of frequency and amplitude of muon spin precession in a series of zero field and weak transverse field μSR measurements, they were able to demonstrate that the relative sample volume occupied by the magnetically ordered phase was shrinking with increasing pressure, going to zero identically at the previously identified critical pressure, but without appreciably changing the ordered moment size. Further, by examining signal relaxation in a series longitudinal field experiments, they were able to show that the spin fluctuations expected at a continuous quantum phase transition were suppressed. Overall, this painted a picture wherein the quantum phase transition in MnSi is of first order and magnetism is destroyed via a "trading-off" between ordered and disordered volumes over an extended range in pressure. Uemura *et al* also demonstrated similar behavior at ambient pressure with doping in ceramic samples of $(Sr_{1-x}Ca_x)RuO_3$. In these materials, the authors were able to reproduce *bulk* magnetization data taken on the same samples using *microscopic* parameters obtained from μSR measurements.

These results have spurred intense debate about the character of quantum phase transitions in metallic systems. Real space phase separation has also been observed in the phase diagrams of many other magnetic systems and, as seen in Figure III-3, has become a major theme in the study of unconventional superconducting compounds.

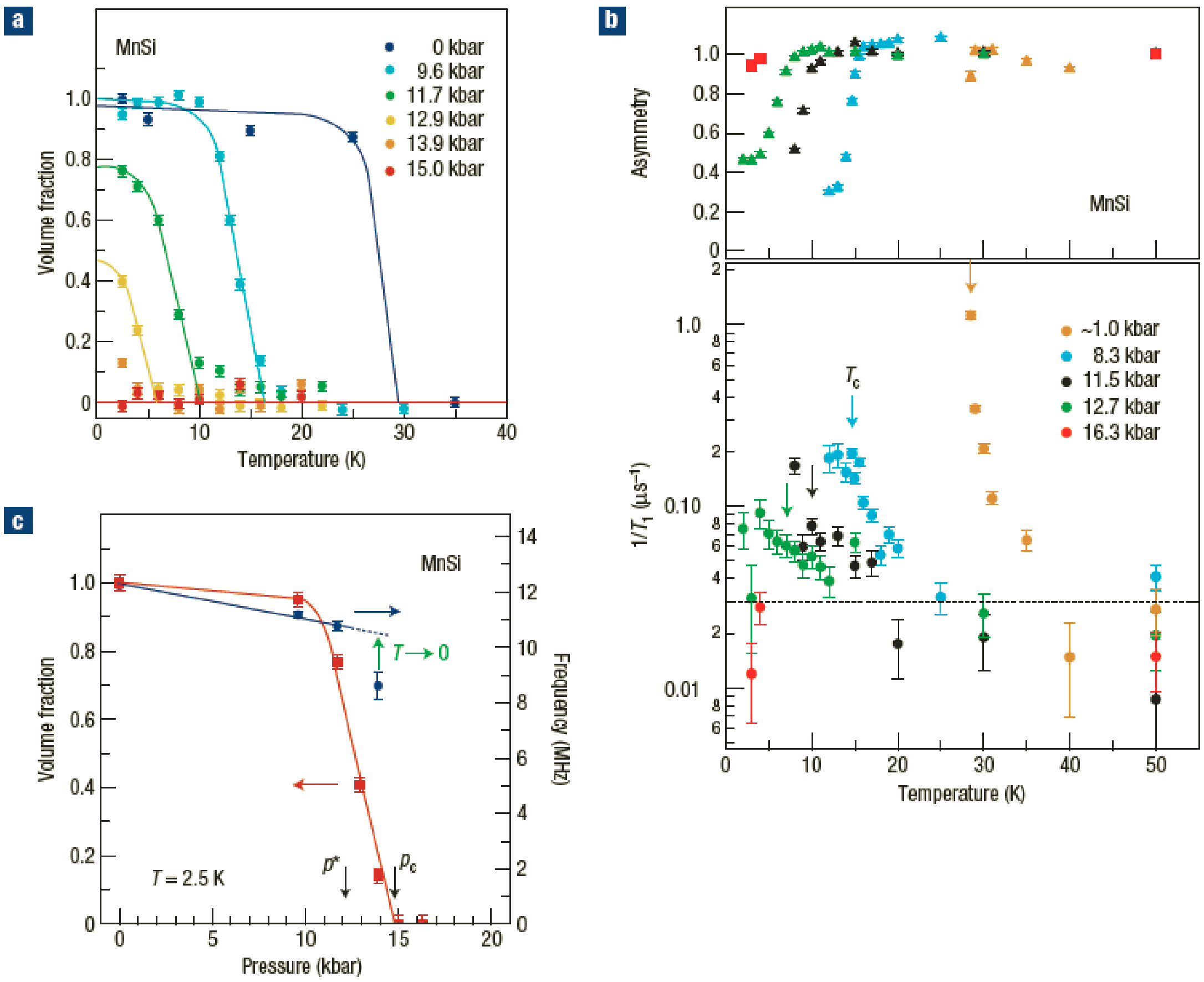

**Figure III-3**: μSR results on the volume fraction, relaxation rate and spin precession frequency in MnSi. (a.) Temperature and pressure dependence of the magnetic volume fraction in single-crystalline MnSi, with static magnetic order determined in weak transverse fields of 100G. $V_f$ remains finite at $T \rightarrow 0$ at the pressure $p$ between 11.7 and 13.9 kbar, indicating phase separation between magnetically ordered and paramagnetic volumes. (b.) The muon spin relaxation rate $1/T_1$ and the relative magnitude of the corresponding muon asymmetry in MnSi in an LF of 200 G. Divergent critical behavior of $1/T_1$ , seen at $p \rightarrow 1$ kbar, is gradually suppressed with increasing pressure. No anomaly of $1/T_1$ is seen at $T_c$ (indicated by arrows) at $p$=12.7kbar ($p^* < p < p_c$ ). At $p$=16.3kbar, $1/T_1$ becomes smaller than the technical limit of detection, indicated by the dashed line. (c.) Pressure dependence of $V_f$ and the zero-field muon spin precession frequency at $T$=2.5 K. The finite frequency near $p_c$ indicates a first-order phase transition. The frequency at $p$=13.9kbar at $T$=2.5K ~ 0.5$T$c is expected to increase for $T \rightarrow 0$ as illustrated by the green arrow.

## *Magnetic Multipoles*

As a purely dipole probe of magnetism, μSR has also been able to meaningfully contribute to the discussion surrounding more unconventional forms of magnetism, such as those involving higher-order magnetic multipoles [Kuramoto2009]. One recent example is $NpO_2$, an actinide compound characterized by a large heat capacity anomaly at $T_0$ = 25.5K [Osborne1953], but without any sign of spin order or structural distortions apparent in neutron scattering [Cariuffo1987], Mossbauer resonance [Friedt1985] or x-ray diffraction data [Mannix1999]. Somewhat surprisingly, μSR demonstrated a well-defined precession signal below $T_0$, indicative of time-reversal symmetry breaking at low temperatures [Kopmann1998, Figure III-4]. It was in attempting to reconcile the seemingly contradictory μSR measurements with previous results that theorists first suggested that the transition at $T_0$ is characterized by the ordering of magnetic octupoles [Santini2000; Santini2006]. This hypothesis has since been confirmed by resonant x-ray scattering. In a similar vein, measurements of the local field by μSR have helped identify octupolar order parameters in the substitutional alloy $Ce_xLa_{1-x}B_6$ [Takagiwa2002; Kubo2003; Kubo2004] and the filled skudderudite $SmRu_4P_{12}$ [Hachitani2006; Ito2007].

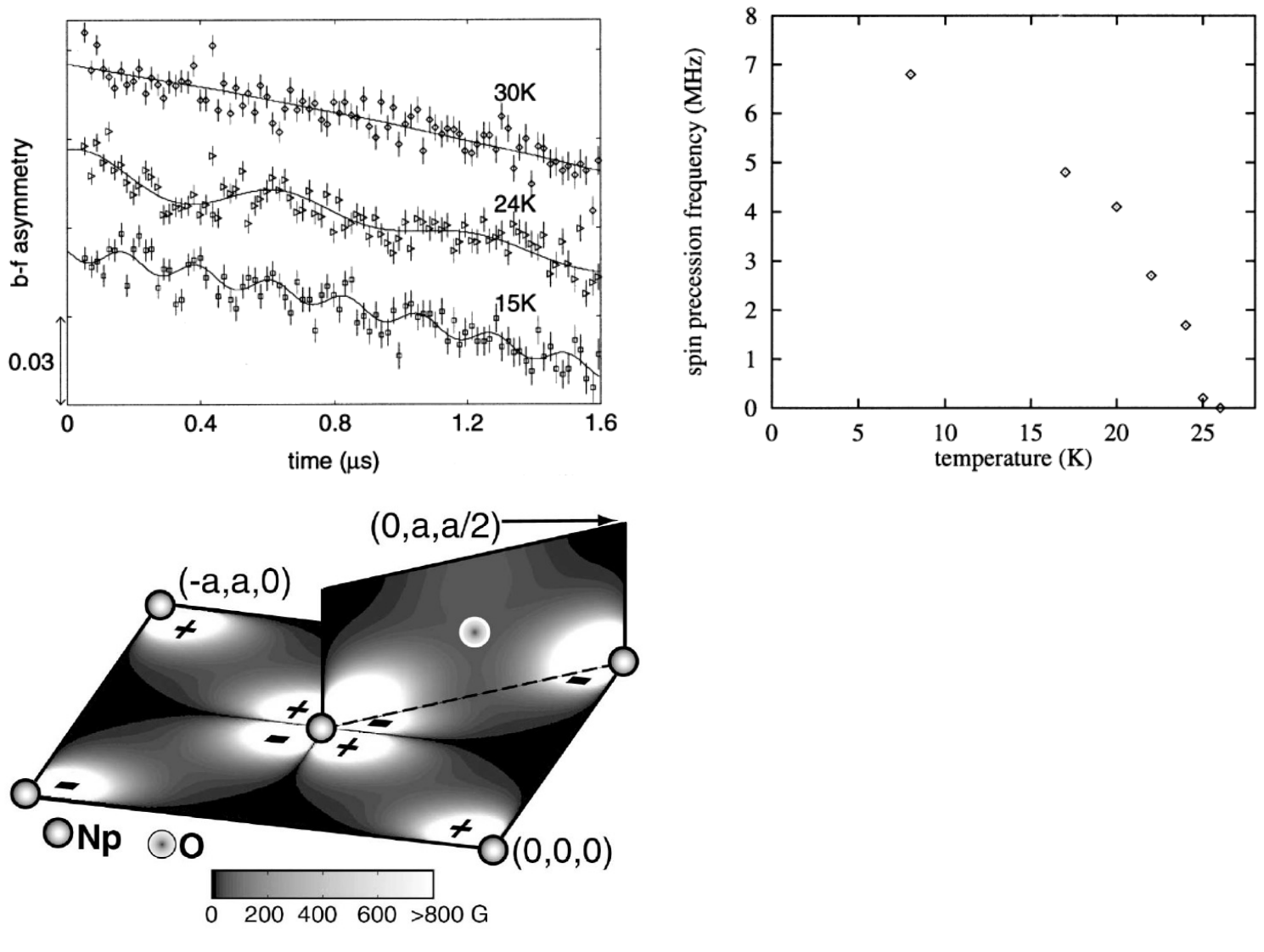

**Figure III-4**: Time-reversal symmetry breaking in $NpO_2$- (left) Some ZF-μSR spectra recorded for $NpO_2$. Whereas the μSR asymmetry decays monotonically at T=30K, the appearance of oscillations at lower temperatures indicates the onset of magnetic order in this sample. The temperature dependence of the oscillation frequency (center), implies a magnetic transition with $T_c$~25K. From [Kopmann1998]. This observation was at odds with existing neutron and Mossbauer measurements, and led to the prediction of an octupolar order parameter in this system (right).

*Reduced Dimensionality and Nano-structures*

Traditional forms of μSR have long been used to study singlet formation, order, fluctuations, and defects in quasi-1D spin chains and ladders, or quasi-2D planar materials. However, with the advent of low-energy μSR, several new fruitful avenues of research have emerged.

Low energy μSR (LEM) is being used extensively to investigate the role of reduced dimensionality in determining magnetic or superconducting properties. In one study, atomic layer-by-layer molecular beam epitaxy was used to construct several superlattices comprising *n* layers of antiferrromagnetic $La_2CuO_4$ separated by non-magnetic spacer layers of $La_{2-x}Sr_xCuO_4$. It was found that, upon approaching the two-dimensional limit, antiferromagnetic long-range order gives way to a novel spin-liquid state characterized by strong quantum fluctuations and reduced spin-stiffness [Suter2011]. In another study, a quasi-1 dimensional "wire" of superconducting $La_{1.94}Sr_{0.06}CuO_4$ was investigated to elucidate the coupling between the superconducting and magnetic order parameters in this system. It was found that the Néel temperature associated with the antiferromagnetic order increased as superconductivity was suppressed by high current density, implying a repulsive interaction between the two order parameters [Shay2009]. There are now several examples of materials, such as $TbPc_2$ nanomagnets [Hofman2012] or spin glasses [Pratt2005; Morenzoni2008], where reduced sample thickness is seen to enhance spin fluctuations over that seen in the bulk.

Heterostructures investigated include Fe/Ag/Fe trilayers [Luetkens2003], and nickel-oxide superstructures [Boris2011]. In the Fe-Ag-Fe study, LEM was able to track the oscillating spin polarization density in the spacer layer, integral to the exchange coupling between the ferromagnetic end layers and leading to long-range order. In a clever study by Boris *et al* [Boris2011], it was demonstrated, through the investigation with LEM of carefully grown superlattices, that correlation effects could induce an insulating and antiferromagnetic ground state in layers of $LaNiO_3$, which is a paramagnetic metal in the bulk (Figure III-5). This study stands as a prototypical example of how a man-made heterostructure can be used to control and manipulate intrinsic material properties such as electron-electron correlations. Through studies such as these, LEM is emerging as one of the most powerful methods in materials science for exploring depth dependent magnetic behavior in artificial systems.

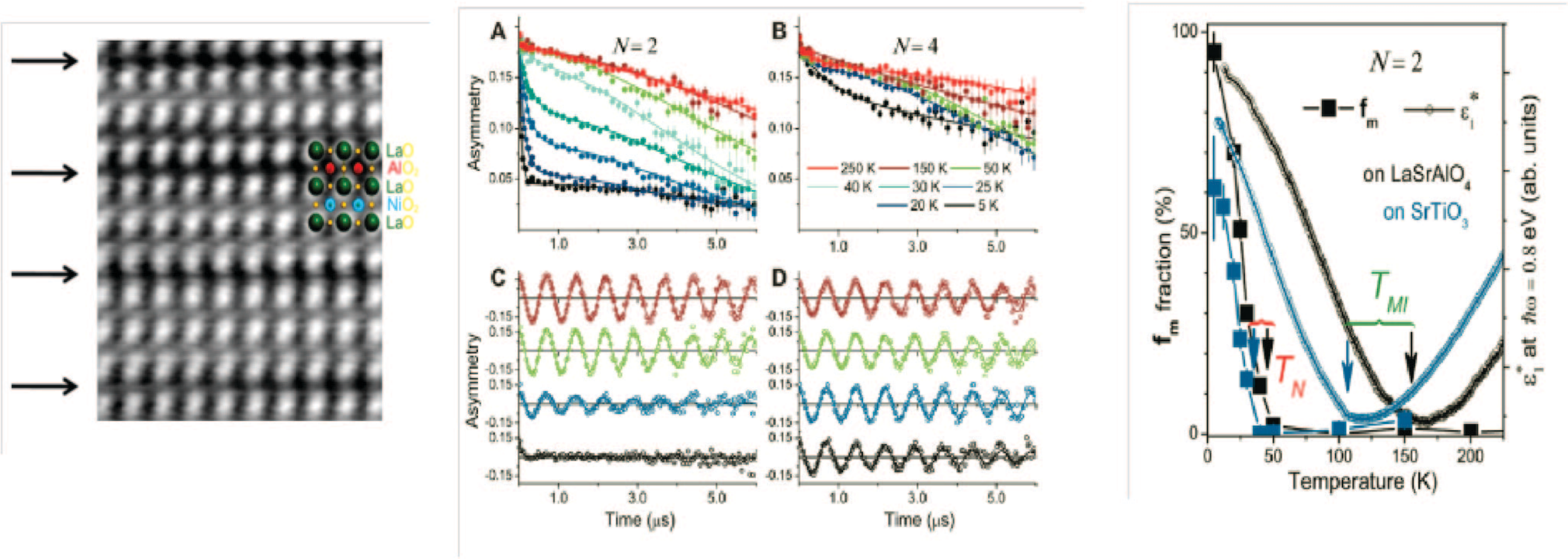

**Figure III-5**: μSR study of nickel oxide superstructures- (left) A high-angle dark field TEM image of one of the samples studies by Boris *et al*. with low-energy μSR [Boris2011]. This sample contains two unit cells of $LaNiO_3$ atop one unit cell of $LaAlO_3$. (middle) A series of zero-field (A,B) and weak transverse field (C,D) μSR spectra on samples with N=2 (A,C) and N=4 (B,D) layered superlattices of $LaNiO_3$, grown on $LaSrAlO_4$. In contrast to N=4, the N=2 samples distinctly show the onset of magnetic order in the entire volume fraction at low temperature. (right) Plots of the magnetic volume fraction from μSR and the real part of the dielectric function for two N=2 samples, each as a function of temperature. The results show a sequence of static charge and spin order in these samples, rather than the paramagnetic metallic behavior seen in the bulk.

### *Superconductivity*

Beyond magnetism, one of the most active areas of μSR research is the investigation of unconventional superconductivity. Details about the vortex lattice in a type-II superconductor can be inferred from the distribution of local magnetic fields, measured directly in a transverse field μSR experiment.

There is a large body of work, not only on the high-transition temperature ($T_c$) copper-oxide superconductors, but also iron arsenide/chalcogenide, heavy fermion, boron carbide and organic superconducting materials, among others. In each of these families, the superconducting ground state is known or thought to exist near a state containing magnetic order, and the interaction between magnetic and superconducting order parameters is a frequent research theme. With its exquisite sensitivity to both magnetic and superconducting order parameters with spatial resolution of tens of Angstroms, μSR has been able to contribute significantly to this body of scientific knowledge.

*Phase separation versus co-existence and the superconducting phase diagram*

Particularly prominent in the literature are instances where μSR has identified static or dynamic magnetic order in a material of interest and investigated phase separation or co-existence with superconductivity at the microscopic scale. The issue of phase coexistence has been addressed in a number of materials, with results that depend on the superconducting family being investigated. For example, in the hybrid ruthenate-cuprate $RuSr_2GdCu_2O_8$, the co-existence of ferromagnetism and superconductivity has been firmly established [Bernhard1999]. In contrast, in the heavy-fermion material $CeCu_{2.2}Si_2$, magnetic order and superconductivity are seen to occupy different volumes of the sample and compete [Luke1998]. Such issues are of fundamental importance and key to unraveling the complex physics underlying unconventional superconductivity. As such, μSR studies of magnetism and the construction of phase diagrams based on this data have played an influential role in conversation surrounding many compounds.

Within the community of superconductivity researchers, μSR is perhaps best known for its many early contributions to the exploration of the copper-oxide high-temperature superconductors, and the significant work that has followed. These contributions were made possible due to the strong signal-to-noise ratio of μSR and the ability of the probe to gain useful information from powder samples. For example, the μSR technique was used to identify a magnetic freezing transition at low temperatures in underdoped high-$T_c$ superconductors shortly after their discovery [Budnick1988; Wiedinger1989; Kiefl1989] and was also the first technique to detect static magnetic order in the parent compounds $Ln_2CuO_{4-y}$ (Ln ≡ Nd, Pr, Sm) of the electron-superconductors [Luke1989]. μSR was the first technique to observe incommensurate magnetic order in the so-called `1/8 compounds' [Luke1991], where superconducting transition temperature is suppressed at dopings where magnetism is stabilized. In related compounds, phase separated incommensurate order and superconductivity were observed simultaneously [A.T. Savici Physical Review B **66** (2002) 014524], and magnetic volume fraction was shown to be controllable with applied magnetic field [Savici2005]. A spin-glass phase identified in underdoped $La_{2-x}Sr_xCuO_4$ and $YBa_{1-x}Ca_xCuO_6$ [Harshman1988; Niedermayer1998] is thought to co-exist with superconductivity on the nanoscale for some range of hole doping. Coexistence of glassy magnetism and superconductivity has also been demonstrated in powder and crystals of $YBaCuO_{7-\delta}$ [Sanna2004; Miller2006], which is of particular significance since these materials are thought to be cleaner than their cation-doped counterparts. In the electron-doped cuprate $Pr_{2-x}Ce_x\ CuO_4$, random magnetic moments are found to grow into long range antiferromagnetism with application of fields as small as 90 Oe [Sonier2003].

More recently, there has been intense interest in the iron arsenide/chalcogenide family of high-temperature superconductors. These materials were discovered in 2008 and found to share much of the phenomenology seen in the widely studied cuprate superconductors, despite some qualitative differences. Almost immediately after their discovery, μSR was quick to identify signals associated with commensurate order in the parent compounds of a number of distinct iron-pnictide families and indications of incommensuration with slight doping in some materials [Klauss2008; Kaneko2008; Aczel2008; Carlo2009; Park2009]. Many early studies suggested the macroscopic phase separation, especially in electron-doped materials [Park2009; Goko2009; Sanna2011; Laplace2012]. However, as with the cuprates, microscopic phase co-existence was eventually confirmed in a number of different iron-arsenide families [Drew2009; Marsik2010; Bernhard2012], and μSR is playing a fundamental role in developing standard phase diagrams [e.g., Drew2009; Leutkens2009; Shermandini2011, Figure III-6]. Work continues to identify and explore the magnetic properties of new families of iron-based superconductors- for example $Li_x(NH_2)_y(NH_3)_{1-y}Fe_2Se_2$, where the superconducting $T_c$ was seen to be enhanced through the addition of a molecular space layer between planes of superconducting FeSe [Burrard-Lucas2013]. Similar phenomenology and important connections to magnetism are being found in other exotic superconductors, such as $Sr_{2-x}Ca_xRuO_4$ [Carlo2012] or the so-called '115' heavy fermion compounds (e.g. $CeCoIn_5$) [Higemoto2002; Schenck2002; Spehling2009].

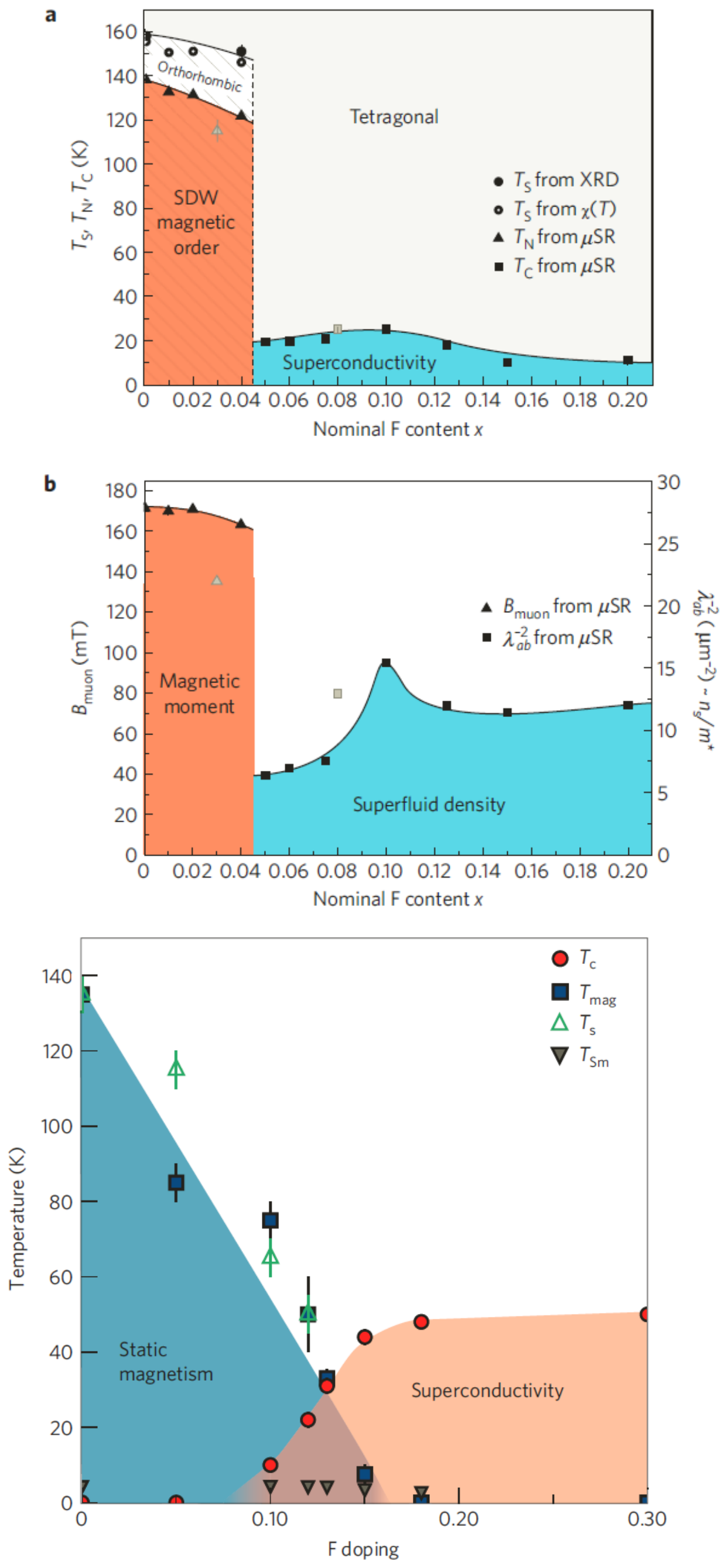

**Figure III-6**: Dueling phase diagrams for the iron based superconducting compounds, published in the same issue of Nature Materials in 2009. The upper is the result of a μSR, x-ray and Mossbauer investigation of LaO1-xFxFeAs and seems to indicate that the transition temperature and moment size associated with a spin-density-wave order drop off precipitously with doping, before superconductivity develops [Luetkins2009]. The lower is a phase diagram resulting from μSR, resistivity and magnetization study of SmFeAsO1-xFx, which seems to indicate a more gradual demise of magnetism with doping and a distinct co-existence region [Drew2009].

*Penetration depth and gap symmetries*

In a μSR experiment, superconductivity is most frequently observed as large internal field inhomogeneities resulting from quantized flux lines (*vortices*) in the mixed state of a type-II superconductor when a transverse field is applied. The details of the μSR lineshape (Fourier transform of field distribution) in this circumstance depends on the spatial arrangement of these vortices in the sample, which itself carries fundamental information about the superconducting state [Sonier2000]. In this sense, μSR is a powerful complement to small-angle neutron scattering (SANS), which can find similar information in reciprocal space when large, clean single crystals are available.

Very common are μSR and SANS measurements of the magnetic penetration depth, λ, in the vortex state of a superconductor. λ is an important characteristic length scale whose behavior is linked to the underlying physical mechanism of superconductivity, and can be extracted from the second moment of the field distribution when a well-ordered vortex lattice exists. A distinct advantage for μSR over SANS in this regard is that useful information can be extracted from either powders or small single crystals, with increased sample purity and homogeneity. A series of now famous μSR experiments [Uemura1991; Uemura1993; Niedermayer1993] performed in the early 1990's on samples of cuprate superconductors with different charge carrier concentrations established universal behavior for the variation of the superconducting transition temperature $T_c$ with $\lambda^{-2}$ in these compounds. In the London limit of type-II superconductors, $\lambda^{-2}$ is proportional to superfluid density. As such, the experimental “Uemura plot”, as it is known today, is recognized by the superconductivity community as a major accomplishment and one of the key properties of high-$T_c$ superconductors requiring an explanation by any ultimate theory (Figure III-7).

As interesting as the absolute value, are studies of the *temperature dependence* of λ, which has proven useful in determining the isotropy of the superconducting gap. For example, the observation of a linear temperature dependence by μSR is in the early 1990's confirmed the *d*-wave pairing symmetry of the Cooper pairs in the vortex state of the high-$T_c$ materials [Sonier1994]. In striking contrast, experiments on the iron-based high-$T_c$ superconductors have revealed the existence of a two *s*-wave pairing gaps [Williams2009], often touted as one of the biggest difference between the two classes of compounds.

A similar study of the ruthenate superconductor $Sr_2RuO_4$ found no nodes in the gap, but rather evidence for a square flux line lattice [Luke2000], in contrast to the usual hexagonal lattice. With its heightened sensitivity to local fields, μSR was additionally able to discern the onset of a weak (~0.5 G) time-reversal symmetry-breaking field at the superconducting $T_c$ [Luke1998_2]. For these reasons, $Sr_2RuO_4$ is now thought to contain a triplet $p_x+ip_y$ pairing symmetry, and is considered one of the strongest candidate materials to exhibit *topological superconductivity* [Qi2011]. Time-reversal symmetry breaking fields seen with muons have similarly been used to identify exotic gap functions in heavy fermion materials (U,Th)$Be_{13}$ [Heffner1990] and $UPt_3$ [Luke 1993], praseodymium-based superconductors $PrOs_4Sb_{12}$ [Aoki2003] and $PrPt_4Ge_{12}$ [Maisuradze2010] and noncentrosymmetric superconductors $LaNiGa_2$ [Hillier2012] and $LaNiC_2$ [Hillier2009].

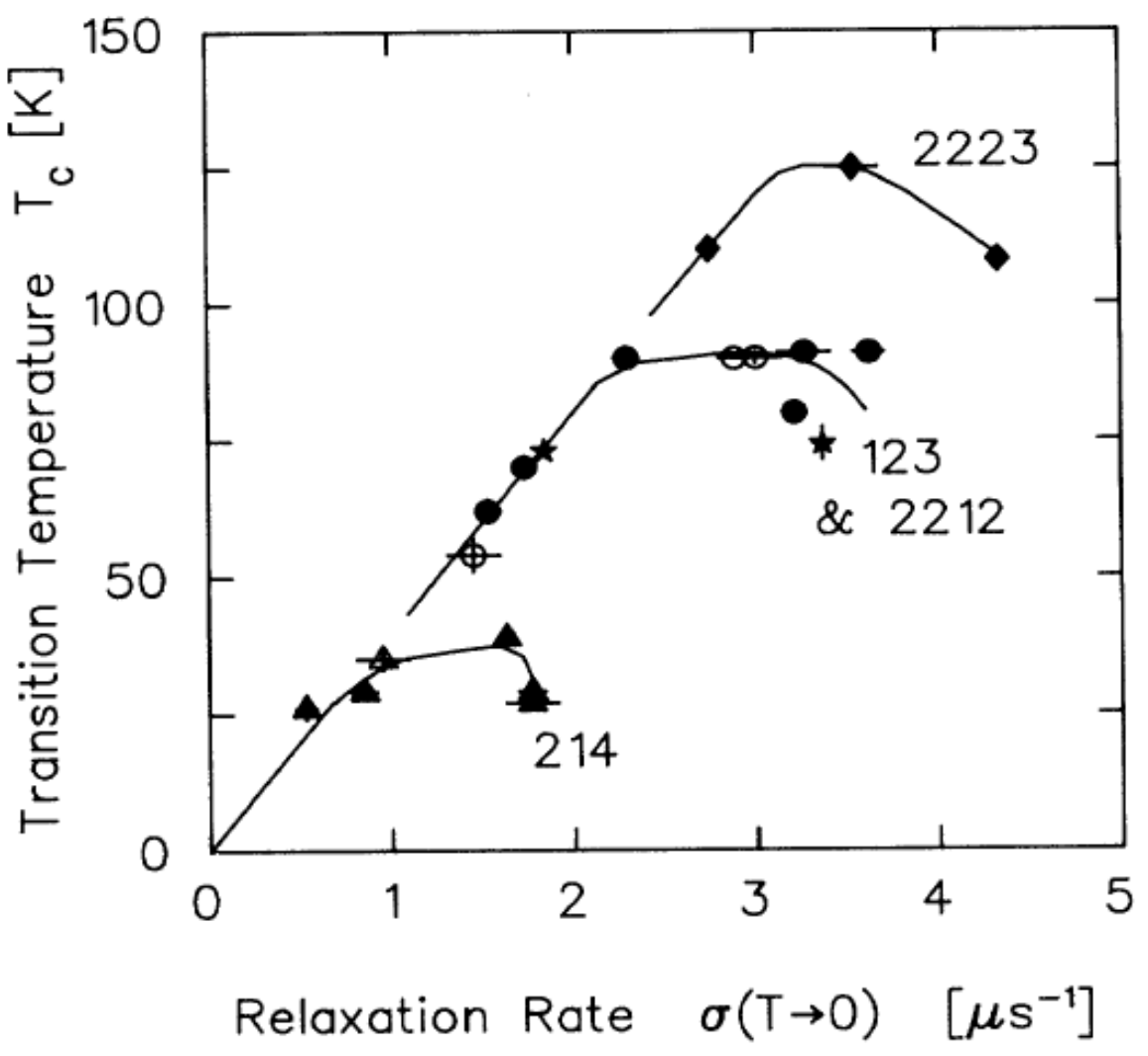

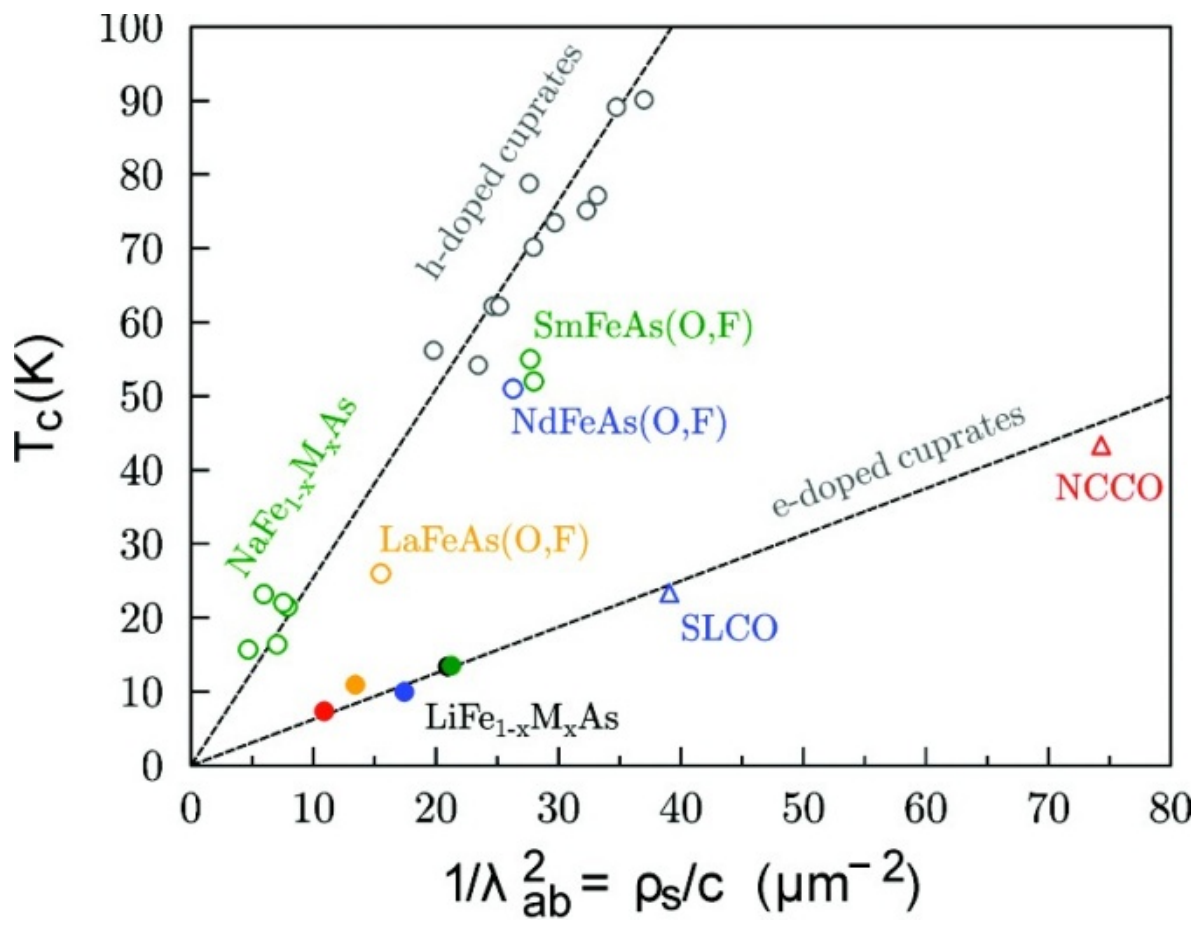

**Figure III-7**: (upper) The original Uemura plot, first published in 1989 [Uemura1989]. Shown is the superconducting transition temperature of several hole-doped cuprate superconducting materials versus the Gaussian relaxation rate of the μSR asymmetry. For the mixed phase of a type-II superconductor, this latter quantity is proportional to the inverse penetration depth squared. (lower) A recent version of the same plot, which also includes electron-doped cuprates and data from the new class of iron-based superconductors [Pitcher2010].

*Vortex matter*

In the past decade, there has been increasing interest in the flux line lattice itself as a mesoscale structure, where the vortex is seen as the fundamental building block. Muon spin rotation (μSR) and small-angle neutron scattering (SANS) have been two very important techniques used in the investigation of the structure of vortices within the bulk of superconductors. Low-energy μSR is also able to explore the character of vortices at surfaces. In addition to studies of the position and symmetry of vortex arrays, these techniques can be used to investigate vortex fluctuations, pinning, melting, and decomposition of the flux lines into 2-dimensional "pancake" vortices in anisotropic materials.

Vortex-lattice *melting* in a high-$T_c$ superconductor was first observed by a group using both μSR and SANS to study the anisotropic material $Bi_{2.15}Sr_{1.85}CaCu_2O_{8+\delta}$ [Lee1993; Cubitt1993]. The μSR study also provided the first indication of melting into a liquid of 2-dimensional "pancake" vortices in independent layers above a crossover field. A later μSR study has provided *solid* proof for the existence of pancake vortices in this material [Kossler1998], and a μSR study of under- and overdoped $Bi_2Sr_2CaCu_2O_{8+\delta}$ [Balsius1999] has provided the first evidence for a *two-stage* melting transition — in which the intralayer coupling of vortices is first overcome by thermal fluctuations, followed by interlayer decoupling of the pancake vortices. Melting of the vortex lattice was clearly observed by μSR deep in the superconducting state of underdoped $YBa_2Cu_3O_{7-\delta}$ [Sonier2000_2], whereas other techniques had only observed melting at low field near the superconducting transition temperature $T_c$.

A vortex line lattice in the organic superconductor κ-(BEDT-TTF)$_2$Cu(SCN)$_2$ was the first clearly identified via an investigation with μSR [Lee1997]. Upon increasing the field a dimensional crossover in the vortex structure was observed. An additional crossover observed at low fields and high temperature was consistent with the theoretical prediction for the thermally induced breakup of vortex lines comprised of weakly coupled pancake vortices.

In recent years, μSR has furnished a great deal of information on the shape and size of the *vortex core size* as a function of temperature and magnetic field [for a timely review, see J.E. Sonier2007, also Figure III-8]. It now appreciated that vortex cores contain bound quasi-particles states, whose character depends on the properties of the underlying superconducting state. It has been seen in several μSR studies of *s*-wave and *d*-wave superconductors that the size of the vortex cores shrink with increasing applied field [Sonier1999; Kadono2001; Ohishi2002; Price2002; Sonier2004; Callaghan2005; Kadono2006], understood to result from the delocalization of highest energy bound core quasiparticle states [Ichioka1999]. It has been shown that the delocalized quasiparticle states inferred from μSR can quantitatively explain measured transport coefficients in some *s*-wave superconductors [Callaghan2005]. In many systems, these quasiparticle interactions are even seen to drive a change in the symmetry of the vortex lattice, for example in the form of a hexagonal to square transition observed with applied field [Ohishi2002; Sonier2004; Kadono2006]. In fact, the opening angle of the vortex lattice structure is useful information and has been used to comment on the form taken by the superconducting gap in some materials [Yaouanc1998]. In case of the cuprates [Miller2002; Sonier2007_2], there is growing evidence in μSR spectra of local antiferromagnetic correlations in vortex cores, further fueling debate about the interplay between magnetism and superconductivity in these materials.

*Depth-dependent studies*

As with magnetism, the advent of low-energy μSR is providing basic and material-specific information about superconductivity in reduced dimensions. Demonstrating the relatively unique role that low-energy μSR plays as a depth sensitive measure of fields, while studying a thin-film of $YBa_2Cu_3O_{7-\delta}$, researchers measured *for the first time, in 2000,* the exponential screening of an applied field at the surface of a superconductor- one of the defining characteristics of the classical theory of superconductivity and first predicted 65 years earlier [Jackson2000]. Building off this historic success, researchers later used deviations from exponential decay in a thin film of superconducting Pb to verify non-local (i.e. Pippard) effects in superconductors, also predicted over 50 years before [Suter2004, Figure III-8]. Separate studies confirmed interesting predictions for the mixed phase of superconductivity, such as the broadening of vortex core sizes at the sample surfaces [Niedermayer1999].

The power of this new probe is also being leveraged to further the ongoing debate surrounding high-temperature cuprate superconductivity- for example by identifying an isotope effect in optimally-doped $YBa_2Cu_3O_{7-\delta}$ [Khasanov 2004] or observing the Giant Proximity Effect in strongly underdoped $La_{2-x}Sr_xCuO_4$ films when placed near optimally-doped samples [Morenzoni2011].

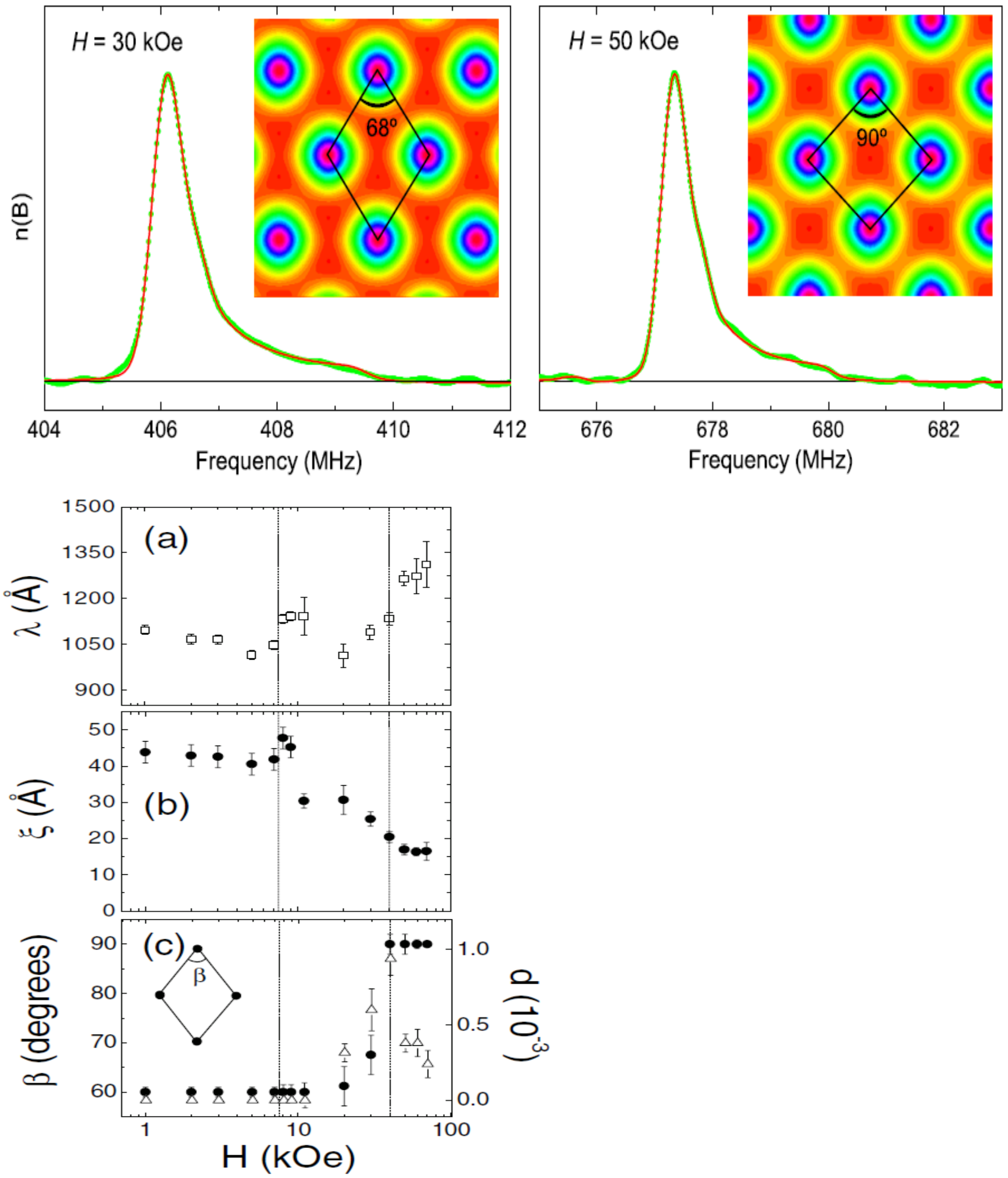

**Figure III-8**: (Upper) Fourier transforms of two μSR spectra taken with TF=30kOe and 50kOe on conventional superconductor $V_3Si$. Green curves represent data and red fit lines Ginzbrug-Landau vortex lattice models. Also shown are contour maps of the field distribution, inferred from fits. (Lower) Material parameters as a function of applied field, extracted from fits within the model. Apparent is a change in vortex lattice symmetry with applied field, accompanied by a field-induced vortex core shrinking. From [Sonier2007; Sonier2004].

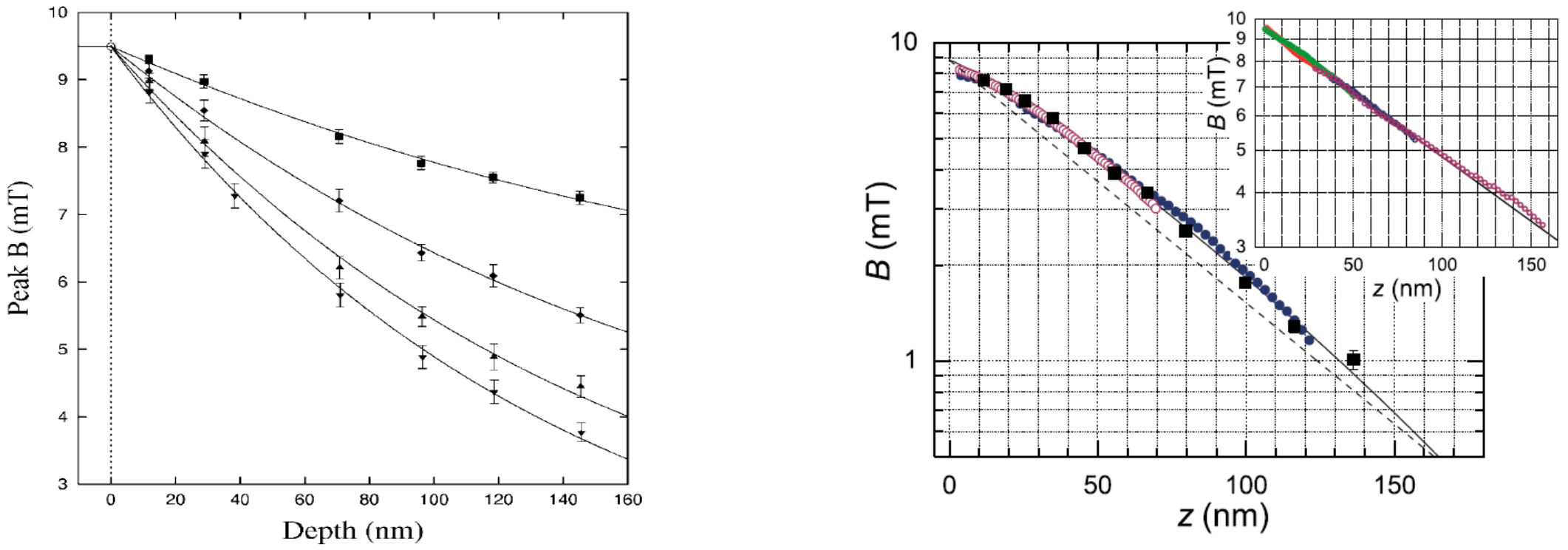

**Figure III-9**: (left) Low-energy μSR measurements of the local magnetic field at the surface of the superconductor $YBa_2Cu_3O_{7-\delta}$ in the Meissner state. Data was taken at T=20K, 50K, 70K and 80K, from top to bottom. Solid lines are fits to expectations from classical theories of superconductivity [Jackson2000]. (right) Main panel shows a plot of field as a function of depth in Pb, known to be a strongly coupled superconductor, at T=3K (main panel).The dotted line represents pure exponential expulsion, and deviation from this line is a signature of non-local (i.e., Pippard) effects. The same measurement for YBCO (inset) shows the field expelled exponentially at the surface [Suter2004].

### *Quantum Diffusion and Battery Materials*

Qualitatively different from the studies above are experiments which use the muon as an *active* probe of material properties. One notable example is the groundbreaking work done using μSR to develop theories of quantum diffusion theory, especially as it relates to light interstitials in solid state materials. The study of hydrogen diffusion in metals has been of interest for well over a century to both pure and applied science communities [Fukai1985]. Starting in the 1970's, μSR was used to test classical theories, in which it was assumed that light interstitials diffused by hopping over barrier potentials between trapping sites. By the early 1980's, it was clear from muon studies at low temperatures that the classical theories were incomplete. The development of modern theories of *quantum tunneling* (i.e. coherent transport) of interstitial atoms in materials stems directly from efforts to understand these early μSR experiments. With the advent of hydrogen storage technologies and battery materials which depend on the diffusion of $Li^+$ cations, quantum tunneling theories are of intense current interest. The muon has a mass of approximately 1/9$^{th}$ the proton's mass, is available in both the positive (muon) and neutral (muonium) forms, and affords the capability to extract detailed microscopic correlation/residency times. As a result of these qualities, μSR has and continues to provide the best and often the sole tests of these theories.

Examples can be found in the recent experimental and theoretical efforts to understand the role of dissipation (i.e., coupling the muon's quantum motion with the electronic and lattice degrees of freedom) or disorder in assisting or limiting diffusive transport. It was understood early on that the temperature dependence of the diffusion rate of muons in materials should vary as $T^{-\alpha}$, with $\alpha$~9. However, experiments on simple metals revealed much smaller values of $\alpha$~0.6-0.7. This was understood as an electron drag effect, where the positive muon moves as a screened charge interacting with both host phonons and electrons, and quantum tunneling processes are hindered by the inability of the electronic screening cloud to follow the muon adiabatically. The integral role of host electrons in the diffusion process was ingeniously demonstrated by a μSR investigation of aluminum at low temperatures. Originally identified as a material with modest diffusion rates, papers by Karlsonn et al. and Kadano *et al* demonstrated that diffusion significantly increased when the material was cooled below its superconducting transition, where electron interactions are "gapped out" and rendered ineffective. They further showed that diffusion can again be hindered in the presence of a modest field, when the delicate superconducting state is destroyed [Karlsonn1995; Kadono1997, Figure III-10].

Experiments that probe the interactions of the tunneling particle (*viz*. neutral muonium in an insulator) with either dynamic phonon or static structural/impurity types of disorder constitute another class of μSR experiments. A study of solid $N_2$ [Storchak1994] is the first confirmed case of the two-phonon mechanism for quantum diffusion of muonium (Mu), showing the previously predicted $T^7$ temperature dependence in the low-$T$ regime and a $T^{-7}$ law at intermediate $T$. Subsequent work on other materials has successfully used neutron scattering measurements of low-energy phonons to predict tunneling rates measured in μSR experiments. The localization of a tunneling particle by static disorder is also revealed by the above work, a theme further explored in alkali halides such as KCl:Na [Kadono1996]. There it was found that for modest impurity concentrations the crystal volume is characteristically (for long range inhomogeneity) divided up into two parts, reflection portions of the sample near and far from "pinning" sites.

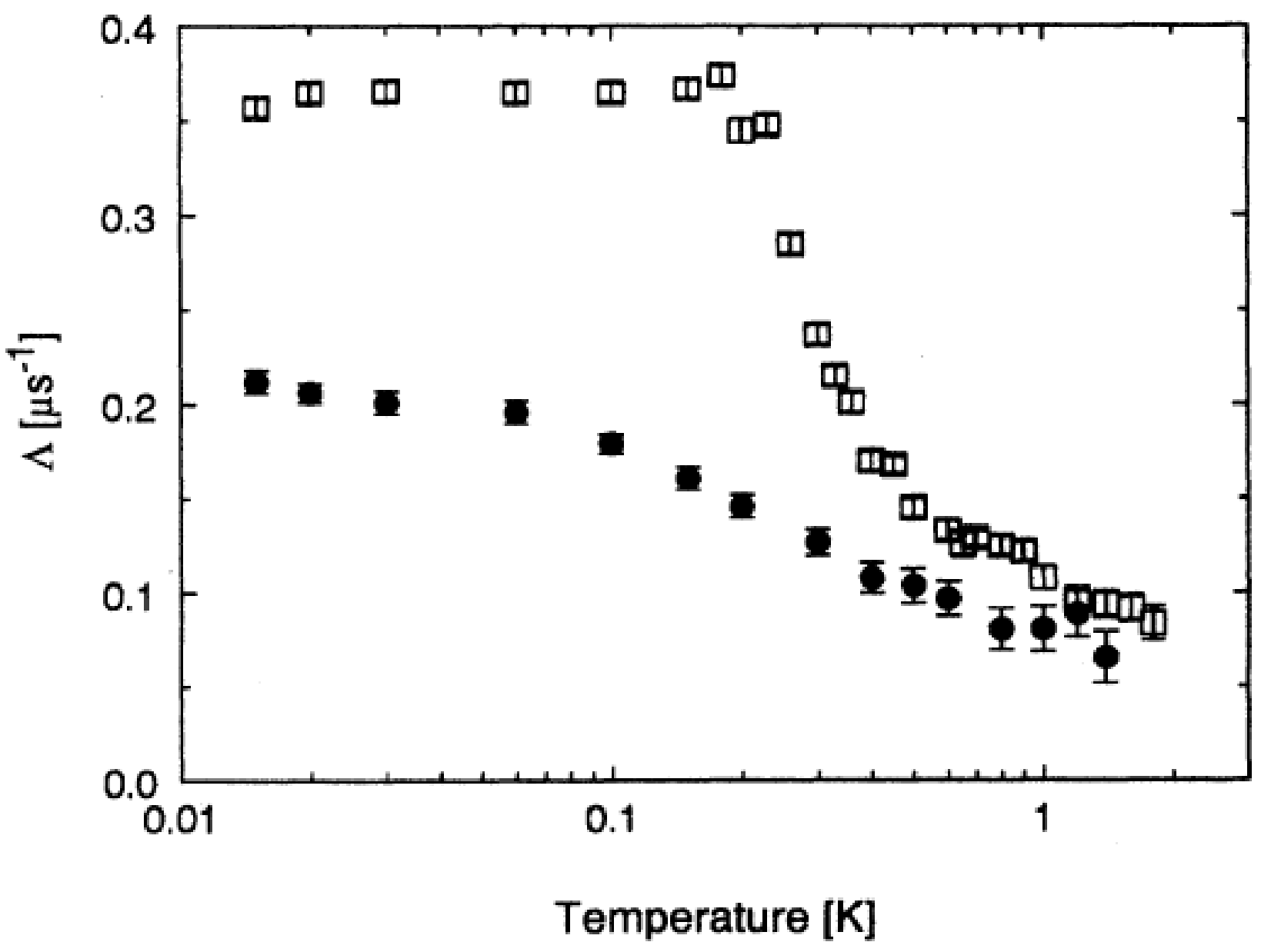

**Figure III-10**: The linewidth (depolarization rate) as a function of temperature for a sample of Li-doped aluminum, which superconducts at 0.5K. Data for the superconducting state (open symbols) are contrasted with data taken in the presence of a weak transverse field (filled symbols), which acts to drive the system back to a normal metallic state. Depolarization rate is related to muon hopping rate, and this data provides strong evidence that drag effects associated with conduction electrons (gapped out in the superconducting state) play a dominant role in diffusion processes in metals at low temperature. From [Karlsonn1995].

μSR-inspired theories of diffusion have found immediate application in the search for reliable hydrogen storage materials, of importance to the development of commercial fuel cell technologies. Furthermore, μSR has been able to comment in a constructive way on the viability of several *next-generation* storage materials, beyond simple hydrogen absorbing metals. One example is Ti-doped sodium alanate, identified recently as a compound with favorable storage and kinetics characteristic for vehicular applications [Bogdanovic1997]. The observation of muon spin oscillations in this non-magnetic materials recently led Kadono *et al.* to suggest that muonium (and by extension hydrogen) forms a radical $H^-$-Mu-$H^-$ bond when at interstitial sites, and it is the forming and breaking of this bond which is the rate-limiting step in hydrogen kinetics in this material [Kadono2008]. Another study identified similar bonds in borohydrides $M(BH_4)_2$ ($M \in \{Li^+, Na^+, K^+]$) and found a distinct correlation between the frequency of H-Mu-H formation and the electronegativity of $M^+$. This observation indicated that μSR can be used as a microscopic indicator of the stability of super-hydogenated $M(BH_4)_2$ states [Sugiyama2010]. Perhaps most surprisingly, one very recent μSR study has suggested the formation of interstitial hydrogen ($CH_2$) groups in sheets of graphene (single-layer graphite) and stable up to 1250 K, identifying this miraculous material as a potentially cheap and effective future hydrogen storage material [Ricco2011, Figure III-11].

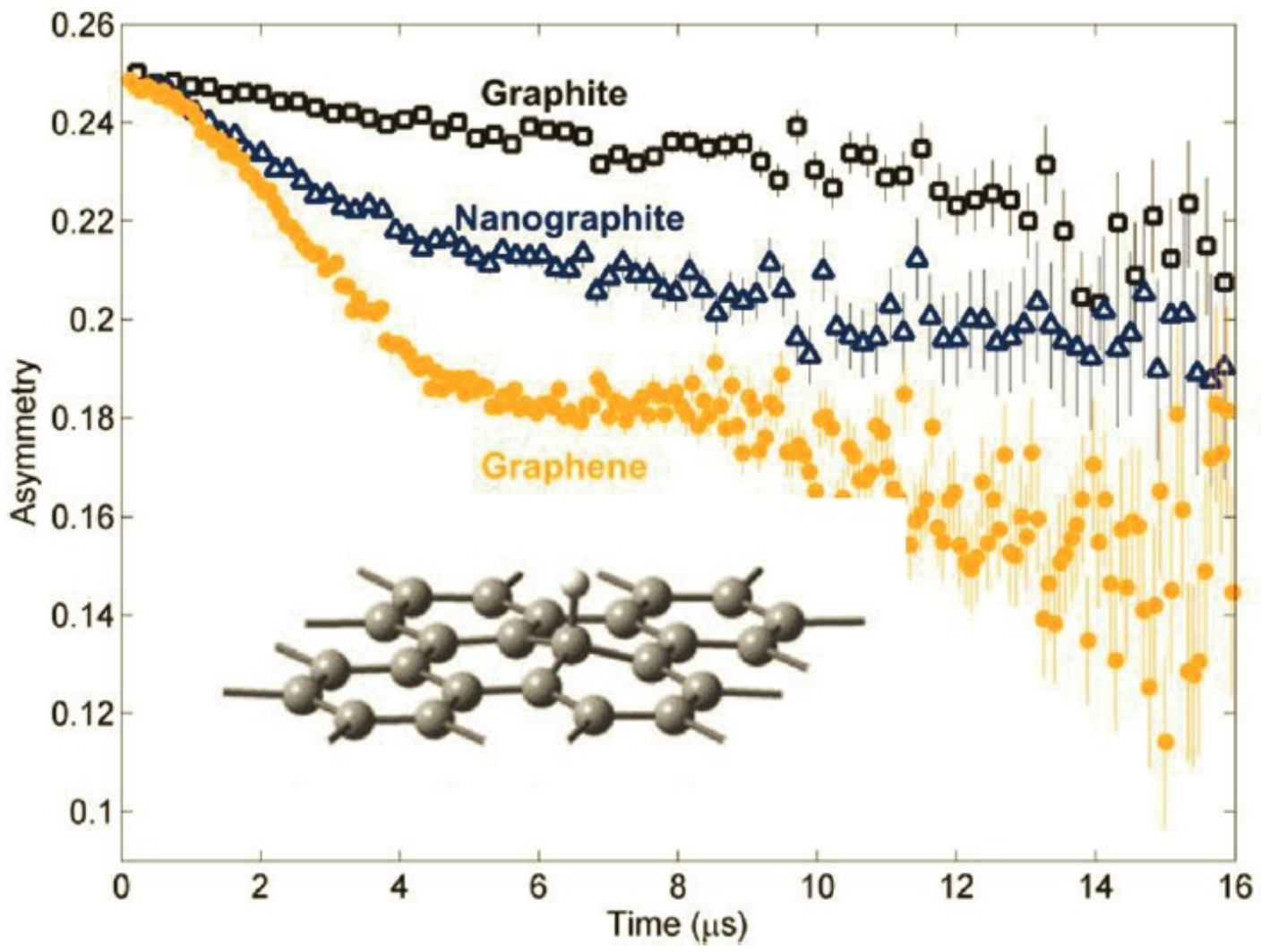

**Figure III-11**: Plots of muon decay asymmetry for three different carbon systems. Of particular note is the clear precession which emerges when the graphite is reduced to a single layer (i.e. graphene). Like a hydrogen atom, on an ideal graphene layer muonium gets chemisorbed, as shown in the inset. From [Ricco2011].

Tangentially related are μSR investigations of lithium-based compounds, which have found a role in the last 15 years as cathodes in the newest generation of reliable rechargeable batteries. In these systems, the operation of the battery depends on the flow of $Li^+$ cations between two insertion electrodes, and the diffusion constant of this cation is of primary importance. Despite the long research history on Li-ion batteries, a proper characterization of lithium diffusion in these materials was only measured in 2000 using μSR. Through a combination of zero-field and longitudinal-field μSR measurements on the spinel material $Li_x[Mn_{1.96}Li_{0.04}]O_4$, Kaiser *et al* were able to measure characteristic hopping rates and ideal "charging temperatures" for two different values of $x$ [Kaiser2000]. This was possible since the muons in these systems hydrogen bond to atomic $O^{2-}$ and are rendered immobile for temperatures of interest. In recent work on the quasi-2D system $Li_xCoO_2$ [Sugiyama2009] and the phospho-olivines [Sugiyama2012], Sugiyama *et al* have developed μSR as a powerful tool for measuring $Li^+$ self-diffusion constants, complementing conventional NMR measurements which are complicated by the presence of transition-metal moments [Tomeno1998; Nakamura1999].

### *Semiconductor Physics*

Introduction of hydrogen in the process of semiconductor fabrication is a very common method of passivating active electrical impurities. The microscopic details of how this occurs are difficult to access with standard magnetic resonance techniques due to the low concentrations of hydrogen typically used. Conversely, the use of muon as a proxy for a light isotope of hydrogen allows researchers to explore the effects of a single dopant in semiconductor samples. In many cases, μSR studies has been the primary source of detailed information on the various charge states and dynamics of isolated hydrogen in semiconductors, often confirmed *a posteriori* with alternate methods. Early μSR investigations of semiconductors identified the static properties of the muon and its paramagnetic state muonium (*i.e.* with its bound electron) charge states. More recent work has focused on ascertaining the dynamics of the charge states and/or site migrations of this hydrogen-like impurity. Current studies focus on a broad range of technologically important semiconductors, such as amorphous silicon where hydrogen mobilization is the major cause of solar cell degradation.

The results of μSR studies in semiconductors are widely recognized by researchers who investigate defects to be the main source of experimental information on *isolated* hydrogen in bulk semiconductors. Hydrogen easily incorporates into semiconductors, such as during crystal growth, film deposition, or device processing steps, and forms stable bound states with intentional dopants, defects and other types of impurities. These interactions result in a dramatic, and often unexpected, modification of the electrical and optical properties of the host. Hydrogen has a high diffusivity and reactivity with other defects, and is usually studied minutes or even days after it is introduced into the sample. Hence, although it is almost always detected as part of a complex with other defects, direct information on the structure of isolated hydrogen is virtually non-existent. In particular, only a single isolated hydrogen center has been characterized, the AA9 center in silicon, first detected using EPR. Indeed, most of the experimental information on isolated hydrogen comes instead from μSR studies of muonium (Mu ≡ $\mu^+e^-$), which is, in essence, a light hydrogen-like atom [Cox2009; Patterson1988; Kiefl1990; Chow1998; Lichti1999].

The majority of experiments on the rich physics of muonium  in semiconductors falls into one of two categories: (1) investigations of the electronic structure of muonium in its three charged states, $Mu^0$, $Mu^+$ and $Mu^-$ or (2) studies of the ``dynamics'' of these centers.

Electronic structure studies of muonium are concerned with determining the crystalline site of the muon, the arrangements of neighboring atoms, and the strength of the interactions between the muon and its surroundings. Thus far, there have been over 20 distinct muonium centers identified in tetrahedrally coordinated Group IV, Group III-V and Group II-VI semiconductors [see, e.g., Gil1999]. These states are denoted as $Mu_{BC}^0$, $Mu_{BC}^+$, $Mu_T^0$ and $Mu_T^-$. The now standard notation labels the charge state of muonium and its general location in the lattice, with the subscript T symbolizing the tetrahedral interstitial site and BC the bond-center location. These experiments and their comparison with theory were crucial in emphasizing the need to include lattice relaxations in calculations of hydrogen (and also other defects) in solid state materials, and continue to be a stringent test of the many theoretical calculations regarding the electronic structures of hydrogenic centers. Furthermore, since the hydrogen analogs of many of the muonium centers are expected to exist in semiconductors, they form the fundamental building blocks upon which any model of hydrogen dynamics must be built.

The second important research area, that of muonium dynamics, include studies of muonium diffusion, cyclic charge state changes, spin exchange scattering with free carriers, interconversion between the various muonium states, and more recently, reactions with impurities in the material. These investigations are clearly crucial to a detailed understanding of hydrogen motion in the crystal, its reactions with free carriers, and ultimately, with defects in the material. Two examples illustrate this point: The first is the observation of the dramatic difference [Chow1996] in the diffusion rates of $Mu_T^0$ and $Mu_T^-$, such as in GaAs, where $Mu_T^-$ is moving about ten orders of magnitude slower than its neutral counterpart. These results argue that although the positive and negative charge states of hydrogen are the dominant ones in p-type and n-type materials respectively, the actual diffusion of hydrogen may well be controlled by the transient $H^0$ species present in the material at high temperatures.

The second example concerns the dynamics of transitions between muonium states: the most detailed model of these transitions exist for muonium in silicon [Hitti1999; Hitti1997; Kreitzman1995]. The transition rates obtained from these measurements compare well with the few available analogous measurements on isolated hydrogen, hence indicating (i) that the conclusions from muonium investigations should be transferable to hydrogen and (ii) that any model that relies on only a single hydrogen species at high temperatures is very likely to be inadequate.

Finally, it should be noted that μSR is poised to take a defining role in the characterization of films and heterostructures constructed using *magnetic* semiconductors, of intense interest to researchers of spintronics. Spintronics (or spin-electronics) involves the control and manipulation of spin instead of charge degrees-of-freedom in solid state systems, thought to be a means of increasing processing times, non-volatility, and reduced power consumption in electronic circuitry [Zutic2004]. In this context, μSR is making valuable contributions to the characterization of dilute magnetic semiconductors, such as $(Ga_{1-x}Mn_x)As$. In these materials, magnetic cations are doped into III-V semiconductors. In one recent study [Dunsiger2010], the unique characteristics of low-energy μSR allowed for probe muons to be implanted in a thin film of $(Ga_{1-x}Mn_x)As$ and determine that the known ferromagnetic ground state was present in 100% of the sample volume, settling a lingering question of whether such a homogeneous state was even possible in these systems [e.g. Storchak2008]. The character of the magnetic ground state also seemed unaffected by the metal-insulator transition on the doping axis, prompting a re-evaluation of the theoretical models being used to discuss these compounds. In a separate study [Deng2011], similar conclusions were reached about the volume fraction of ferromagnetic order in bulk magnetic semiconducting compound, Li(Zn,Mn)As confirming theoretical prediction [Masek2007] and opening up new avenues of research.

On a different tack, Drew *et al.* have also shown the usefulness of low-energy μSR as a near unique, depth-resolved measure of the spin polarization of charge carriers within *buried* layers of real devices [Drew2008, Figure III-12]. In this study, researchers implanted probe muons at different depths in a fully functional spin valve device, made up of a spacer layer of organic semiconductor tris-(8-hydroxyquinoline) aluminum (Alq3) between two ferromagnetic electrode layers. By measuring the shape of the local magnetic field distribution in the presence and absence of a spin-polarized current and as a function of depth, they were able to get a quantitative measure of the spin diffusion length, cleanly separated from interface effects. By correlating their data with bulk magnetoresistance measurements at different temperatures, they conclusively showed that spin diffusion is a key parameter of spin transport in organic materials. Similar studies are now being used to explore new device concepts for future spintronics applications [Schulz2010].

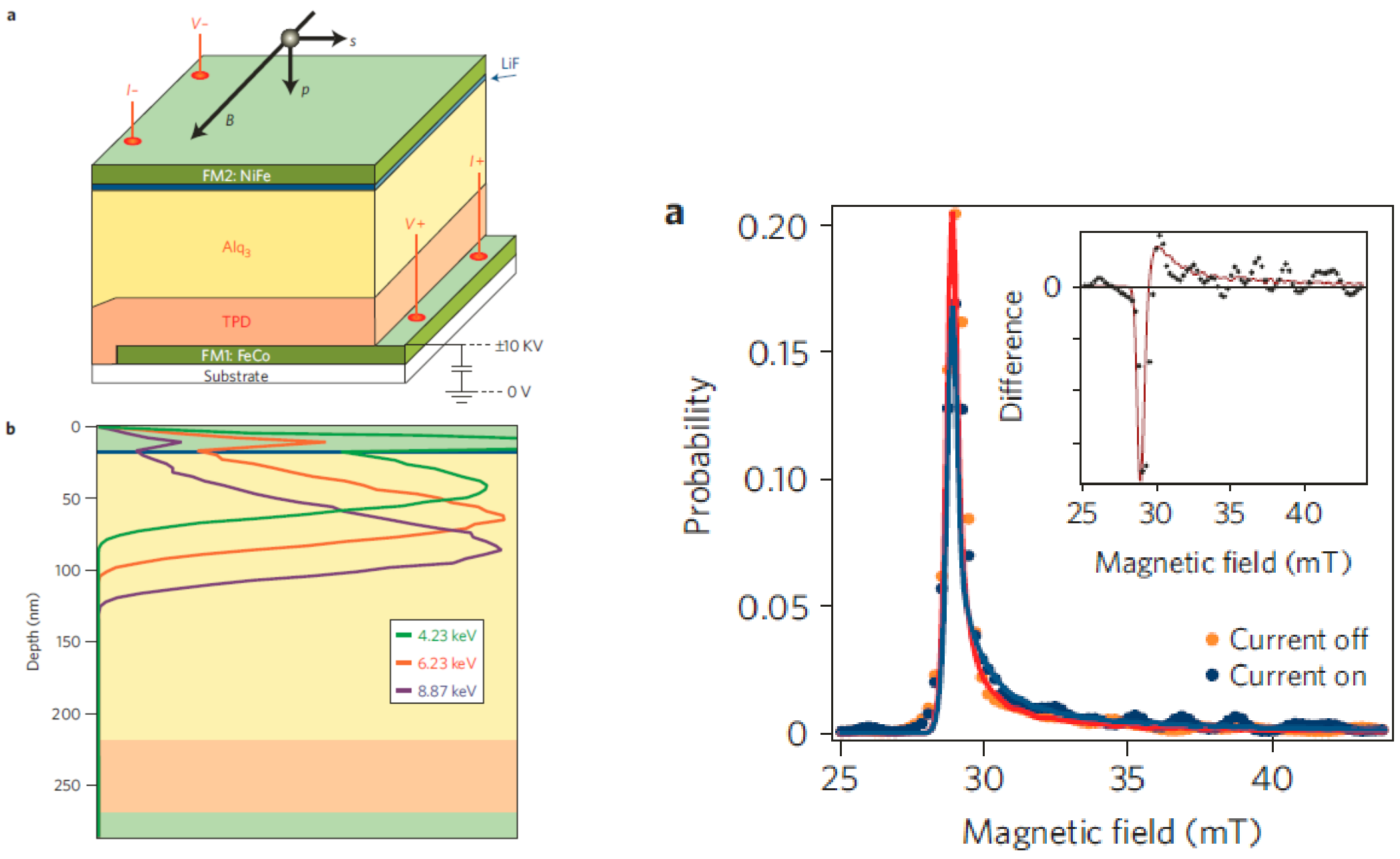

**Figure III-12**: (left) Schematic diagram of the spin valve system investigated by Drew *et al*. (top), and the depth profile of implanted muons with different initial energies (right) An example spectrum taken with 6.23 keV muons in the presence of a 29mT applied field. Data with a current density of 0 and 3mA are shown in red and blue, respectively. Circles show small, but statistically significant difference between the two datasets. Lines are the result of model calculations for a dipolar field distribution due to rough interfaces and due to current-induced injection of polarized spin carriers. The small oscillation apparent in the data is understood as a finite size effect. From [Drew2009_2].

### III.2.2 **Chemistry**

Applications of muons in chemistry are mostly related to hydrogen atom and free radical chemistry. When a positive muon stops in a non-metallic material it binds an electron to form the atom muonium (Mu). Since the reduced mass of Mu is close to H, their atomic properties are very similar and Mu can be considered a light isotope of hydrogen. Reaction of Mu with an unsaturated molecule results in a free radical incorporating a muon in place of a proton. Thus free radical properties and reactions can be studied by utilizing Mu as an isotopic tracer. Applications of muonium range from the study of kinetic isotope effects in fundamental gas-phase reactions to the use of Mu as a probe of chemistry under extreme conditions, such as those in the primary cooling-water cycle of a nuclear reactor or the high pressure/temperature reaction vessel of a facility designed to destroy hazardous waste. A key aspect of such studies is that Mu can be (and has been) studied under conditions not accessible by more conventional H atom and free radical studies.

An example of free radical chemistry involving muonium is the study of the chemistry of guest molecules inside gas hydrates similar to the infamous "ice crystals" that were involved in the Deepwater Horizon oil rig disaster in the Gulf of Mexico. Flames and explosions propagate through free radical reactions, yet almost nothing is known about the diffusion and interactions of atoms and radicals through the cavities of gas hydrate crystal structures. An experiment at TRIUMF has recently detected and characterized for the first time organic free radicals in clathrate hydrate structures at close to ambient temperatures.

The most fundamental muonium chemistry program is the work of Fleming et al. on gas-phase kinetics [Fleming1976; Reid1987; Gonzalez1989]. The data produced in these experiments are invaluable to theorists working in the area of chemical reaction dynamics, either because they are more precise than equivalent data on H atom reactions, or because prediction of Mu/H kinetic isotope effects is easier than calculation of absolute rate constants. In more recent work the Fleming group have extended the bounds of $H + H_2$ studies by employing muonic helium $^4He\mu$ as a heavy hydrogen isotope [Fleming2011]. A general interest article in Science [Fleming2011_2, Figure III-13] generated much media interest and editorial commentary in science magazines such as Nature News, New Scientist and Chemistry World.

A more technological application of muonium chemistry involves measuring Mu reaction rates in supercritical water. This is needed because accurate modeling of aqueous chemistry in the heat transport systems of pressurized water-cooled nuclear reactors (PWRs) requires data on the rate constants of reactions involved in the radiolysis of water. Unfortunately, available experimental data do not extend to the high temperatures used in current PWRs, typically around 320°C; and the next generation design employs higher, supercritical temperatures (~650°C). It has been common practice to extrapolate experimental data on diffusion coefficients and rate constants from their measured ranges (mostly less than 200°C), but it would be dangerous to rely on this for the supercritical regime [Ghandi2003]. Furthermore, experiments at TRIUMF have shown that bimolecular rate constants exhibit extreme non-Arrhenius behaviour, with a maximum followed by a minimum as the temperature rises past the critical point [Percival2007].

Muonium-substituted free radicals were first detected by $\mu$SR in 1978 [Roduner1978], but full characterization of muoniated free radicals requires the complementary technique of muon avoided-level crossing resonance (LCR) [Kiefl1986, Percival1987]. LCR makes possible the determination of nuclear hyperfine constants other than that of the muon, providing key data for radical identification and thence, for example, analysis of intramolecular dynamics [Percival1988,Yu1990]. A recent example is the study of the free radical reactivity of novel low-valent organosilicon, organogermanium, and organophosphorus compounds [McCollum2009; West2010; Percival2011; Percival2012].

Such materials are at the forefront of modern inorganic chemistry, mostly because of their applications in the design of catalysts (e.g. the work of Chauvin, Grubbs and Schrock, Nobel Prize in Chemistry 2005).

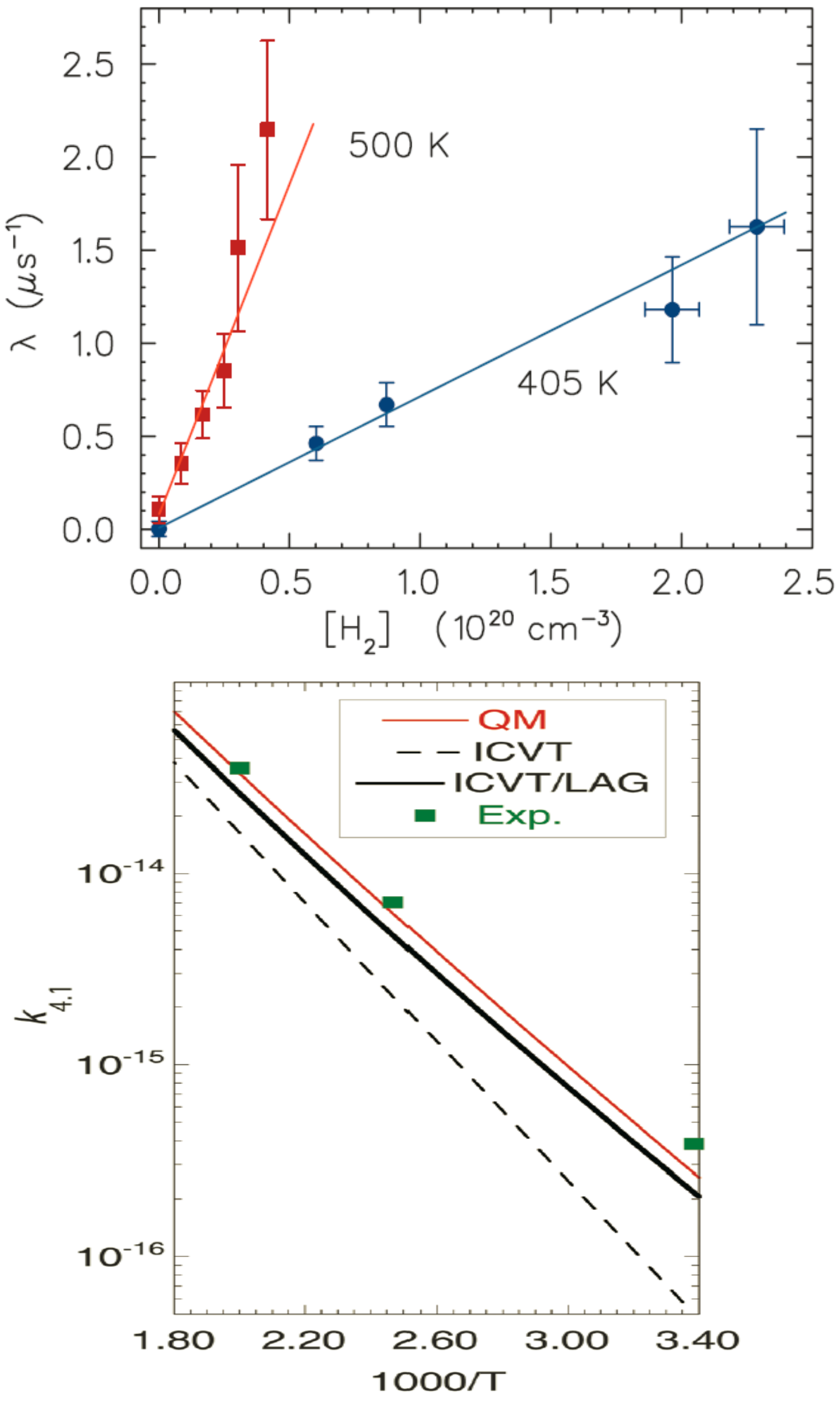

**Figure III-13**: Data from the study of Fleming et al., wherein researchers provided an unprecedented test of chemical kinetic theories through the use of muonic helium 4He$\mu$ as a heavy hydrogen isotope (denoted 4.1H). The upper panel shows measured relaxation rates versus concentration of H2 at two different temperatures. The lower shows rate constants of the 4.1H+1H2 reaction, extracted from slopes in the previous plot, as compared to expectations from theory. From [Fleming2011_2].

## III.3 Techniques of Muon Spin Rotation

### III.3.1 Brief Historical Summary

The technique of muon spin relaxation, rotation or resonance, known collectively as $\mu$SR, was first suggested in an historic 1957 paper in Physical Review by Garwin, Lederman and Weinrich, in which parity non-conservation in the weak decay of the muon was demonstrated. These authors wrote that "it seems possible that polarized positive and negative muons will become a powerful tool for exploring magnetic fields in nuclei ..., atoms and interatomic regions." Pioneering efforts at the old cyclotron facilities (LBL, SREL and NEVIS) spawned developments in the technique and its scientific reach. Attempts to realize the original vision of Garwin et al. on a practical scale - where high data rates with relatively clean backgrounds would be available - would have to await construction of the high-intensity meson factories at LAMPF (Los Alamos, USA, 1972), SIN (now PSI, Villigen, Switzerland, 1974) and TRIUMF (Vancouver, Canada, 1974), which could deliver high luminosity muon beams. A major breakthrough occurred in 1985 when Bowen *et al*. built the first 100% polarized surface muon beam (originally called an Arizona beam) at LBL.

The vision of Garwin et al. has been carried out in Canada, Switzerland, the UK (ISIS/RAL) and Japan (first at KEK and now at J-PARC). There is a plan to construct a facility in South Korea (RISP). All existing facilities are heavily subscribed, and must reject many good proposals. There has been no capability for experiments utilizing μSR in the United States since the closure of the LAMPF muon facility in the 1990's.

### III.3.2 Beams and timing structures

#### *Types of μSR beams*

The production of muons appropriate for μSR begins with pion production in a proton beam. Pion decay to polarized muons is common to all forms of μSR beams. The polarization of the muons is a rather significant application of the fundamentally parity-violating nature of the weak interaction.

*Surface muons* are produced by the decay of these pions after they have come to rest in the target material. The low energy beams of polarized muons produced are collected by magnetic optics and transported to the sample stations, with appropriate care being taken to maintain the polarization which is the essence of the technique.

*Low Energy Muons, (LEM)*

In this technique, a beam of surface muons of kinetic energy 4 MeV are cooled to quasi-thermal energies of 1-30 keV. There are two methodologies currently in use or under investigation. In the first, the surface muons are directed onto a moderator, typically a thin metal plate covered with a layer of a solid noble gas. The emerging muons have been shown [ Morenzoni1994; Morenzoni2000] to retain a high degree of polarization, and the technique is in regular use at the PSI μE4 line.

A second methodology relies on resonant laser ionization of muonium created by surface muons in a moderator. This technique is used at pulsed sources such as ISIS and J-PARC, and creates a beam of muons with selectable energies in the approximate range 0.1-1.0 keV, with very good energy precision. The overall polarization of the beam is less because of depolarizing interactions in the muonium atoms, which dilutes the available asymmetries.

*Decay-in-Flight Muons*

Lastly, muon beams created by the decay-in-flight of pions produce muons which can penetrate containers, and are thus optimized for studies of chemical or high pressure phenomena. An example of such a beams is μE1 at PSI, with available muon momenta up to 125 GeV/c. Such a beam is likely to be an attractive option for Project X and will be the subject of further studies. It is not part of the design concept outlined in this document.

### *Time structures for experimental μSR*

In *conventional continuous (CW) μSR*, the arrival of a muon at the sample serves as a start signal to a fast timing circuit, with the stop signal being provided by detection of the decay positron from the embedded muon – see Figure III-14. Because of the need to associate each detected positron with one and only one specific stopped muon (pile-up rejection) , this technique is traditionally limited to muon input rates of approximately 50 kHz. Another way of thinking of this is that only one muon at a time is accepted and allowed to decay.

To increase the rates of detected events, and to optimize experiments with existing accelerators, *pulsed μSR* may be used. In this technique, the time resolution is the convolution of the pulsed beam structure with the spread caused by the 26 ns lifetime of the parent pion. No attempt is made to determine the individual start times of arriving muons at the sample. This method gives increased statistics and shorter data-taking times, but has certain limitations when large internal or external magnetic fields cause rapid precession of the embedded muon spins.

As already noted, conventional μSR is limited to a single detected muon at a time. In addition, there are backgrounds associated with muons stopping near the experimental sample being studied. This background limits the sensitivity of the technique. An elegant solution to this problem has been developed [Abela1999], referred to as *MORE* (Muons on Request). In this system, fast electrostatic kickers send muons to each experiment only when it signals its readiness to accept a pulse, hence limiting the exposure of the equipment to spurious muons. This technique can be very advantageous, and is expected to be implemented in the relevant parts of the proposed Project X system.

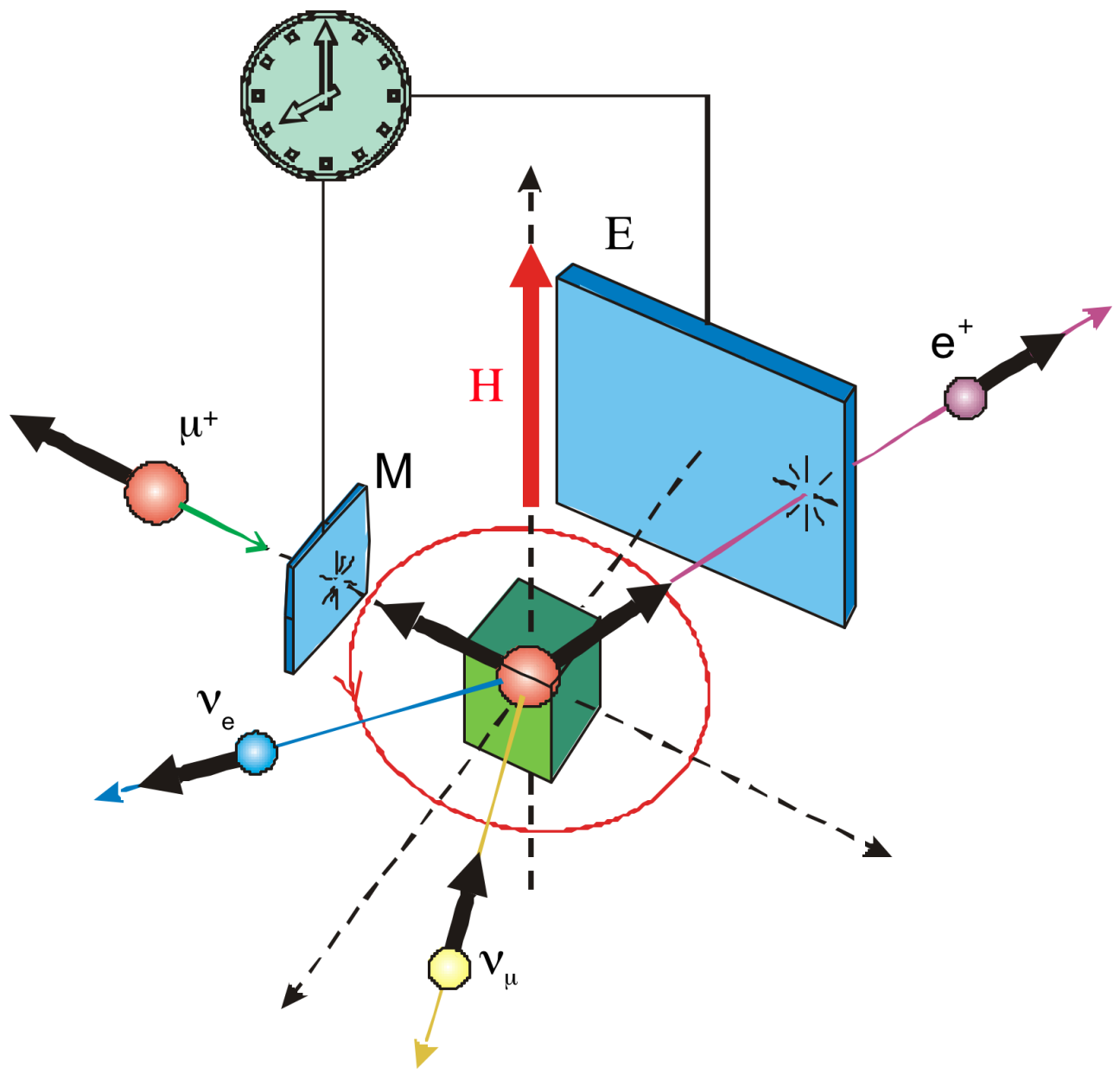

**Figure III-14**: Illustration of the timing setup for conventional μSR

## III.4 World Facilities, Capabilities, and Needs

At the time of this writing (2013) there are four μSR facilities in operation worldwide: two in Europe (PSI and ISIS), one in North America (TRIUMF), and one in Asia (J-PARC). Two facilities (PSI, TRIUMF) are based on the pseudo-continuous muon sources, while ISIS and J-PARC are pulsed sources. Both continuous and pulsed beams have their pros and cons as discussed in the previous section. Key parameters of the existing facilities are summarized in the table below. A discussion of the individual facilities follows. All of the facilities are heavily used by physicists from all-over the world with a significant (~20-30%) US-based researcher fraction.

| Facility | Muon beams | Time structure | Beam parameters | Upgrade plans |
|---|---|---|---|---|
| **PSI** | High energy (4-50 MeV), surface (4 MeV), low energy (0-30 keV) | Pseudo-continuous (CW),<br><br>MORE available | 600 MeV, 2.2mA ≅<br><br>1.4 $\times 10^{16}$ protons/sec<br><br>~$6.5\times10^{8}$ surface μ+/sec at the most intense beamline<br><br>LEM:$1.5\times10^{8}$ μ+/sec, 4500 μ+/sec to sample | N/A |
| **ISIS** | High energy (20-120 MeV/c) surface (27 MeV/c) | Double pulse width ~80 ns separated by 300ns, repeats every 20 ms (50 Hz) | 800 MeV protons, 200 μA, $2.5\times10^{13}$ protons/pulse<br><br>surface (27 MeV/c): $1.5\times10^{6}$ μ+/sec<br><br>decay (60 MeV/c): $4\times10^{5}$ μ+/sec, $7\times10^{4}$ μ-/sec | N/A |
| **TRIUMF** | High energy, surface | Pseudo-continuous (CW) | Cyclotron,<br><br>1.5-$2\times10^{6}$ μ+/sec | N/A |
| **J-PARC** | High energy, surface | Pulsed, ~50 ns pulse every 20 ms | Cyclotron, $1.8\times10^{6}$ μ+/sec (2009) at 120 kW, planned $1.5\times10^{7}$ μ+/sec at 1 MW | 3 more beamlines under construction (one is low energy muons) |

**Table III-1**: Worldwide μSR facilities operational today.

### *Paul Scherrer Institute (PSI), Switzerland*

Paul Scherrer Institute is a leading Swiss research institution with multiple facilities including neutron source, synchrotron light source, and muon source. There are 6 μSR instruments in 5 beam lines and diverse sample environments (17 cryostats, 2 furnaces, special setups for pressure, photon irradiation, E-Fields). The personnel at the μSR facility comprises 9 staff (tenure and tenure track), 4 postdocs, 5 PhD students, and ~ 3 technicians and computer support staff.

For the Laboratory for Muon Spectroscopy (LMU), typically there are 360 visits/year, ~200 new proposals, >700 beam days for users. The oversubscription factor is 2 to 3.5.

### *ISIS, UK*

The ISIS pulsed neutron and muon source at the Rutherford Appleton Laboratory in Oxfordshire, UK, is a world-leading center for research in the physical and life sciences. Recent proposal rounds for the facility gave the following structure:

- 19 different countries (UK + 11 European + 7 outside Europe)
- 67 separate research groups
- 738 days applied for: 417 available (1.8 oversubscription)
- ~35% of applicants are regular neutron source users

### *TRIUMF, Vancouver, Canada*

For muon spectroscopy TRIUMF has multiple beamlines, 7 spectrometers, 7 cryostats, 2 furnaces. Recently the facility has added two new beamlines (M9A, M20).The facility has beta-NMR with beam time extremely limited (~ 1-2 weeks/year). Staffing includes 2 FTE from TRIUMF and 6 FTE on research grants.

### *J-PARC, Tokai, Japan*

The MUSE facility D-line is operational, with 3 more beamlines under construction. One is low-energy (U-line) and is commissioning. At J-PARC, low energy muons are intended to be produced by a laser-based technique. Figure III-15 compare beam requests to beam availability over the last five years.

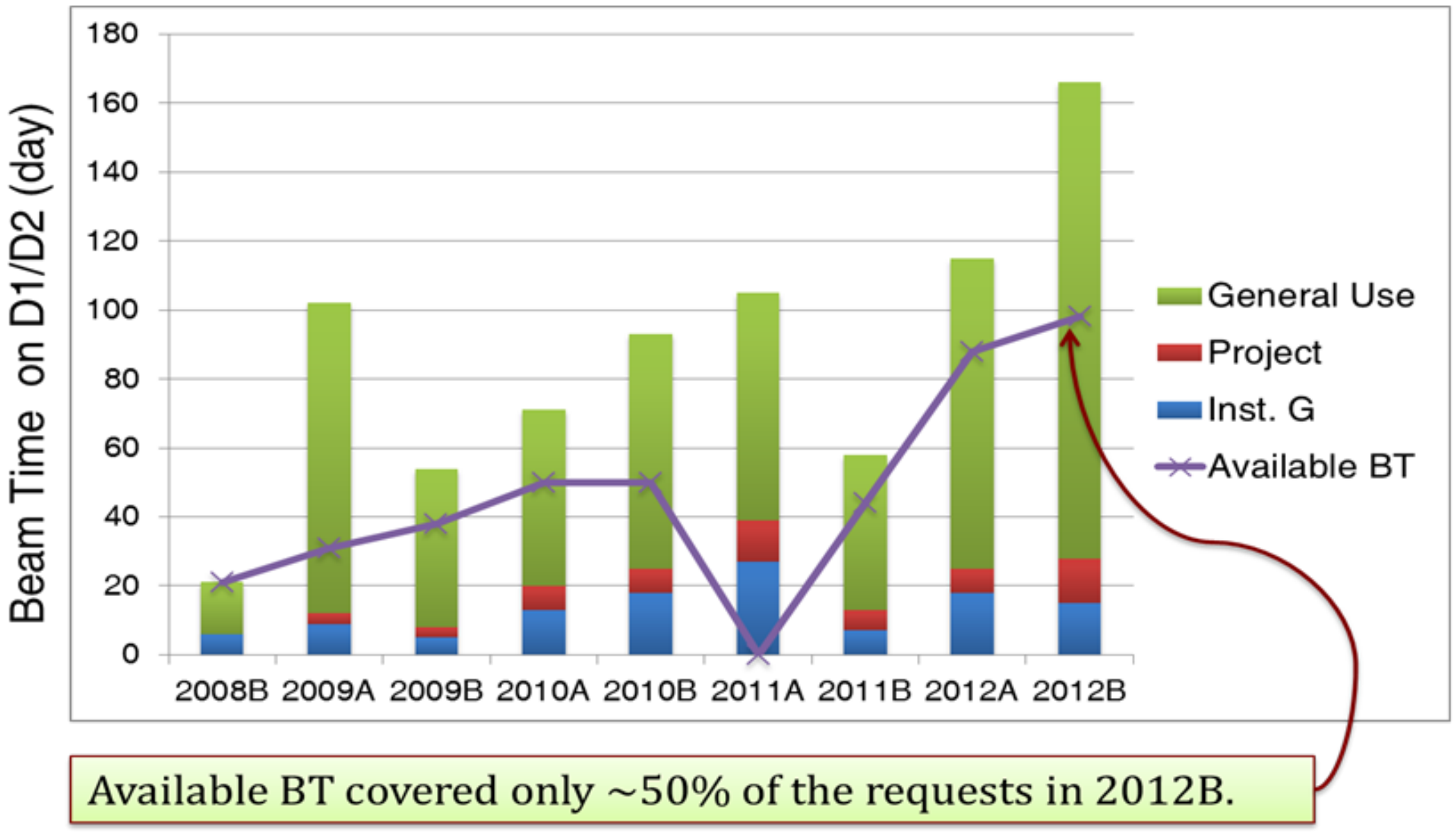

**Figure III-15**: Experimental requests at J-PARC D-line μSR facility [Kadono2013]

### III.4.1 Overview of current and expected needs

Both pulsed and CW μSR facilities are in high demand as evidenced by an average factor of 2 overbooking at all of the facilities with as high as 3.5 factors reported at PSI for low energy muons. Therefore, another state-of-the-art facility is likely to be of high international demand even without extending the existing user base. If the new facility is beyond state-of-the-art it is expected to be of extremely high demand.

Even though mesoscale phenomena at surfaces and interfaces is one of the fastest developing branches of the condensed matter physics and material science, there is only one low energy muon facility in operation world-wide (PSI), and one more commissioning. An obvious need exists to extend this application of $\mu$SR.

### III.4.2 Considerations specific to North America

The USA represents the world's largest condensed matter physics community with a current share of about 20% of all μSR experiments worldwide. However, no state-of-the-art μSR facility is present on the US soil. World demand for this technique is expected to be high, strengthening the need for national expertise.  There is a high degree of complementarity with neutron scattering techniques as discussed in section (A). The implementation of a LEM facility will provide capability never before seen in the United States.

## III.5 Technical Capabilities of Project X for μSR

### III.5.1 Stage 1 Project X and μSR

The technical capabilities of Stage 1 Project X have been discussed in Part One of this book. For the purposes of this document we are primarily interested in 1) the extreme *flexibility* in timing structure offered by Project X, and 2) the *high power* available at 1 GeV in a quasi-CW configuration of bunches. As we will see in the following, it is possible to exploit these features of the accelerator complex to create a user facility for μSR which provides all the available technologies for experimenters, often in a way which can be reprogrammed to adjust to technological developments or scientific needs. This will expand the scientific capacity of the complex in a way which benefits the U.S. scientific program with special emphasis on the needs of the DOE/Basic Energy Sciences user community. In this section we will review the unique timing structures available at Project X, and give estimates for the intensity of the muon source(s) that can result from exploitation of the accelerator.

### III.5.2 Beam Structure

We propose a two-pronged beam structure for μSR. For the purposes of LEM μSR the achieved efficiencies of the thermalization process are small, – hence the total beam power is a relevant figure of merit. For this purpose we propose to add a surface muon target and associated beamlines for LEM in the high-power beamline discussed in Volume One of this book. Such beamlines typically consume 10% of the available proton power, a parameter that can be tuned by detailed target design. Current or future LEM technologies can be used with the surface muons to create the desired thermalization. The needs of the user community must determine the exact configuration of beamlines.

Independently of the spallation beamline, we propose a set of μSR beamlines which use low-power targeting. By programming the high-frequency chopper at the low-energy end of the Project X Linac, we can produce the beam structure shown in Figure III-16. This structure can simultaneously serve the needs of the spallation program, the muon experimental program, and μSR. The μSR pulses are groups of pulses separated by 24 ns. Repeating the structure every 5 μs gives the possibility of using a series of programmable kickers [Abela1999] to separate the resulting surface muon beams to service 4 end stations with a repetition rate per station of 50 kHz (20 μs between pulses). The number of pulses given to each endstation per 20 μS interval is programmable, and the endstation receiving each of the 4 independent pulse trains can also be selected by programming the sequence of kicker pulses. The combination of programmable pulse structure and selectable endstations gives a high, perhaps unprecedented, level of flexibility to the facility. A possible overall layout is shown in Figure III-17.

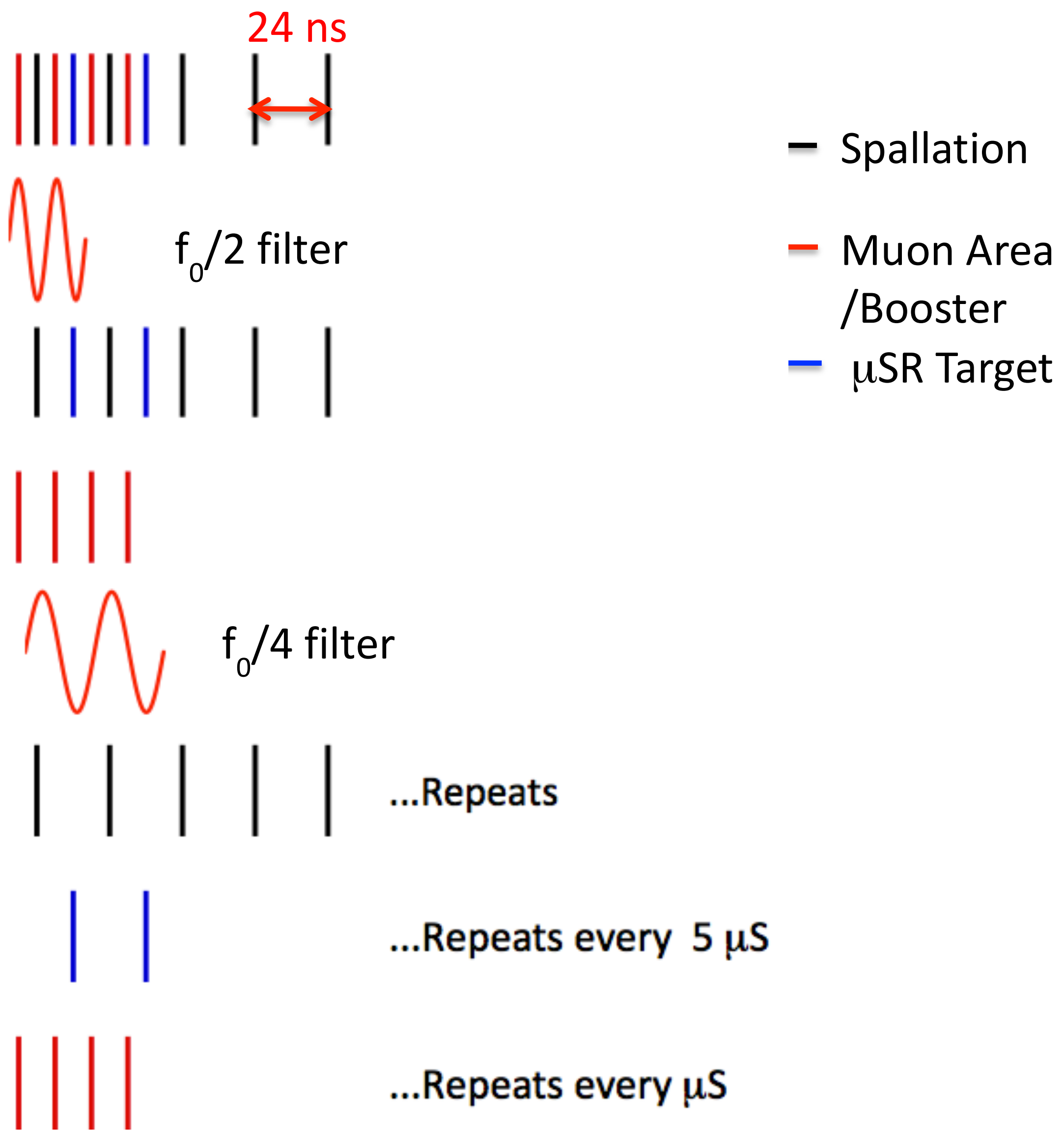

**Figure III-16**: Conceptual beam structure including dedicated µSR beamline. See text for details.

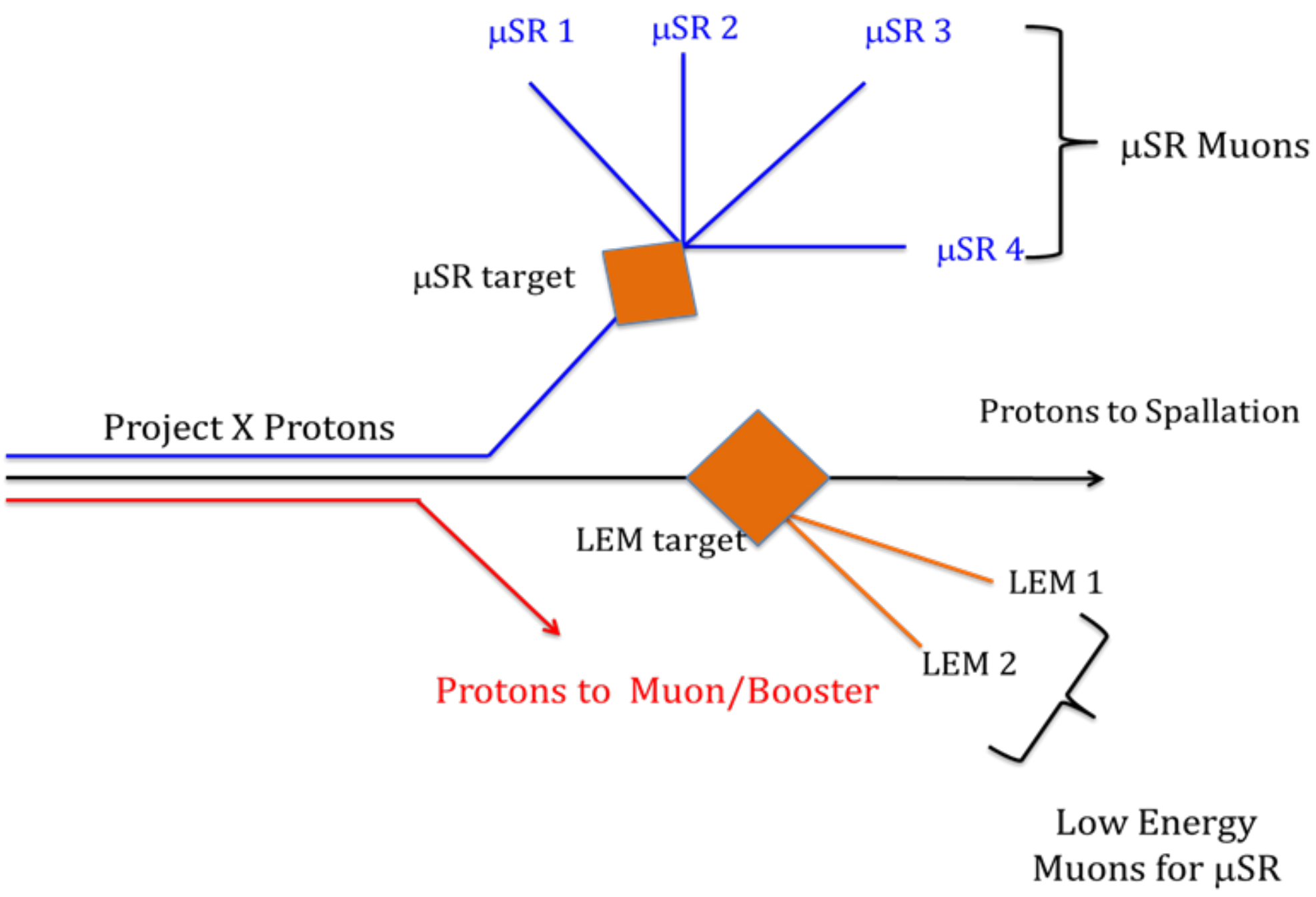

**Figure III-17**: Design concept for Stage-1 beam layout servicing μSR endstations.

### III.5.3 **Beam Intensity**

For the purposes of this document, we assume a dedicated beam structure with two sets of pulses as shown in Figure III-16, and two single pulses. The single pulses are appropriate for single-muon μSR, while the double pulses can service a pulsed μSR experiment. Importantly, this beam is independent of the high power pulse train which goes to the spallation area and services LEM experiments. We assume 2 LEM beamlines will be made available.

For the purposes of estimating intensity, we use the following parameters, based on experience at the PSI μE4 beamline:

i) A conversion factor between protons on target and muons at the experiment of $2 \times 10^{-7}$. This factor consists of $5 \times 10^{-8}$ achieved performance and a factor of 4 from anticipated possible optimizations of a new, dedicated system [Morenzoni2012].

ii) A number of protons per Project X bunch of $1.3 \times 10^{8}$. This is consistent with a of 0.9 mA and 40 bunches/μs.

iii) A conversion efficiency for creation of LEM muons of $1.0 \times 10^{-5}$.

With these assumptions we obtain:

i) $2.6 \times 10^{6}$ muons/sec for each of 2 pulsed μSR beamlines.

ii) A saturated rate of $5 \times 10^4$ muons/sec for each of 2 conventional single-muon mSR beamlines.

iii) $1.1 \times 10^4$ LEM muons for each of 2 stations.

It should be noted that all of these can run simultaneously. The total power demands of the dedicated μSR proton lines in this reference concept are 6.8 kW, with each pulsed target receiving 2.25 kW and each conventional station 1.1 kW.

### III.5.4 **Discussion**

The need for μSR facilities is well-established, and can be expected to grow in the next decade, for example in the use of LEM to explore finite size and interface effects in thin magnetic films. Figure III-18 shows the wide variety of user applications that have been seen at the ISIS facility in recent years. There is every reason to expect that a μSR facility as outlined above would attract a similar spectrum of interest from the Materials Science and Applied Physics Communities, both inside the United States and worldwide.

LEM muons are of very high current interest, and require MW scale hadron facilities to produce the requisite fluxes of keV muons, because of the inherent inefficiencies in the know conversion processes. The LEM beamlines discussed in the previous section would add invaluable capacity to a very limited set of world facilities in this area. Including such a capability from an early point in the Project X design will allow us to take advantage of optimizations in acceptance and targetry, an exciting possibility not always available to older facilities. In this way the intensity of the LEM beams we discuss can be made very high, on a scale that is competitive with or exceeds beamlines in Europe or Asia. Project X μSR provides an opportunity for the United States to become a leader in this research methodology.

The design concept is extremely flexible. The beamlines μSR1-μSR4 can be programmed by a combination of the driving proton time structure from the Project X linac, the system of muon kickers downstream of the muon production target, and the demand for muons from the experiments (MORE). The intensity of the pulsed muon beamlines will be as high or higher than similar lines in ISIS and J-PARC.

It should be pointed out that the μSR beam area will also be available for fundamental physics experiments with surface muons. It will be possible to leave space for additional beamlines if desired, which can be fed from more frequent pulses in the Project X linac as the stages of that machine evolve.

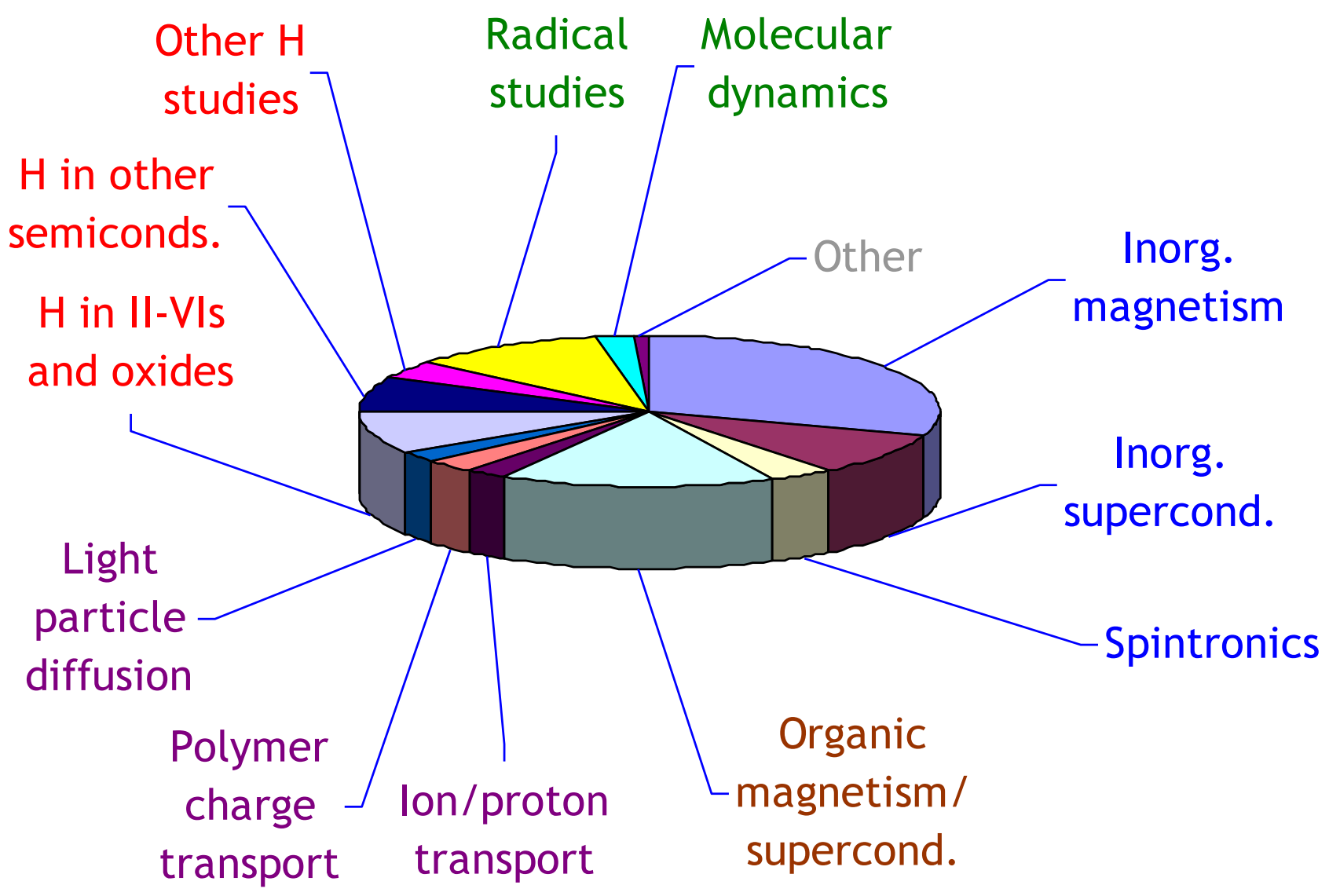

**Figure III-18**: Variety of user applications at the ISIS facility [Kilcoyne2012]

## III.6 **Research and Development**

No research field is static. In this section we point out some areas in which the science of μSR may be enhanced by participation of Fermilab in the ongoing development of the technique, and which may prove of eventual use in the program.

### III.6.1 **Targeting**

Fermilab has extensive experience in design and implementation of targets for high intensity proton beams. Projects have been executed for diverse environments, ranging from the AP0 antiproton creation facility to the underground NuMI long-baseline neutrino beam. The target group at Fermilab is involved in state-of-the-art projects such as the LBNE long-baseline neutrino project, whose final goals are in the multi-MW range for power on target. Moreover, the group has extensive professional connections with other targetry groups around the world, including the UK and RAL. Advancements in targetry design, beginning with simulation studies (typically using MARS) and proceeding through detailed engineering, can be expected to produce optimized target designs for Project X applications, including for μSR, with concomitant improvements in the efficiency and intensity of the beamlines.

### III.6.2 Multi-muon CW Capability with Advanced Detectors

The inability of bulk CW μSR to benefit from high intensities has been discussed above. One widely discussed approach to reducing the pile-up problems, which limit the usable rate is higher segmentation of both the incoming muon and outgoing positron detectors. Pattern recognition using such detectors may prove able to increase the usable bulk CW μSR rate by as much as one or two orders of magnitude.

Fermilab's technical expertise in the area of precision position-sensitive particle detectors is immense. Projects as diverse as silicon strip and pixel detectors for collider experiments (CDF, CMS and D0), to scintillating fiber trackers (D0), to precision CCD instruments for astronomy (DES) have been designed, prototyped, assembled, installed, and operated by lab staff, both on the site and remotely. There is an active research group engaged in advancing knowledge about such detectors, as well as advanced test and fabrication equipment and on-site access to a test beam. It is certain that an interest in creating a μSR facility at Fermilab would benefit from cross-fertilization with this ongoing effort.

### III.6.3 Storage Rings

Fermilab has a long history of interest in the possible development of muon storage rings. These efforts include studies of muon colliders, long-baseline neutrino factories, and more recently the NuSTORM proposal for a short-baseline neutrino source located on the Fermilab site. The collection of low-energy muons in a small storage ring would allow the manipulation of the time structure of extracted pulses in such a way as to compress hundreds of Project X pulses into a 10-20 ns pulse. The intensity of such a pulse would be very large (possibly $> 10^8$ muons/sec). This would allow experiments to be carried out very rapidly and accurately. While conceivably a transformational technology, much further work will need to be done to establish the workability of this concept.

## III.7 Developing a μSR community for Fermilab

### III.7.1 General Considerations

As accelerator technology has developed, its uses in both the general scientific community and the overall economy have multiplied. Perhaps the best example of reuse of knowledge from particle physics accelerator science is the proliferation and abundant utility of synchrotron light sources, technology derived from electron-positron colliders. Applications of spallation neutron scattering from proton machines has also found wide application and a dedicated community of users within Material Science.

The μSR community in the United States is reliant on off-shore facilities. It is to be expected that this community will naturally take advantage of a modern, flexible facility, as proposed here, and will grow to include new users and ideas. New local capacity at Fermilab will be part of a change facilitated by Project X and should prove highly attractive to users. However, there are actions that should be carried out in order to maximize the impact of this technology.

### III.7.2 The Path Forward

In order to make μSR in the United States as effective as possible, three paths should be pursued: Adequate user facilities, appropriate laboratory culture and support, and outreach.

*User facilities* should include technical capacity to handle samples, easily accessible end station setup, staging areas, and technical capacity to make necessary modifications quickly. It is worth noting that the Fermilab test beam effort has already pioneered many of these concepts for users worldwide. Importantly, the facility proposed in this document gives a dedicated user environment which co-exists with other uses of the extremely flexible Project X architecture.

*Laboratory culture* includes an open and welcoming environment with well-defined hardware and administrative interfaces which do not require years of experience to accomplish a program of experiments, or a single experiment. This has been pioneered by synchrotron light sources, and their example clearly can be adapted. Steps to move towards this model should be taken early, during design and construction of Project X.

*Outreach* is necessary to bridge the divide between the research areas of fundamental particle physics and user-driven applications. Workshops, colloquia, personal visits and visitor support will all play an important role in both designing the μSR facility and in understanding its exploitation. The early input of the user community is vital to the success of the undertaking.

## III.8 Conclusions

In this document we have pointed out the strong science available via the μSR technique. Project X Stage 1 at Fermilab provides an opportunity to create a world-class facility using the already planned beam timing structure. This facility fits naturally into the expected uses of the high-intensity Project X beams, and can produce a suite of beamlines and techniques which is compelling. There is adequate time to solicit the input of the Materials Science community to ensure that the facility will meet the needs of the users.

The proposed μSR facility provides a window on the materials world which will contribute significantly to the scientific arsenal of the United States. Project X provides a unique opportunity to establish such a facility in well-designed way which complements the broad range of fundamental and applied science available at a high power proton linear accelerator.

## III.9 References

[Abela1999] R. Abela et al., Hyperfine Interactions **120/121** (1999) 575.

[Aczel2008] A.A. Aczel *et al.*, Physical Review B **78** (2008) 214503.

[Aoki2003] Y. Aoki *et al.*, Physical Review Letters **91** (2003) 067003.

[Balsius1999] T. Balsius *et al.*, Physical Review Letters **82** (1999) 4926.

[Bernhard1999] C. Bernhard *et al.*, Physical Review B **59** (1999) 14099.

[Bernhard2012] C. Bernhard, Physical Review B **86** (2012) 184509.

[Bogdanovic1997] B. Bogdanovic and M. Schwickardi, J. Alloys Cmpd. **253**, 1 (1997).

[Boris2011] A.V. Boris *et al.*, Science **332** (2011) 937.

[Budnick1988] J.I. Budnick *et al.*, Europhysics Lett. **5** (1988) 651.

[Burrard-Lucas2013] M. Burrard-Lucas *et al.*, Nature Materials **12** (2013) 15]..

[Callaghan] Callaghan F D, Laulajainen M, Kaiser C V and Sonier J E, Physical Review Letters **95** (2005) 197001.

[Cariuffo1987] R. Cariuffo *et al.*, Solid State Communications **64** (1987) 149.

[Carlo2009] J.P. Carlo *et al.*, Physical Review Letters **102** (2009) 087001.

[Carlo2012] J.P. Carlo *et al.*, Nature Materials **11** (2012) 323.

[Carretta2011] P. Carretta and A. Keren in *Introduction to Frustrated Magnetism: Materials, Experiment, Theory*. Springer Series in Solid-State Sciences **164** (2011) 79

[Choi2012] S.K. Choi *et al.*, Physical Review Letters **108** (2012) 127204.

[Chow1996] K.H. Chow *et al.*, Physical Review Letters **76** (1996) 3790.

[Chow1998] K.H. Chow, B.Hitti and R.F. Kiefl in *Identification of Defects in Semiconductors*, ed. M. Stavola (Academic Press, New York 1998), p.137.

[Cox2009] S.F.J. Cox, Reports on Progress in Physics **72** (2009) 116501.

[Cubitt1993] R. Cubitt *et al.*, Nature **365** (1993) 407.

[Dalmas de Reotier 1997] P. Dalmas de Réotier and A Yaouanc, Journal of Physics: Condensed Matter **9** (1997) 9113–9166.

[Dalmas de Reotier2006] P. Dalmas de Réotier *et al.*, Physical Review Letters 96 (2006) 127202.

[Dalmas de Reotier2012] P. Dalmas de Réotier *et al.*, Physical Review B 86 (2012) 104424.

[Deng2011] Z. Deng *et al.*, Nature Communications **2** (2011) 1.

[Disseler2012] S.M. Disseler *et al.*, Physical Review B **85** (2012) 174441.

[Drew2009] A.J. Drew *et al.*, Nature Materials **8** (2009) 310.

[Drew2009] A.J. Drew *et al.*, Nature Materials **8** (2009) 109.

[Dunsiger1996] S. R. Dunsiger *et al.*, Physical Review B **54** (1996) 9019.

[Dunsiger2006] S. R. Dunsiger *et al.*, Physical Review B 73 (2006) 172418.

[Dunsiger2010] S.R. Dunsiger, Nature Materials **9** (2010) 299.

[Fleming1976] D.G. Fleming, *et al*, . J. Chem. Phys. **64** (1976) 1281.

[Fleming2011] D.G. Fleming, *et al.*, J. Chem. Phys. **135** (2011) 184310.

[Fleming2011_2] D.G. Fleming *et al.*, Science **331** (2011) 448.

[Friedt1985] J. M. Friedt, F. J. Litterst and J. Rebizant, Physical Review B **32** (1985) 257.

[Fukai1985] Y. Fukai and H. Sugimoto, Advances in Physics **34** (1985) 263.

[Gardner1999] J. S. Gardner *et al.*, Physical Review Letters **82** (1999) 1012.

[Gil1999] J.M. Gil *et al*., Physical Review Letters **83** (1999) 5294.

[Ghandi2003] K. Ghandi and P.W. Percival.. *J. Phys. Chem*. **A107** (2003) 3005.

[Goko2009] T. Goko *et al*., Physical Review B **80** (2009) 024508.

[Gonzalez1989] A.C. Gonzalez, *et al*., J. Chem. Phys. **91** (1989) 6164.

[Hachitani2006] K. Hachitani, H. Fukazawa and Y. Kohori, Physical Review B **73** (2006) 052408.

[Harshman1988] D.R. Harshman *et al*., Physical Review B **38** (1988) 852.

[Heffner1990] R.H. Heffner *et al*., Physical Review Letters **65** (1990) 2816.

[Hillier2009] A. D. Hillier, Physical Review Letters **102** (2009) 117007.

[Hillier2012] A.D. Hillier *et al.,* Physical Review Letters **109** (2012) 097007.

[Hitti1997] B. Hitti *et al*., Hyperfine Interactions **105** (1997) 321.

[Hitti1999] B. Hitti *et al*., Physical Review B **59** (1999) 4918.

[Hofman2012] A. Hofmann *et al*., ACS Nano **6** (2012) 8390.

[Ichioka1999] M. Ichioka, A. Hasegawa and K. Machida, Physical Review B B **59** (1999) 184; M. Ichioka, A. Hasegawa and K. Machida, Physical Review B B **59** (1999) 8902.

[Ito2007] T.U. Ito *et al*., Journal of the Physical Society of Japan **76** (2007) 053707.

[Jackson2000] T.J. Jackson *et al*., Physical Review Letters **84** (2000) 4958.

[Kadono1996] R. Kadono *et al*., Physical Review B **53** (1996) 3177.

[Kadono1996] R. Kadono *et al*., Physical Review B **54** (1996) R9628.

[Kadono1997] R. Kadono *et al*., Physical Review Letters **79** (1997) 107.

[Kadono2001] R. Kadono *et al*., Physical Review B **63** (2001) 224520.

[Kadono2006] R. Kadono *et al*., Physical Review B **74** (2006) 024513.

[Kadono2013] R. Kadono, presentation to J-PARC International Advisory Committee, February 25-26, 2013, Tokai, Japan.

[Karlsson1995] E. Karlsson *et al.*, Physical Review B **52** (1995) 6417.

[Kaiser2000] C.T. Kaiser *et al.*, Physical Review B **62** (2000) R9236.

[Kaneko2008] K. Kaneko *et al.*, Physical Review B **78** (2008) 212502.

[Khasanov2004] R. Khasanov *et al.*, Physical Review Letters **92** (2004) 057602.

[Kiefl1986] R.F. Kiefl,, *et al.*, Phys. Rev. A. **34** (1986) 681.

[Kiefl1989] R.F. Kiefl *et al.*, Physical Review Letters **63** (1989) 2136.

[Kiefl1990] R.F. Kiefl and T.L. Estle in *Hydrogen in Semiconductors*, ed. by J. Pankove and N.M. Johnson (Academic, New York 1990), p.547.

[Kilcoyne2012] S. Kilcoyne, presentation at Fermilab Muon Spin Rotation Workshop, October 2012.

[Klauss2008] H.H. Klauss *et al*. Physical Review Letters **101** (2008) 077005.

[Kojima1995] K. M. Kojima *et al.*, Physical Review Letters **74** (1995) 281.

[Kojima1997] K. M. Kojima *et al.*, Physical Review Letters **78** (1997) 1787.

[Kopmann1998] W. Kopmann *et al.*, Journal of Alloys and Compounds **271-273** (1998) 463.

[Kossler1998] W.J. Kossler *et al.*, Physical Review Letters **80** (1998) 592.

[Kreitzman1995] S.R. Kreitzman *et al.*, Physical Review B **51** (1995) 13117.

[Kubo2003] K. Kubo and Y. Kuramoto, Journal of the Physical Society of Japan **72** (2003) 1859.

[Kubo2004] K. Kubo and Y. Kuramoto, Journal of the Physical Society of Japan **73** (2004) 216.

[Kuramoto2009 J. Kuramoto, H. Kusunose and A. Kiss, Journal of the Physical Society of Japan **78** (2009) 072001.

[Laplace2012] Y. Laplace *et al.,* Physical Review B **86**, 020510(R) (2012).

[Lee1993] S. L. Lee *et al.*, Physical Review Letters **71** (1993) 3862.

[Lee1997] S. L. Lee *et al.*, Physical Review Letters **79** (1997) 1563.

[Lichti1999] R.L. Lichti, *in Hydrogen in Semiconductors, II*, ed. by N. Nickel (Academic Press, New York 1999), p.311.

[Leutkens2003] H. Luetkens *et al.*, Physical Review Letters **91** (2003) 017204.

[Leutkens2009] H. Leutkens *et al.*, Nature Materials **8** (2009) 305.

[Luke1989] G.M. Luke *et al.*, Nature **338** (1989) 49.

[Luke1991] G.M. Luke *et al.*, Physica C **185-189** (1991) 1175.

[Luke1993] G.M. Luke *et al.*, Physical Review Letters **71** (1993) 1466.

[Luke1998] G.M. Luke *et al.*, Physical Review Letters **80** (1998) 3843.

[Luke1998_2] G.M. Luke *et al.*, Nature **394** (1998) 558.

[Luke2000] G.M. Luke *et al.*, Physica B **289** (2000) 373.

[Maisuradze2010] A. Maisuradze *et al.,* Physical Review B **82** (2010) 024524.

[Mannix1999] D. Mannix, G. H. Lander and J. Rebizant, Physical Review B **60** (1999) 15187.

[Marsik2010] P. Marsik *et al.*, Physical Review Letters **105** (2010) 057001.

[Masek2007] J. Masek *et al.*, Physical Review Letters **98** (2007) 067202.

[Matsuda1997] M. Matsuda *et al.*, Physical Review B **55** (1997) 11953.

[McCollum2009] B.M. McCollum, *et al.*, . Chem. Eur. J. 15 (2009) 8409.

[Miller2002] R. I. Miller *et al.,* Physical Review Letters **88** (2002) 137002.

[Miller2006] R.I. Miller *et al.,* Physical Review B **73** (2006) 144509.

[Morenzoni1994] E. Morenzoni, et al., Physical Review Letters **72** (1994) 2794.

[Morenzoni2000] E. Morenzoni*, et. al.*, Physica **B289-290** (2000) 653.

[Morenzoni2008] E. Morenzoni *et al.*, Physical Review Letters **100** (2008) 147205.

[Morenzoni2011] E. Morenzoni *et al.*, Nature Communications **2** (2011) 272.

[Morenzoni2012] E. Morenzoni, private communication.

[Nakamura1993] K. Nakamura *et al.*, Solid State Ionics **121** (1999) 301.

[Niedermayer1993] Ch. Niedermayer *et al.*, Physical Review Letters **71** (1993) 1764.

[Niedermayer1998] Ch. Niedermayer *et al.* Physical Review Letters **80** (1998) 3843.

[Niedermayer1999] Ch. Niedermayer *et al.*, Physical Review Letters **83** (1999) 3932.

[Ofer2009] O. Ofer and A. Keren, Physical Review B **79** (2009) 134424.

[Ohishi2002] Ohishi K *et al*. Physical Review B **65** (2002) 140505.

[Olariu2006] A. Olariu *et al.*, Physical Review Letters **97** (2006) 167203.

[Osborne1953] D.W. Osborne and E.F. Westrum, J. Chem. Phys. **21** (1953) 1884.

[Park2009] J.T. Park *et al.*, Physical Review Letters **102** (2009) 117006.

[Patterson1988] B.D. Patterson, Reviews of Modern Physics **60** (1988) 69.

[Percival1987] P.W. Percival *et al.*, Chem. Phys. Lett. **133** (1987) 465.

[Percival1988] P.W. Percival, *et al.,* Chem. Phys. **127** (1988) 137.

[Percival2007] P.W. Percival, *et al.*, Rad. Phys. Chem., **76** (2007) 1231.

[Percival2011] P.W. Percival, *et al.,* Chem. Eur. J. **17** (2011) 11970.

[Percival2012] P.W. Percival et al., Organometallics, **31** (2012) 2709.

[Pitcher2010] M.J. Pitcher *et al.*, Journal of the American Chemical Society **132** (2010) 10467.

[Pratt2005] Pratt *et al.*, Physical Review B **72** (2005) 121401R.

[Pratt2011] F.L. Pratt *et al.*, Nature **471** (2011) 612.

[Price2002] A. N. Price *et al.,* Physical Review B **65** (2002) 214520.

[Qi2011] X.-L Qi and S.-C Zhang, Reviews of Modern Physics **83** (2011) 1057.

[Reid1987] I.D. Reid *et al.*, J. Chem. Phys. **86** (1987) 5578.

[Ricco2011] M. Ricco *et al*, Nano Letters **11** (2011) 4919.

[Roduner1978] E. Roduner *et al.,*Chem. Phys. Lett., **57** (1978) 37.

[Sanna2004] S. Sanna *et al*. Physical Review Letters **93** (2004) 207001.

[Sanna2011] S. Sanna *et al*. Physical Review Letters **107**, (2011) 227003.

[Santini2000] P. Santini and G. Amoretti, Physical Review Letters **85** (2000) 2188.

[Santini2006] P. Santini *et al*., Physical Review Letters **97** (2006) 207203.

[Savici2002] A.T. Savici Physical Review B **66** (2002) 014524.

[Savici2005] A.T. Savici *et al*., Physical Review Letters **95** (2005) 157001.

[Schulz2010] L. Schulz *et al*., Nature Materials **10** (2010) 39.

[Shay2009] M. Shay, Physical Review B **80** (2009) 144511.

[Shermandini2011] Z. Shermandini *et al*., Physical Review Letters **106** (2011) 117602.

[Shimizu2006] Y. Shimizu *et al*., Physical Review B **73** (2006) 140407R.

[Sonier2003] J.E. Sonier *et al*., Physical Review Letters **91** (2003) 147002.

[Sonier1994] J.E. Sonier *et al*., Physical Review Letters **72** (1994) 744.

[Sonier1999] J.E. Sonier *et al* Physical Review Letters **83** (1999) 4156.

[Sonier2000] J.E. Sonier, J.H. Brewer and R.F. Kiefl, Reviews of Modern Physics **72** (2000) 769.

[Sonier2000_2] J.E. Sonier *et al*., Physical Review B **61** (2000) R890.

[Sonier2004] J.E. Sonier *et al*., Physical Review Letters **93** (2004) 017002.

[Sonier2007] J.E. Sonier *et al*., Reports on Progress in Physics **70** (2007) 1717–1755.

[Sonier2007_2] J.E. Sonier *et al*, Physical Review B **76** (2007) 064522.

[Storchak1994] V. Storchak *et al*., Physical Review Letters **74** (1994) 3056.

[Storchak2008] V.G. Storchak, Physical Review Letters **101** (2008) 027202.

[Sugiyama2009] J. Sugiyama. Physical Review Letters **103** (2009) 147601.

[Sugiyama2010] J. Sugiyama *et al*., Physical Review B **81** (2010) 092103.

[Sugiyama2012] J. Sugiyama, Physical Review B **85** (2012) 054111.

[Suter2004] A. Suter *et al.*, Physical Review Letters **92** (2004) 087001.

[Suter2011] A. Suter *et al.*, Physical Review Letters **106** (2011) 237003.

[Takagiwa2002] H. Takaguwa *et al.*, Journal of the Physical Society of Japan **71** (2009) 31.

[Tomeno1998] I. Tomeno and M Oguchi, Journal of the Physical Society of Japan **67** (1998) 318.

[Uemura1991] Y.J. Uemura *et al.*, Physical Review Letters **66** (1991) 2665.

[Uemura1993] Y.J. Uemura *et al.*, Nature **364** (1993) 605.

[West2010] R. West and P.W. Percival. Dalton Transactions, **39** (2010) 9209.

[Wiedinger1989] A. Wiedinger *et al.*, Physical Review Letters **62** (1989) 102.

[Williams2009] T.J. Williams, Physical Review B **80** (2009) 094501.

[Yamashita2008] S. Yamashita *et al.*, Nature Physics **4** (2008) 459–462.

[Yaouanc1998] A. Yaouanc *et al*¸ Journal of Physics: Condensed Matter **10** (1998) 9791.

[Yu1990] D. Yu, *et al.,* Chem. Phys. **142**, 229.

[Zhao2011] S.R. Zhao *et al.*, Physical Review B **83** (2011) 180402.

[Zutic2004] I. Zutic *et al.*, Reviews of Modern Physics **76** (2004) 323.

# APPENDIX 1 – Accelerator Driven Systems

Since the early 1990's, accelerator driven systems (ADS) – subcritical assemblies driven by high power proton accelerators through a spallation target which is neutronically coupled to the core – have been proposed for addressing certain missions in advanced nuclear fuel cycles. There are several programs at laboratories around the world evaluating the role of ADS in nuclear waste transmutation and energy production. A summary of Accelerator Driven Systems history, technology and technical readiness can be found in [1].

Outside of the US, research into ADS for both transmutation and power generation has been accelerating. In 2001 the European Technical Working Group evaluated the state of ADS technologies and recommended the construction of an experimental ADS. In 2002 an expert group, convened by the Organization for Economic Cooperation and Development's Nuclear Energy Agency (OECD/NEA), authored a comprehensive report entitled *Accelerator Driven Systems (ADS) and Fast Reactors in Advanced Nuclear Fuel Cycles* [2]. In it, they conclude

"On the whole, the development status of accelerators is well advanced, and beam powers of up to 10 MW for cyclotrons and 100 MW for linacs now appear to be feasible. However, further development is required with respect to the beam losses and especially the beam trips to avoid fast temperature and mechanical stress transients in the reactor."

Technology demonstration has gained momentum with the Belgian government's announcement of its intention to construct MYRRHA [3], an 85-MW prototype ADS at the Belgian Nuclear Research Centre, SCK•CEN. The government has committed to finance 40% of the construction cost, and is preparing for a construction start in 2015. In addition, ADS Program plans have recently been formulated in China. These plans [4] call for a development program leading to a very high power accelerator driving a GW-scale subcritical core by the early 2030s. To reach this ambitious goal a dedicated ADS laboratory is being established. Likewise, the Indian government is considering construction of a prototype ADS facility at a similar scale of MYRRHA. ADS technology development programs exist in Europe, Japan, South Korea, India, China and Russia which are focused on both waste transmutation and power generation.

Accelerator Driven Systems may be employed to address several missions, including:

- Transmuting selected isotopes present in nuclear waste (e.g., actinides, fission products) to reduce the burden these isotopes place on geologic repositories.
- Generating electricity and/or process heat.
- Producing fissile materials for subsequent use in critical or sub-critical systems by irradiating fertile elements.

The principal advantages that accelerator-driven sub-critical systems have relative to critical reactors are twofold: greater flexibility with respect to fuel composition, and potentially enhanced safety. Accelerator driven systems are ideally suited to burning fuels which are problematic from the standpoint of critical reactor operation, namely, fuels that would degrade neutronic characteristics of the critical core to unacceptable levels due to small delayed neutron fractions and short neutron lifetimes, such as $^{233}U$ and minor actinide fuel. Additionally, ADS allows the use of non-fissile fuels (e.g.. Th) without the incorporation of U or Pu into fresh fuel. The enhanced safety of ADS is due to the fact that once the accelerator is turned off, the system shuts down. If the margin to critical is sufficiently large, reactivity-induced transients can never result in a super-critical accident with potentially severe consequences. Power control in accelerator-driven systems is achieved through the control of the beam current, a feature that can be utilized for fuel burnup compensation.

To date no country employs a fuel cycle that destroys the minor actinides (MA) present in used LWR fuel. Minor actinide destruction through transmutation is one mission that ADS are well suited to address. Unlike critical fast reactors which generally incorporate uranium or thorium in the fuel for safe operation, ADS can potentially operate on a pure MA feed stream, meaning a smaller number of ADS can be deployed to burn a fixed amount of minor actinides. ADS can recycle the MA multiple times until it is completely fissioned, such that the only actinide waste stream from these systems would derive from the recycling residuals, which could yield a significant reduction (by a factor of hundreds) in the amount of actinide waste per kW-hr of electricity generated, as compared to a once-through fuel cycle. Because accelerator driven systems do not require fuels containing uranium or thorium, they are more efficient at destroying MA waste – up to seven times more efficient according to one study [1] – than critical reactors, based on grams of minor actinides fissioned per MW-hr of energy generated.

A facility for transmutation of waste would also generate substantial power. An ADS generates high-quality process heat, can be operated at high temperature which could be utilized to produce another form of energy (e.g. biofuels or diesel fuel) or could be used to generate electrical power.

Several proposed ADS concepts with the goal of power production utilize thorium-based fuel to take advantage of some of its benefits, including greater natural abundance (3-4 times greater than uranium), proliferation resistance, and significantly reduced production of transuranics which are a major source of radiotoxicity and decay heat relative to uranium-based fuel. An ADS system based on Th fuel would not require incorporation of fissile material into fresh fuel, and could operate almost indefinitely in a closed fuel cycle.

An accelerator driven system consists of a high-power proton accelerator, a heavy-metal spallation target that produces neutrons when bombarded by the high-power beam, and a sub-critical core that is neutronically coupled to the spallation target. To achieve good neutronic coupling the target is usually placed at the center of a cylindrical core. The core consists of nuclear fuel, which may be liquid (e.g., molten salt) or solid as in conventional nuclear reactors.

ADS technology has evolved considerably since the last National Research Council study and report nearly two decades ago [5]. There have been several key advances in the last two decades, which make ADS a viable technology that is ready to proceed to the demonstration phase:

- The construction, commissioning and operation of a high-power continuous wave front-end system that meets the beam current performance required for up to 100 mA ADS accelerator system (the Low-Energy Demonstration Accelerator (LEDA) at Los Alamos)
- The construction, commissioning and MW-level operation with acceptable beam loss rates of a modern linear accelerator based on independently-phased superconducting accelerating structures (the Spallation Neutron Source at ORNL)
- The construction and deployment of a wide variety of pulsed and continuous-wave superconducting accelerating structures for proton/ion acceleration over a wide range in particle velocities, which is a key ingredient to achieving high reliability operation
- The high-power beam test of a liquid Pb-Bi eutectic spallation target loop at the Paul Scherrer Institute in Switzerland (the MEGAPIE project), and the operation of a MW-class liquid metal spallation target system at SNS.

Perhaps more important, recent analyses of subcritical reactor response to beam interruptions reveal greater tolerance to and therefore more relaxed requirements for beam trips, which had been a key criticism of previous ADS concepts.

The principal mission of Project X is particle physics. However, the continuous wave MW-class beam in the GeV energy range that is produced is precisely the beam which is needed to demonstrate and further develop key ADS technologies. Should the priorities for externally driven reactor technologies change in the US, Project X would be an ideal research and development platform.

An optimized Target Station would provide the flexibility for supporting key R&D with an emphasis on spallation neutron target and transmutation studies [6]. The R&D focus is on developing, demonstrating and verifying several critical aspects of neutron spallation target systems for ADS:

- Lead-bismuth target R&D including oxygen control, cleanup chemistry, safety in in-beam conditions
- Development and testing of windowless concepts
- Materials irradiation studies relevant to the ADS environment
- Characterization of neutron yield, spectra, spatial distributions, etc.

Fuel studies as described elsewhere in this report are also relevant to the ADS mission. The flagship experiment, which can be carried out with Project X, involves the transmutation of nuclear fuel coupled with reliable accelerator operation.

Finally, Project X provides a platform for the exploration and demonstration of key accelerator technology and accelerator physics solutions that are required for ADS. These include the development and demonstration of very high reliability accelerator operation through automated fault recovery, deployment of specialized diagnostic and control systems, accelerator-target coupling studies, and beam-loss control and mitigation.

**References**

[1] Accelerator and Target Technology for Accelerator Driven Transmutation and Energy Production: http://science.energy.gov/~/media/hep/pdf/files/pdfs/ADS_White_Paper_final.pdf

[2] "Accelerator Driven Systems (ADS) and Fast Reactors in Advanced Nuclear Fuel Cycles," 2002: http://www.nea.fr/ndd/reports/2002/nea3109.html

[3] H. Ait Abderrahim et. al., Nuclear Physics News, vol. 20, no. 1, 2010, p. 24.

[4] W.-L. Zhan , "ADS Programme and Key Technology R&D in China," Proc. 2013 International Particle Accelerator Conference, Shanghai, May 2013.

[5] "Nuclear Wastes: Technologies for Separations and Transmutation," National Research Council, National Academies Press, 1996.

[6] Y. Gohar, D. Johnson, T. Johnson, S. Mishra, Fermilab Project X Nuclear Energy Application: Accelerator, Spallation Target and Transmutation Technology Demonstration: https://indico.fnal.gov/getFile.py/access?resId=1&materialId=8&confId=3

## APPENDIX 2 – Project X Muon Spin Rotation Forum

A) The Project X Muon Spin Rotation Forum occurred at Fermilab October 17–19, 2012

B) Charge: The muon spin rotation (μSR) forum will engage μSR leaders in a discussion of current facilities, the current worldwide μSR science program, opportunities and requirements of a next generation facility and a discussion on how Project X could serve this community. In particular, opportunities for future bulk μSR studies with high energy (>4 MeV) and surface muon beams (4 MeV), and low energy (0-30 keV) beams will be discussed.

C) List of presentations and presenters:

October 17th, regular Fermilab colloquium:
'Muon Spin Rotation Spectroscopy-Utilizing Muons in Solid State Physics'
A. Suter (PSI)

October 18th , first day of forum:

Welcoming remarks,
P. Oddone (FNAL director)

'Current Plans and Future Vision of the Fermilab Accelerator Complex',
S. Henderson (FNAL Associate director for accelerators.

Discussion of PX Stage 1 and R&D Plan',
S. Holmes (FNAL, PX Project Manager)\

'Goals and Charge to the μSR Forum',
R. Tschirhart (FNAL)

'Overview of TRIUMF Facilities',
P. Percival (SFU/TRIUMF)

'Overview of PSI Facilities'
E. Morenzoni (PSI)
'Overview of ISIS Facilities',
S. Kilcoyne (University of Huddersfield)

'Possibilities with the Fermilab Muon Campus',
Chris Polly (FNAL)

'Towards a next generation muon facility',

R. Cywinski (University of Huddersfield)

‘Making the Case for a Muon Materials Science Facility at FNAL, A personal view.’,
R. Heffner (LANL)

‘μSR Studies of Superconductivity and Magnetism at a Next Generation Muon Source’,
G. Luke (McMaster University)

October 19th, second day of forum:

‘J-PARC muon beam facilities and slow muon production with laser resonant ionization’
Yasuhiro Miyake (KEK)

‘Science/Condensed Matter’
Robert Kiefl (TRIUMF)

‘Surface Muon Production at PSI’,
Andreas Suter (PSI), Daniela Kiselev (PSI)

‘Low Energy Muon Production by Moderation’,
Elvezio Morenzoni (PSI)

‘Low energy muon production estimates at Project X’,
Sergei Striganov (FNAL)

# APPENDIX 3 – DOE/BES Priorities in materials science

We present in this appendix, lists of known national priorities in DOE funded materials science. Areas of relevancy for muon science are in bold italic type.

**2008 DOE/BES Grand Challenges - from observational to control science**

- ***Control of materials' processes at electronic (quantum) level.***
- Design new forms of matter with tailored properties.
- ***Understand and control emergent, collective phenomena.***
- Master energy/information technology on ***nanoscale*** to rival living systems.
- ***Characterize and control matter very far from equilibrium.***

**2012 BESAC Subcommittee on Mesoscale Science**

- ***Master defect mesostructure and its evolution***.
- Optimize transport and response properties by design/control of mesoscale structure.
- ***Elucidate non-equilibrium and many-body physics of electrons.***
- Harness fluctuations and degradation for control of metastable ***mesoscale*** systems.
- Directing assembly of hierarchical functional materials.